\documentclass[%
 reprint,
superscriptaddress,
 amsmath,amssymb,
 aps,
]{revtex4-2}

\usepackage{graphicx}
\usepackage{dcolumn}
\usepackage{bm}
\usepackage{mathrsfs}

\newcommand{\bo}{\raise-1mm\hbox{\Large$\Box$}}
\newcommand{\expva}[1]{\langle #1 \rangle}

\newcommand{\UFPA}{\affiliation{Universidade Federal do Pará, Campus Salinópolis,
68721-000 Salinópolis, Pará, Brazil}}
\newcommand{\UFCG}{\affiliation{Unidade Acadêmica de Física, Universidade Federal de Campina Grande,
Rua Aprígio Veloso, 58429-900 Campina Grande, Paraíba, Brazil}}
\newcommand{\UESC}{\affiliation{Laborat\'orio de Astrof\'{\i}sica Te\'orica e Observacional, 
Departamento de Ci\^encias Exatas,  Universidade Estadual de
Santa Cruz, 45650-000 Ilh\'eus, Bahia, Brazil}}
\newcommand{\UFRB}{\affiliation{Centro de Ci\^encias Exatas e Tecnol\'ogicas, Universidade Federal
do Rec\^oncavo da Bahia, Rua Rui Barbosa 710, 44380-000 Cruz das Almas, Bahia, Brazil}}
\newcommand{\UFU}{\affiliation{Instituto de Física, Universidade Federal de Uberlândia,
Avenida João Naves de Ávila 2121, 38400-902 Uberlândia, Minas Gerais, Brazil}}
\newcommand{\UFABC}{\affiliation{Centro de Ci\^encias Naturais e Humanas, Universidade Federal
do ABC, Avenida dos Estados 5001, 09210-580 Santo Andr\'e, SP, Brazil}}

\begin{document}

\preprint{APS/123-QED}

\title{Third-order relativistic fluid dynamics at finite\\ density
in a general hydrodynamic frame}

\author{Saulo Diles}
\email{smdiles@ufpa.br}\UFPA\UFCG


\author{Alex S. Miranda}
\email{asmiranda@uesc.br}\UESC

\author{Luis A. H. Mamani}
\email{luis.mamani@ufrb.edu.br}\UFRB\UESC

\author{Alex M. Echemendia}
\email{alex.enchemendia@ufu.br}\UFU\UESC

\author{Vilson T. Zanchin}
\email{zanchin@ufabc.edu.br}\UFABC


\date{\today}

\begin{abstract}
The motion of water is governed by the Navier-Stokes equations, which are complemented by the continuity
equation to ensure local mass conservation. In this work, we construct the relativistic generalization of
these equations through a gradient expansion for a fluid with conserved charge in a curved $d$-dimensional
spacetime. We adopt a general hydrodynamic frame approach and introduce the Irreducible-Structure (IS)
algorithm, which is based on derivatives of both the expansion scalar and the shear and vorticity tensors.
By this method, we systematically generate all permissible gradients up to a specified order and derive the
most comprehensive constitutive relations for a charged fluid, accurate to third-order gradients.
These constitutive relations are formulated to apply to ordinary, non-conformal, and conformally invariant
charged fluids. Furthermore, we examine the hydrodynamic frame dependence of the transport coefficients for
a non-conformal charged fluid up to the third order in the gradient expansion. The frame dependence of the scalar,
vector, and tensor parts of the constitutive relations is obtained in terms of the field redefinitions of the
fundamental hydrodynamic variables. Managing these frame dependencies is challenging due to their non-linear
character. However, in the linear regime, these higher-order transformations become tractable, enabling the
identification of a set of frame-invariant coefficients. An advantage of employing these coefficients is the
possibility of studying the linear equations of motion in any chosen frame and,  hence, we apply this approach
to the Landau frame. Subsequently, these linear equations are solved in momentum space, yielding dispersion
relations for shear, sound, and diffusive modes for a non-conformal charged fluid, expressed in terms of the
frame-invariant transport coefficients.
\end{abstract}

\maketitle


\section{Introduction}

The modern understanding of electric charge incorporates the principle of gauge invariance inherent
in the equations of motion for the electromagnetic field, which ensures that the relativistic current
$J^\alpha = (\rho,\vec{J})$ is conserved: $\partial_\alpha J^\alpha=0$. This means that electric charge
is locally conserved and a change in its density at a local level can occur only due to current flow.
Naturally, most macroscopic distributions of matter exist in an electrically neutral, or uncharged, state.
However, when we examine the microscopic structure of matter, at the quantum scale, particles are in fact
represented by fermionic electrically charged fields. In the case of hadronic matter, these fermionic fields
are related to quarks as per the quark model \cite{Eichmann:2016yit}. It has been observed that the total
baryon number, which is defined by a balance between the total number of quarks and antiquarks, regardless
of flavors, is locally conserved \cite{Halzen:1984mc, FileviezPerez:2010gw}. In the early stages of the
development of Quantum Chromodynamics (QCD), this concept was termed heavy particle conservation
\cite{Wigner:1952ps, Lee:1955vk}.

A fermionic field, represented by $\psi$, defines a conserved current $J^\alpha = \bar{\psi}\gamma^\alpha \psi$
in such a way that $J^0$ is interpreted as a local charge density. The fact that electrons, as fermions, carry
electric charge blends the concepts of electric charge conservation and electron number conservation. Although
these two conservation laws are consistently observed in nearly all elementary particle processes,
they are fundamentally different; the former pertains to
the gauge symmetry of the electromagnetic field, while the latter is related to the fermionic nature of electrons.

In this paper, our objective is to establish the equations of motion for a fluid in the presence of conserved
charges associated with matter, typically referred to as charged fluids in the literature of relativistic hydrodynamics.
Specifically, we focus on relativistic theories characterized by a conserved energy-momentum tensor $T^{\alpha\beta}$
and a global $U(1)$ current $J^{\alpha}$. In the hydrodynamics nomenclature, fluids devoid of the corresponding $U(1)$
current, such as those arising from a $\phi^4$ field theory of a real scalar field $\phi$ or a $SU(N)$ Yang-Mills
theory of pure glue, are termed ``uncharged'', while the $|\phi^4|$ field theory of a complex scalar field
$\phi$ or quantum chromodynamics (QCD) result in ``charged fluids '', where the relevant $U(1)$ charge for
QCD is the baryon number \cite{Kovtun:2019hdm}.

The conservation of a matter current captures the essential physics of a diffusion process \cite{Landau:1987}.
The relevance of finite baryon density in the formation of a quark-gluon plasma (QGP) during heavy-ion collisions,
as well as the diffusion of heavy quarks in this medium \cite{He:2022ywp}, provides experimental reasons for
investigating diffusion processes in the context of relativistic hydrodynamics. Traditionally, diffusion has
been modeled through a constitutive phenomenological relation for the current, specifically by considering the
diffusive current as proportional to the concentration gradient, as stipulated by Fick's first law \cite{Fick:1855}, and incorporating this information into a continuity equation. Nowadays, the gradient expansion approach of hydrodynamics enhances this modeling method for diffusion by accounting for the influences of higher-order derivatives when characterizing the diffusive current.

In nonrelativistic fluid mechanics, our understanding of matter conservation finds its mathematical foundation in the continuity equation, which is nonrelativistic in nature but resembles the conservation of a relativistic vector current.
Transitioning to the relativistic formulation of fluid equations, the continuity equation governing matter density evolves into the zeroth component of the conservation equation for the relativistic energy-momentum tensor, which contains essential information about the content of matter, energy, momentum, and stress distributions within the fluid.
The Einstein theory of relativity establishes a connection between mass and energy, thereby significantly altering our understanding of hydrodynamics.

In a relativistic theory, one must adequately account for the momentum and energy conveyed by pure radiation,
which directly contributes to the energy-momentum tensor $T^{\alpha\beta}$. This tensor also represents the energy
and momentum carried by the matter content of the fluid. In addition to the energy-momentum tensor, the transport
of matter via the flow of a conserved matter species, such as the total number of baryons, must also be considered.
Information on the flow of matter is encoded in a vector current $J^\alpha$, which represents the diffusion of matter
and has its own constitutive relation. One of the purposes of this study is to formulate the gradient expansion of the
relativistic charge by providing a precise recipe to derive the general class of constitutive relations for both the
energy-momentum tensor and the conserved matter current.

In the gradient expansion approach, viscosity phenomena are seen as consequences of corrections in the constitutive
relations of an ideal fluid by a finite number of derivatives of its macroscopic degrees of freedom. The idea of
developing effective theories through a series of field derivatives has previously been utilized in various contexts
such as effective string theory \cite{Luscher:2004ib, Aharony:2009gg}, Lovelock gravity \cite{Lovelock:1971yv, Mardones:1990qc}, and Horndeski theory for dilaton-gravity systems \cite{Horndeski:1974wa}. In these examples,
the derivative expansion is implemented at the action level. However, there is generally no effective action principle
for viscous fluid dynamics. We then consider, in practice, series expansions of the conserved currents $T^{\alpha\beta}$
and $J^\alpha$. This concept of defining the theory of relativistic fluid mechanics through a gradient expansion
generalizes the Muller-Israel-Stewart (MIS) formalism \cite{Israel:1976tn,Israel:1979wp,Muller:1967zza}. The MIS
theory considers a construction of the entropy current of a fluid using up to second-order gradients and has
motivated the establishment of a second-order and, more recently, third-order theory of hydrodynamics in the gradient
expansion scheme \cite{Romatschke:2009kr, Betz:2009zz, Grozdanov:2015kqa, Diles:2019uft}.
The introduction of second-order gradients in MIS theory was
deemed necessary to resolve the causality problem that arose in the first-order formulation. Fortunately, a new strategy
for formulating the theory of relativistic hydrodynamics has emerged recently, capable of achieving a causal and stable
first-order hydrodynamic theory, known as General Frame Hydrodynamics or Benfica-Disconzi-Noronha-Kovtum (BDNK) theory \cite{Bemfica:2017wps,Bemfica:2019knx,Bemfica:2020zjp,Kovtun:2019hdm,Hoult:2020eho}. In the current discussion, we
examine the third-order gradient expansion of a charged fluid in light of the general hydrodynamic frame approach.

This work is structured as follows. In Section \ref{non-conformal}, we focus on the general case of charged nonconformal
fluids in $d$-dimensional spacetimes. We employ a systematic approach based on two computational algorithms to expand the energy-momentum tensor and the current to the third order in the gradient expansion. Within this context, we identify the
set of hydrodynamic frame-invariant coefficients that are retained during the linearization process. In Section
\ref{conformal-fluid}, we investigate charged conformal fluids, implementing conformal symmetry through minimal
coupling. This approach enables us to directly derive third-order conformal corrections. Section \ref{dispersion_relations}
is dedicated to the study of linearized fluctuations of the charged fluid, leading to the determination of the dispersion relations in terms of a reduced set of hydrodynamic frame-invariant transport coefficients for nonconformal fluids.
Conclusions and final remarks are presented in Section \ref{conclusions}. Supplementary materials are discussed in
Appendices \ref{sec:D_thermo_vairables} and \ref{equiv_symmetry}. An alternative derivation of the dispersion relations
in a general hydrodynamic frame can be found in the Appendix \ref{general-frame-derivation}.

\section{Non-conformal charges and gradient expansion}
\label{non-conformal}

The comprehension of hydrodynamics as an effective theory for low frequencies and long wavelength modes in any given
field theory forms the foundation of the gradient expansion approach \cite{Kovtun:2012rj}. Fluid dynamics is considered
a macroscopic representation of a system of many (quantum) interacting particles, and we expect the underlying symmetries
of particle interactions to be manifest in macroscopic dynamics. This macroscopic manifestation of symmetries is
incorporated in the gradient expansion formulation, where the conservation of global currents supplies the equations
of motion for macroscopic quantities.

In Refs.~\cite{Grozdanov:2015kqa, Diles:2019uft}, the hydrodynamic gradient expansion has been explored in detail up
to the third order for neutral (uncharged) fluids, which means those without any conserved matter current, for both
conformal and ordinary (nonconformal) systems. Uncharged fluids exemplify a system devoid of matter fields, which in
any microscopic theory signifies the absence of fermions in the fundamental representation or the absence of quarks
for particular cases of QCD-like theories.

The standard model of particle physics includes six quark flavors in a single conserved current, the baryon current.
All these quark flavors are produced during high-energy collisions and play a role in the QGP droplet. Even at lower
energies, the presence of the three lighter quark flavors is observed. In addition to quarks, there are leptons with
an associated conserved current. Hence, the potential existence of a range of distinct conserved matter charges in
the gradient expansion warrants consideration.

For the sake of simplicity, throughout the majority of this study, we will make the assumption that there is only
one conserved matter current, associated with a single chemical potential. Discussions regarding the extension of
our results to encompass an arbitrary number of conserved charges will be addressed whenever appropriate.

\subsection{Ideal fluids and the fundamental degrees of freedom}

Ideal fluid dynamics establishes the zeroth-order theory in the gradient expansion framework and characterizes
configurations in thermodynamic equilibrium. The relativistic equations governing a perfect fluid are defined by
the vanishing divergence of the ideal fluid energy-momentum tensor and the vanishing divergence of its matter current:
\begin{equation}
\nabla_\beta T^{\alpha\beta}_{\text{ideal}} = 0, \qquad\quad
\nabla_{\alpha} J^{\alpha}_{\text{ideal}} =0.
\label{ideal_equations}
\end{equation}
The energy-momentum tensor is the same as that of a neutral fluid, given by
\begin{equation}
T^{\alpha\beta}_{\text{ideal}} = \epsilon u^{\alpha}u^{\beta} + p\Delta^{\beta},
\label{Tideal}
\end{equation}
where $\epsilon$ and $p$ are respectively the local energy density and pressure, and
$\Delta^{\alpha\beta} = g^{\alpha\beta}+u^{\alpha}u^{\beta}$ is the projector onto the
hypersurface orthogonal to the vector field $u^{\alpha}$ corresponding to the flow
velocity of a fluid element. The matter current is expressed as
\begin{equation}
J^{\alpha}_{\text{ideal}} = n u^{\alpha},
\label{Jideal}
\end{equation}
with $n$ representing the number density of the matter species. From a quantum point of view,
$n = n(x^{\alpha})$ is the expected value of a number operator averaged over a small volume in the
vicinity of the point $x^{\alpha}$.

The foregoing equations for ideal fluids should be considered as approximations, since they undergo corrections through
higher-order derivative terms and establish relationships between first-order derivatives of the fundamental gradients.
In practice, these ideal fluid equations allow the elimination of the longitudinal derivatives of the mechanical and thermodynamic degrees of freedom. This is achieved by the following process. The divergenceless condition of the
energy-momentum tensor yields a vector equation that can be decomposed into longitudinal and transverse components.
The former can be used to replace the longitudinal derivative of temperature (or entropy) by the divergence of
$u^{\alpha}$, the expansion, which in turn equals the transverse divergence of the velocity. The latter allows us to
replace the longitudinal derivative of velocity with a transverse derivative of temperature (or entropy). Consequently,
only transverse derivatives of temperature and velocity account for the gradient expansion of the uncharged fluid.

In the presence of one global charge, there is an additional scalar degree of freedom,
namely the number density $n$ or equivalently, its chemical potential $\mu$, along with an additional equation:
the second equation in \eqref{ideal_equations}. Employing the constitutive relation of the ideal fluid, as given
by \eqref{Jideal}, we obtain
\begin{equation}
Dn = - n\Theta,
\label{derivada_temporal_n}
\end{equation}
where the longitudinal derivative operator $D$ and the expansion $\Theta$ are, respectively,
defined by $D\equiv u^{\alpha}\nabla_{\alpha}$ and $\Theta \equiv \nabla_{\alpha} u^{\alpha}$.

The longitudinal and transverse projections of the conservation of the ideal energy-momentum tensor yield
\begin{equation}
D \epsilon = -h \Theta, \qquad\quad
D u^\alpha = - \frac{1}{h}\nabla_{\perp}^\alpha p,
\label{eq_adicionais_ordem_zero}
\end{equation}
where $h = \epsilon + p$ denotes the enthalpy density of the fluid, and
$\nabla_{\perp}^{\alpha} \equiv\Delta^{\alpha\beta} \nabla_{\beta} $ represents the transverse derivative operator.

These equations of motion for the ideal fluid map the longitudinal derivatives of $\epsilon$, $n$ and $u^\alpha$
into the transverse derivatives of $p$ and $u^\alpha$. Taking into account that $\epsilon = \epsilon (T,\mu)$,
$n = n(T,\mu)$, and $p = p(T,\mu)$, the equations effectively map the longitudinal derivatives of $T$, $\mu$ and
$u^\alpha$ to their respective transverse derivatives (see Appendix \ref{sec:D_thermo_vairables} for further details).
In the gradient expansion framework, one has the freedom to select any pair of thermodynamic variables that are not
canonically conjugated to each other. In most of this work, we use the temperature $T$ and the chemical potential $\mu$.

The proposition demonstrated in \cite{Grozdanov:2015kqa} for neutral fluids remains valid in the case of charged fluids,
and only transverse derivatives of $T, \mu, u^\alpha$ are present in the gradient expansion. The generalization of this proposition for a multicomponent fluid is straightforward: for each additional density $n_k$, there is a corresponding continuity equation $Dn_k = - n_k \Theta$, and only transverse gradients of $n_k$ will be present in the (on-shell)
gradient expansion. Subsequently, we can replace densities with chemical potentials employing an analogous procedure
to that outlined in the Appendix \ref{sec:D_thermo_vairables}, with the exception that each new component increases
the dimension of the relevant linear system by one.

The possibility of eliminating the longitudinal derivatives of all dynamical degrees of freedom, with the exception
of geometric ones, considerably simplifies the construction of the gradient expansion, as will become evident in
the subsequent discussion. Here, we will use the following set of fundamental degrees of freedom: temperature $T$,
chemical potential $\mu$, velocity $u^\mu$, and metric $g^{\mu\nu}$. We confine our analysis to torsion-free
geometries, where the connection is given by the Christoffel symbols and does not entail independent degrees of freedom.

A crucial aspect in constructing a gradient expansion is to ensure that the symmetries of the system are preserved.
For a relativistic fluid in curved space, only structures that are covariant under diffeomorphisms are permitted,
and we seek the simplest covariant derivatives of the fundamental degrees of freedom. In $d$ spacetime dimensions,
the $d$ independent components of the gradients of the scalars, $\nabla_\alpha T$ and $\nabla_\alpha \mu$, are not
all inequivalent, as their longitudinal derivatives can be eliminated. To represent this, we decompose the covariant
derivative into longitudinal and transverse components, $\nabla_\alpha = u_\alpha D + \nabla_{\perp \alpha}$.
Consequently, only the transverse derivatives $\nabla_{\perp \alpha} T$, $\nabla_{\perp \alpha} \mu$, and
$\nabla_{\perp \alpha} u^\beta$ appear in the gradient expansion. These transverse gradients vanish identically
upon contraction with the velocity field, rendering only contractions with the metric relevant.

\subsection{Hydrodynamic frames and on-shell equivalences}

The construction of a relativistic theory for viscous fluids in the presence of a conserved current is well
discussed by Landau and Lifshitz in their seminal book \cite{Landau:1987}. The gradient expansion up to the
$n$th order may be written in the form
\begin{equation}
T^{\alpha \beta} = T^{\alpha \beta}_{\text{ideal}} + \sum_{i=1}^n\Pi^{\alpha\beta}_{\scriptscriptstyle{(i)}}(\partial^i),
\quad J^\alpha = J^{\alpha}_{\text{ideal}} + \sum_{i=1}^n\Upsilon^{\alpha}_{\scriptscriptstyle{(i)}}(\partial^i),
\label{T_J_decomposition}
\end{equation}
where $\partial^i$ represents any combination involving the $i$th-order derivatives of the fundamental degrees of freedom.

For the progression of the work, it is instructive to introduce the transverse, symmetric, and traceless (TST) part of a general second-rank tensor $A^{\alpha\beta}$, denoted as $A^{\expva{\alpha\beta}}$. This is defined through the projection operator $\Delta^{\alpha\beta}$ as follows:
\begin{equation}
A^{\expva{\alpha\beta}}= \Delta^{\alpha\gamma}\Delta^{\beta\delta}A_{(\gamma\delta)}
- \frac{1}{d-1}\Delta^{\alpha\beta}\Delta^{\gamma\delta}A_{\gamma\delta},
\end{equation}
where the parentheses in $A_{(\gamma\delta)}$ indicate the operation of symmetrization,
\begin{equation}
A_{(\gamma\delta)} = \frac{1}{2}(A_{\gamma\delta}+A_{\delta\gamma}).
\end{equation}

To implement the derivative expansion in practice, we decompose the corrections to the energy-momentum
tensor and to the matter current into their most general form:
\begin{equation}
\begin{split}
\Pi^{\alpha\beta}_{\scriptscriptstyle{(i)}} & = \mathcal{E}_{\scriptscriptstyle{(i)}}u^\alpha u^\beta +
\mathcal{P}_{\scriptscriptstyle{(i)}}\Delta^{\alpha \beta} + \mathcal{Q}_{\scriptscriptstyle{(i)}}^\alpha u^\beta
+ \mathcal{Q}_{\scriptscriptstyle{(i)}}^\beta u^\alpha + \tau_{\scriptscriptstyle{(i)}}^{\alpha\beta},\\
\Upsilon_{\scriptscriptstyle{(i)}}^\alpha & = \mathcal{N}_{\scriptscriptstyle{(i)}}u^\alpha +
\mathcal{J}_{\scriptscriptstyle{(i)}}^\alpha,
\end{split}
\end{equation}
where $\mathcal{E}_{\scriptscriptstyle{(i)}}$, $\mathcal{P}_{\scriptscriptstyle{(i)}}$,
$\mathcal{N}_{\scriptscriptstyle{(i)}}$ are scalars, $\mathcal{Q}_{\scriptscriptstyle{(i)}}^\alpha$ and $\mathcal{J}_{\scriptscriptstyle{(i)}}^\alpha$ are vectors transverse to the fluid velocity $u^\alpha$,
and $\tau_{\scriptscriptstyle{(i)}}^{\alpha\beta}$ are TST tensors.

The general expressions for the conserved energy-momentum tensor and current can then be written as:
\begin{equation}
\begin{split}
    T^{\alpha\beta} & = \mathcal{E}u^\alpha u^\beta +  \mathcal{P} \Delta^{\alpha \beta} +
    \mathcal{Q}^\alpha u^\beta+ \mathcal{Q}^\beta u^\alpha + \tau^{\alpha \beta},\\
    J^\alpha & = \mathcal{N}u^\alpha + \mathcal{J}^\alpha,
\end{split}
\label{general_TJ}
\end{equation}
where
\begin{equation}
\begin{aligned}
\mathcal{E} & = \epsilon + \sum_{i=1}^n\mathcal{E}_{\scriptscriptstyle{(i)}}, &
\mathcal{P} & = p + \sum_{i=1}^n\mathcal{P}_{\scriptscriptstyle{(i)}}, &
\mathcal{Q}^\alpha & = \sum_{i=1}^n\mathcal{Q}_{\scriptscriptstyle{(i)}}^{\alpha},\\
\mathcal{N} & = n + \sum_{i=1}^n\mathcal{N}_{\scriptscriptstyle{(i)}}, &
\mathcal{J}^\alpha & = \sum_{i=1}^n\mathcal{J}_{\scriptscriptstyle{(i)}}^{\alpha}, &
\tau^{\alpha \beta} & = \sum_{i=1}^n \tau_{\scriptscriptstyle{(i)}}^{\alpha\beta}.
\end{aligned}
\end{equation}
In the case of multicomponent fluids, there is an additional scalar, $\mathcal{N}_k$, and an additional
vector current, $\mathcal{J}_{k}^\alpha$, for each new fundamental component with equilibrium density $n_k$.

The corrections to the ideal fluid equations represent slight deviations from thermodynamic equilibrium,
altering the notion of variables that parameterize the equilibrium state, such as energy and number density.
To precisely define the equilibrium quantities, it is essential to select a hydrodynamic frame. In the context
of the gradient expansion, fixing a hydrodynamic frame entails defining the thermodynamic quantities (see, e.g., \cite{Kovtun:2019hdm}), and this should not be confused with the choice of a frame of reference in the spacetime.

In our previous work \cite{Diles:2019uft}, the Landau(-Lifshitz) frame was utilized to simplify the formulation
of the third-order gradient expansion for neutral (uncharged) fluids. For a charged fluid, the Landau frame is
characterized by the following conditions:
\begin{equation}
u_\beta \Pi^{\alpha\beta}=0, \qquad\quad u_\alpha\Upsilon^\alpha=0.
\label{Landau}
\end{equation}
These conditions imply that the projection of the energy-momentum tensor along the fluid velocity defines the
equilibrium energy, while the projection of the current defines the equilibrium baryon density. In this frame,
all corrections are transverse; consequently, there are neither heat flow corrections in the energy-momentum
tensor nor scalar corrections to the baryon density in the matter current.

Another common choice in the context of charged fluids is the Eckart frame \cite{Eckart:1940te}, which was employed
in \cite{Lahiri:2019lpk} for the second-order gradient expansion of a nonconformal charged fluid. In the Eckart frame,
the charge flow in the local rest frame of the fluid is absent, which is represented as $\mathcal{J}^{\alpha} = 0$.
The defining properties of this frame are complemented by $\mathcal{E} = \mathcal{N} = 0$. With respect to the total
number of transport coefficients, the Landau frame proves to be equivalent to the Eckart frame. This equivalence is
supported by the observation that, in both frames, the gradient expansion includes one set of scalars, one set of
transverse vectors, and one set of TST tensors for each conserved charge \cite{Monnai:2019jkc}. Recent studies on
relativistic local thermodynamic equilibrium (LTE) have established the equivalence between the ``energy states''
associated with the Landau frame and the ``particle states'' associated with the Eckart frame
(see \cite{Salazar:2022yud} and references therein). 

A discussion of fluid dynamics in an arbitrary frame has recently emerged as a means of addressing the
long-standing issue of stability in first-order relativistic hydrodynamics and has been intensively developed since
its inception \cite{Bemfica:2017wps,Bemfica:2019knx,Bemfica:2020zjp,Kovtun:2019hdm,Hoult:2020eho,Taghinavaz:2020axp,
Noronha:2021syv,Hoult:2021gnb,Heller:2022ejw}. The basic idea of general frame hydrodynamics (GFH) involves initially formulating the most comprehensive energy-momentum tensor and gauge currents, and then writing the equations of motion by taking into account all the transport coefficients. The challenge of stability is assessed within this extensive parameter space, where stable regions can be identified, and only after this step does one select a hydrodynamic frame, thereby constraining this large set of transport coefficients to a more concise set of independent ones. This discussion has implications for the construction of the gradient expansion, which also needs to be firmly established in the context of GFH.
  
To investigate the frame dependence of the coefficients in \eqref{general_TJ}, we extend the analysis of \cite{Kovtun:2019hdm} from the first to the $n$th order in the gradient expansion. Following Ref.~\cite{Israel:1976tn}, we consider a generalized frame transformation (or redefinition) of the thermodynamic degrees of freedom:
\begin{equation}
u'^{\alpha}=Au^{\alpha}+\bar{\delta} u^\alpha, \quad T'=T+\bar{\delta} T,\quad \mu'=\mu + \bar{\delta}\mu,
\end{equation}
where $\bar{\delta} T = \sum_{i=1}^n \bar{\delta} T_{\scriptscriptstyle{(i)}}(\partial^i)$,
$\bar{\delta} \mu = \sum_{i=1}^n\bar{\delta}\mu_{\scriptscriptstyle{(i)}}(\partial^i)$ and the vector
$\bar{\delta} u^\alpha = \sum_{i=1}^n\bar{\delta} u_{\scriptscriptstyle{(i)}}^\alpha(\partial^i)$ is
orthogonal to $u^\alpha$, such that $u_\alpha\bar{\delta} u_{\scriptscriptstyle{(i)}}^\alpha = 0$ for
all values of $i$. In the equilibrium state, the transformed quantities $u'^{\alpha}$, $T'$, and $\mu'$
coincide with the original variables. Due to the normalization condition $u'_{\alpha}u'^{\alpha} = -1$,
the scalar function $A$ is given by
\begin{equation}
A = (1+\bar{\delta}^2 u)^{1/2}
\end{equation}
with $\bar{\delta}^2 u \equiv \bar{\delta} u^\alpha\bar{\delta} u_\alpha$.

Both the energy-momentum tensor and the current remain invariant under
the transformation to the new (primed) variables \cite{Kovtun:2012rj}, which means
\begin{equation}
\begin{split}
T^{\alpha\beta}(u'^\gamma, T',\mu') & = T^{\alpha\beta}(u^\gamma, T,\mu)\\
J^\alpha(u'^\gamma, T',\mu') & = J^\alpha(u^\gamma, T,\mu).
\end{split}
\end{equation}
The transformed (primed) and original (unprimed) coefficients in the decomposition
\eqref{general_TJ} are related as follows:
\begin{equation}
\mathcal{E}' = \mathcal{E} + B,\quad
\mathcal{P}' = \mathcal{P} + \frac{B}{d-1},\quad
\mathcal{N}' = A\,\mathcal{N} - \mathcal{J}_{\alpha}\bar{\delta} u^{\alpha},
\label{relacoes_escalares}
\end{equation}
\begin{equation}
\begin{split}
\mathcal{Q}'^\alpha = A\mathcal{Q}^\alpha - \mathcal{H}\bar{\delta} u^\alpha
- (u^\alpha \mathcal{Q}_{\beta} & + \tau^{\alpha}_{\;\;\beta})\bar{\delta} u^\beta\\
& - B\left(A u^\alpha + \bar{\delta} u^\alpha\right),
\end{split}
\label{relacao_vetorial1}
\end{equation}
\begin{equation}
\begin{split}
\mathcal{J}'^\alpha  = \mathcal{J}^\alpha - \bar{\delta} u^\alpha(A\,\mathcal{N}
& - \mathcal{J}^{\beta}\bar{\delta} u_{\beta})\\
& + u^\alpha \left(A\,\mathcal{J}^\beta - \mathcal{N}\bar{\delta} u^\beta\right)\bar{\delta} u_\beta,
\end{split}
\label{relacao_vetorial2}
\end{equation}
\begin{equation}
\begin{gathered}
\tau'^{\alpha\beta} = \tau^{\alpha\beta} + 2\mathcal{Q}^{(\alpha}u^{\beta)} 
+ \mathcal{H}\bar{\delta} u^\alpha \bar{\delta} u^\beta +
\tau^{\mu\nu}\bar{\delta} u_{\mu}\bar{\delta} u_{\nu}\,u^\alpha u^\beta \\
- 2\big[A\mathcal{Q}^{(\alpha} - \bar{\delta} u_{\mu}\tau^{\mu(\alpha}\big]\big(Au^{\beta)}
+ \bar{\delta} u ^{\beta)}\big)
+ 2\mathcal{Q}^{\mu}\bar{\delta} u_{\mu} u^{(\alpha}\bar{\delta} u^{\beta)}\\
- B\frac{\Delta^{\alpha\beta}}{d-1} - B\frac{d-2}{d-1}\left[u^\alpha u^\beta -
(A u^{\alpha}+\bar{\delta} u^\alpha)(A u^{\beta} + \bar{\delta} u^\beta)\right],
\end{gathered}
\label{relacao_tensorial}
\end{equation}
where $\mathcal{H}\equiv\mathcal{E}+\mathcal{P}$,
$B\equiv \mathcal{H}\,\bar{\delta}^2 u - 2A \mathcal{Q}^\alpha\bar{\delta} u_\alpha
+ \tau^{\alpha\beta}\bar{\delta} u_\alpha \bar{\delta} u_\beta$ and the above equations
should be considered, order by order, in the gradient expansion. As discussed in detail
by Kovtun \cite{Kovtun:2019hdm} for first-order hydrodynamics, the most general field
redefinition is given by gradient expansions of $\bar{\delta} T_{\scriptscriptstyle{(i)}}$,
$\bar{\delta} u_{\scriptscriptstyle{(i)}}^\alpha$, and $\bar{\delta}\mu_{\scriptscriptstyle{(i)}}$.
Denoting $N_{\scriptscriptstyle{\mathcal{S}}}^{\scriptscriptstyle{(i)}}$ and $N_{\scriptscriptstyle{\mathcal{V}}}^{\scriptscriptstyle{(i)}}$ as the number of
independent scalar and vector structures of the $i$th order, these expansions are expressed as
\begin{equation}
\begin{split}
\bar{\delta} T_{\scriptscriptstyle{(i)}} & = \sum_{j=1}^{N_{\scriptscriptstyle{\mathcal{S}}}^{\scriptscriptstyle{(i)}}}
a_{j}^{\scriptscriptstyle{(i)}}\mathcal{S}_{j}^{\scriptscriptstyle{(i)}},\\
\bar{\delta} u_{\scriptscriptstyle{(i)}}^\alpha & = \sum_{j=1}^{N_{\scriptscriptstyle{\mathcal{V}}}^{\scriptscriptstyle{(i)}}}
b_{j}^{\scriptscriptstyle{(i)}}(\mathcal{V}_{j}^{\scriptscriptstyle{(i)}})^{\alpha},\\
\bar{\delta}\mu_{\scriptscriptstyle{(i)}} & = \sum_{j=1}^{N_{\scriptscriptstyle{\mathcal{S}}}^{\scriptscriptstyle{(i)}}}
c_{j}^{\scriptscriptstyle{(i)}}\mathcal{S}_{j}^{\scriptscriptstyle{(i)}},
\end{split}
\label{expansions_delta}
\end{equation}
where the coefficients $a_{j}^{\scriptscriptstyle{(i)}}$, $b_{j}^{\scriptscriptstyle{(i)}}$, and $c_{j}^{\scriptscriptstyle{(i)}}$ are unspecified functions of $T$ and $\mu$ that are to be chosen
during the frame-fixing process.

Since our focus is on the gradient expansion up to the third order, we expand the relations
\eqref{relacoes_escalares}--\eqref{relacao_tensorial}, retaining terms of the third order and below:
\begin{equation}
\begin{aligned}
\mathcal{E}' & = \epsilon + \mathcal{E}_{\scriptscriptstyle{(1)}}
+\Big[\mathcal{E}_{\scriptscriptstyle{(2)}} - 2 \Big(\mathcal{Q}_{\scriptscriptstyle{(1)}}^\alpha - \frac{1}{2}h \bar{\delta} u_{\scriptscriptstyle{(1)}}^\alpha\Big)\bar{\delta} u_{{\scriptscriptstyle{(1)}}\alpha}\Big]\\
&+\Big[\mathcal{E}_{\scriptscriptstyle{(3)}}- 2\Big(\mathcal{Q}_{\scriptscriptstyle{(1)}}^\alpha - h\bar{\delta} u_{\scriptscriptstyle{(1)}}^\alpha\Big)\bar{\delta} u_{{\scriptscriptstyle{(2)}}\alpha}\\
&-2\Big(\mathcal{Q}_{\scriptscriptstyle{(2)}}^\alpha
-\frac{1}{2}\mathcal{H}_{\scriptscriptstyle{(1)}} \bar{\delta} u_{\scriptscriptstyle{(1)}}^\alpha
-\frac{1}{2} \tau_{\scriptscriptstyle{(1)}}^{\alpha\beta}\bar{\delta} u_{{\scriptscriptstyle{(1)}}\beta}\Big)\bar{\delta} u_{{\scriptscriptstyle{(1)}}\alpha}\Big],
\end{aligned}
\label{rel_transf_para_E}
\end{equation}
\begin{equation}
\begin{aligned}
\mathcal{P}' & = p + \mathcal{P}_{\scriptscriptstyle{(1)}}
+ \Big[\mathcal{P}_{\scriptscriptstyle{(2)}} - \frac{2}{d-1} \Big(\mathcal{Q}_{\scriptscriptstyle{(1)}}^\alpha
- \frac{1}{2}h \bar{\delta} u_{\scriptscriptstyle{(1)}}^\alpha\Big)\bar{\delta}
u_{{\scriptscriptstyle{(1)}}\alpha}\Big]\\
&+\Big[\mathcal{P}_{\scriptscriptstyle{(3)}} - \frac{2}{d-1} \Big\{\Big(\mathcal{Q}_{\scriptscriptstyle{(1)}}^\alpha
- h\bar{\delta} u_{\scriptscriptstyle{(1)}}^\alpha\Big)\bar{\delta} u_{{\scriptscriptstyle{(2)}}\alpha}\\
&+\Big(\mathcal{Q}_{\scriptscriptstyle{(2)}}^\alpha-\frac{1}{2}\mathcal{H}_{\scriptscriptstyle{(1)}}
\bar{\delta} u_{\scriptscriptstyle{(1)}}^\alpha -\frac{1}{2} \tau_{\scriptscriptstyle{(1)}}^{\alpha\beta}\bar{\delta} u_{{\scriptscriptstyle{(1)}}\beta}\Big)\bar{\delta} u_{{\scriptscriptstyle{(1)}}\alpha}\Big\}\Big],
\end{aligned}
\label{rel_transf_para_P}
\end{equation}
\begin{equation}
\begin{aligned}
\mathcal{N}' & = n + \mathcal{N}_{\scriptscriptstyle{(1)}}
+ \Big[\mathcal{N}_{\scriptscriptstyle{(2)}} - \Big(\mathcal{J}_{\scriptscriptstyle{(1)}}^\alpha - \frac{1}{2}n \bar{\delta} u_{\scriptscriptstyle{(1)}}^\alpha\Big)\bar{\delta} u_{{\scriptscriptstyle{(1)}}\alpha}\Big]\\
&+\Big[\mathcal{N}_{\scriptscriptstyle{(3)}} - \Big(\mathcal{J}_{\scriptscriptstyle{(1)}}^\alpha - n\bar{\delta} u_{\scriptscriptstyle{(1)}}^\alpha\Big)\bar{\delta} u_{{\scriptscriptstyle{(2)}}\alpha} \\
& -\Big(\mathcal{J}_{\scriptscriptstyle{(2)}}^\alpha-\frac{1}{2}\mathcal{N}_{\scriptscriptstyle{(1)}}
\bar{\delta} u_{\scriptscriptstyle{(1)}}^\alpha\Big)\bar{\delta} u_{{\scriptscriptstyle{(1)}}\alpha}\Big],
\end{aligned}
\label{rel_transf_para_N}
\end{equation}
\begin{equation}
\begin{aligned}
\mathcal{Q}'^\alpha & = \big(\mathcal{Q}_{\scriptscriptstyle{(1)}}^\alpha - h\bar{\delta} u_{\scriptscriptstyle{(1)}}^\alpha\big) + \big(\mathcal{Q}_{\scriptscriptstyle{(2)}}^\alpha - h\bar{\delta} u_{\scriptscriptstyle{(2)}}^\alpha\big)-\Big\{\mathcal{H}_{\scriptscriptstyle{(1)}}\Delta^{\alpha\beta}\\
& + \tau_{\scriptscriptstyle{(1)}}^{\alpha\beta} - u^\alpha\big(\mathcal{Q}_{\scriptscriptstyle{(1)}}^\beta -
h\bar{\delta} u_{\scriptscriptstyle{(1)}}^\beta\big)\Big\}\bar{\delta} u_{{\scriptscriptstyle{(1)}}\beta}
+ \Big[\big(\mathcal{Q}_{\scriptscriptstyle{(3)}}^\alpha - h\bar{\delta} u_{\scriptscriptstyle{(3)}}^\alpha\big)\\
&-\Big(\mathcal{H}_{\scriptscriptstyle{(2)}}\Delta^{\alpha\beta} + \tau_{\scriptscriptstyle{(2)}}^{\alpha\beta}
- \frac{1}{2}\mathcal{Q}_{\scriptscriptstyle{(1)}}^\alpha
\bar{\delta} u_{\scriptscriptstyle{(1)}}^\beta\Big)\bar{\delta} u_{{\scriptscriptstyle{(1)}}\beta}
- \big(\mathcal{H}_{\scriptscriptstyle{(1)}}\Delta^{\alpha\beta}\\
& +\tau_{\scriptscriptstyle{(1)}}^{\alpha\beta}\big)\bar{\delta} u_{{\scriptscriptstyle{(2)}}\beta}
+ 2\bar{\delta} u_{\scriptscriptstyle{(1)}}^\alpha\Big(\mathcal{Q}_{\scriptscriptstyle{(1)}}^\beta
-\frac{1}{2}h\bar{\delta} u_{\scriptscriptstyle{(1)}}^\beta\Big)
\bar{\delta} u_{{\scriptscriptstyle{(1)}}\beta}\\
&- u^\alpha\big(B_{\scriptscriptstyle{(3)}} + \mathcal{Q}_{\scriptscriptstyle{(2)}}^\beta
\bar{\delta} u_{{\scriptscriptstyle{(1)}}\beta} + \mathcal{Q}_{\scriptscriptstyle{(1)}}^\beta
\bar{\delta} u_{{\scriptscriptstyle{(2)}}\beta}\big)\Big],
\end{aligned}
\label{rel_transf_para_Q}
\end{equation}
\begin{equation}
\begin{aligned}
\mathcal{J}'^\alpha &= \big(\mathcal{J}_{\scriptscriptstyle{(1)}}^\alpha
- n\bar{\delta} u_{\scriptscriptstyle{(1)}}^\alpha\big) + \big(\mathcal{J}_{\scriptscriptstyle{(2)}}^\alpha
- n\bar{\delta} u_{\scriptscriptstyle{(2)}}^\alpha\big)
- \Big\{\mathcal{N}_{\scriptscriptstyle{(1)}}\Delta^{\alpha\beta}\\
&- u^\alpha\big(\mathcal{J}_{\scriptscriptstyle{(1)}}^\beta -
n\bar{\delta} u_{\scriptscriptstyle{(1)}}^\beta\big)\Big\}\bar{\delta} u_{{\scriptscriptstyle{(1)}}\beta}
+\Big[\big(\mathcal{J}_{\scriptscriptstyle{(3)}}^\alpha
- n\bar{\delta} u_{\scriptscriptstyle{(3)}}^\alpha\big)\\
&-\mathcal{N}_{\scriptscriptstyle{(1)}}\bar{\delta} u_{\scriptscriptstyle{(2)}}^\alpha
-\mathcal{N}_{\scriptscriptstyle{(2)}}\bar{\delta} u_{\scriptscriptstyle{(1)}}^\alpha
+\bar{\delta} u_{\scriptscriptstyle{(1)}}^\alpha\Big(\mathcal{J}_{\scriptscriptstyle{(1)}}^\beta- \frac{1}{2} n
\bar{\delta} u_{\scriptscriptstyle{(1)}}^\beta\Big)\bar{\delta} u_{{\scriptscriptstyle{(1)}}\beta}\\
&+ u^\alpha\Big\{\big(\mathcal{J}_{\scriptscriptstyle{(2)}}^\beta
- \mathcal{N}_{\scriptscriptstyle{(1)}}\bar{\delta} u_{\scriptscriptstyle{(1)}}^\beta - n\bar{\delta} u_{\scriptscriptstyle{(2)}}^\beta\big)\bar{\delta} u_{{\scriptscriptstyle{(1)}}\beta}\\
& +\big(\mathcal{J}_{\scriptscriptstyle{(1)}}^\beta - n\bar{\delta} u_{\scriptscriptstyle{(1)}}^\beta\big)
\bar{\delta} u_{{\scriptscriptstyle{(2)}}\beta}\Big\}\Big],
\end{aligned}
\label{rel_transf_para_J}
\end{equation}
\begin{equation}
\begin{aligned}
\tau'^{\alpha\beta} & = \tau_{\scriptscriptstyle{(1)}}^{\alpha\beta} + \tau_{\scriptscriptstyle{(2)}}^{\alpha\beta}
- 2\Big(\mathcal{Q}_{\scriptscriptstyle{(1)}}^{\langle\alpha} - \frac{1}{2}h\bar{\delta} u_{\scriptscriptstyle{(1)}}^{\langle\alpha}\Big)\bar{\delta} u_{\scriptscriptstyle{(1)}}^{\beta\rangle}\\
& + 2\bar{\delta} u_{{\scriptscriptstyle{(1)}}\mu}\tau_{\scriptscriptstyle{(1)}}^{\mu(\alpha}u^{\beta)}
+ \tau_{\scriptscriptstyle{(3)}}^{\alpha\beta} - 2 \Big(\mathcal{Q}_{\scriptscriptstyle{(1)}}^{\langle\alpha}
- h\bar{\delta} u_{\scriptscriptstyle{(1)}}^{\langle\alpha}\Big)\bar{\delta} u_{\scriptscriptstyle{(2)}}^{\beta\rangle}\\
& -2 \Big[\bar{\delta}^2 u_{\scriptscriptstyle{(1)}}\Big(\mathcal{Q}_{\scriptscriptstyle{(1)}}^{(\alpha}
-\frac{d-2}{d-1}h\bar{\delta} u_{\scriptscriptstyle{(1)}}^{(\alpha}\Big)
- \bar{\delta} u_{{\scriptscriptstyle{(1)}}\mu}\tau_{\scriptscriptstyle{(2)}}^{\mu(\alpha}\\
&- \bar{\delta} u_{{\scriptscriptstyle{(2)}}\mu}\tau_{\scriptscriptstyle{(1)}}^{\mu(\alpha}
+ \bigg(\frac{d-3}{d-1}\bigg)\bar{\delta} u_{{\scriptscriptstyle{(1)}}\mu}\mathcal{Q}_{\scriptscriptstyle{(1)}}^\mu
\bar{\delta} u_{\scriptscriptstyle{(1)}}^{(\alpha}\Big]u^{\beta)}\\
& -2\Big[\mathcal{Q}_{\scriptscriptstyle{(2)}}^{\langle\alpha}
- \frac{1}{2}\mathcal{H}_{\scriptscriptstyle{(1)}}\bar{\delta} u_{\scriptscriptstyle{(1)}}^{\langle\alpha}
- \bar{\delta} u_{{\scriptscriptstyle{(1)}}\mu}\tau_{\scriptscriptstyle{(1)}}^{\mu\langle\alpha}
\bigg]\bar{\delta} u_{\scriptscriptstyle{(1)}}^{\beta\rangle}\\ 
&+\Big(u^{\alpha}u^{\beta}+\frac{\Delta^{\alpha\beta}}{d-1}\Big)\bar{\delta} u_{{\scriptscriptstyle{(1)}}\mu}
\bar{\delta} u_{{\scriptscriptstyle{(1)}}\nu}\tau_{\scriptscriptstyle{(1)}}^{\mu\nu},
\end{aligned}
\label{rel_transf_para_Tau}
\end{equation}
where $B_{\scriptscriptstyle{(3)}}$ represents the third-order term in the gradient expansion of $B$, given by
\begin{equation}
\begin{split}
B_{\scriptscriptstyle{(3)}} = 2\big[& h\bar{\delta} u_{\scriptscriptstyle{(1)}}^\alpha - \mathcal{Q}_{\scriptscriptstyle{(1)}}^\alpha\big]
\bar{\delta} u_{{\scriptscriptstyle{(2)}}\alpha}\\
&+ \big[\mathcal{H}_{\scriptscriptstyle{(1)}}\bar{\delta} u_{\scriptscriptstyle{(1)}}^\alpha + \tau_{\scriptscriptstyle{(1)}}^{\alpha\beta}\bar{\delta} u_{{\scriptscriptstyle{(1)}}\beta}
- 2\mathcal{Q}_{\scriptscriptstyle{(2)}}^\alpha\big] \bar{\delta} u_{{\scriptscriptstyle{(1)}}\alpha}.
\end{split}
\end{equation}
These equations will be used later in this study to derive the relationships between the transport coefficients of
a general frame gradient expansion and a set of transport coefficients that remain invariant under field redefinition
up to the $n$th order in a derivative expansion. This process will play a pivotal role in simplifying the linear
dispersion relations in a general hydrodynamic frame.

In addition to the constraints on the constitutive relations imposed by hydrodynamic frame fixing, there are equivalences
when the equations of motion hold, known as on-shell equivalences. In the recent literature of GFH, these constraints on
the constitutive relations are implemented at the final stages, allowing the constitutive relations to be expressed
off-shell for the first time. On the other hand, when dealing with higher-order gradient expansions, it is highly
favorable to implement constraints arising from the equations of motion as early as possible. The motivation behind
this is that, in higher orders, the number of corrections grows exponentially, and any simplification that reduces
this extensive list is considered beneficial. Moreover, the concept of gradient expansion arises when we formulate hydrodynamics at the level of the dissipative equations of motion.

Gradient expansion represents an on-shell formulation of fluid dynamics' effective theory. Consequently, the validity
of the equations of motion can be assumed at the outset of elaborating the gradient expansion. This viewpoint was
extensively detailed in Ref.~\cite{Grozdanov:2015kqa}. Here, the conservation principle inherent in the equations
of motion dictates the substitution of time (longitudinal) derivatives with spatial (transverse) derivatives within
the gradient expansion. In this work, we enforce on-shell equivalences from the onset and replace time derivatives
with spatial derivatives in the constitutive relations.

It is important to note that the existence of on-shell equivalences requires additional choices in the construction of constitutive relations. This implies that frame fixing, accomplished by establishing out-of-equilibrium definitions of thermodynamic functions, does not uniquely define constitutive relations. Instead, it results in an equivalence class of constitutive relations. Stability conditions must remain independent of the selected element within this equivalence class, since they are defined on-shell. When assessing the dispersion relations derived from the gradient expansion, we limit the equations of motion to linear perturbations. This reduces the number of relevant transport coefficients involved, without incorporating any new dynamical information. Thus, even if the gradient expansion is expressed off-shell, any additional information is ruled out in the linearized equations, as can be verified by analyzing the results of~\cite{Bemfica:2020zjp}.

\subsection{Algorithms to generate the tensorial structures}

In Ref.~\cite{Grozdanov:2015kqa}, Grozdanov and Kaplis (GK) developed a systematic algorithm capable of producing a gradient expansion for relativistic hydrodynamics up to any desired order. The extension of the GK algorithm, as it will be referred to henceforth, to charged systems involves generating dissipative corrections to the conserved currents, $T^{\alpha\beta}$ and $J^{\alpha}$, by applying the gradient operator to the fundamental degrees of freedom $\{T,\mu,u^\alpha,g^{\alpha\beta}\}$ in order to obtain all the relevant ingredients for the construction of the expansion up to a specific order. The resulting quantities are systematically multiplied and contracted in every conceivable combination. The equations of motion, along
with other algebraic identities and symmetries, are employed to eliminate redundancies.

In this study, we implement an extended version of the methodology presented in Ref.~\cite{Diles:2019uft}, serving as an alternative approach to the one delineated in \cite{Grozdanov:2015kqa}, for the construction of higher-order corrections. Rather than adopting the fundamental set of degrees of freedom as our starting point, we turn to the lowest-order covariant tensors derived from these fundamental quantities, specifically: $\{\nabla_{\perp \alpha}T,\nabla_{\perp \alpha}\mu, \nabla_{\perp \alpha} u^\beta, R^{\alpha \beta \sigma \gamma}\}$. In this set, there are two first-order transverse vectors, one first-order transverse rank-2 tensor, and one second-order rank-4 tensor. The Riemann tensor distinguishes itself from the other structures due to several unique aspects: it is not built with a linear covariant derivative but with a non-linear combination of partial derivatives of the metric; it is inherently second-order; and it figures on the left-hand side of the Einstein equations.

A distinguishing feature of hydrodynamics is the separation of velocity gradients into three structures via Weyl decomposition,
each having a distinct and well-defined physical interpretation. We thus have
\begin{equation}
\nabla^\alpha u^\beta = \sigma^{\alpha \beta}+\Omega^{\alpha \beta} + \frac{1}{d-1}\Theta \Delta^{\alpha\beta}.
\end{equation}
In the above equation, $\sigma^{\alpha \beta}$ is the transverse, symmetric, and traceless (TST) tensor that encodes shear information. Simultaneously, $\Omega^{\alpha \beta}$ is an antisymmetric tensor that captures vorticity information, and $\Theta$ is a scalar, or trace, that encapsulates expansion information. It is thus beneficial to employ $\sigma, \Omega, \Theta$ as substitutes for $\nabla_\perp u$ when formulating the gradient expansion. Furthermore, using $\sigma, \Omega, \Theta$ brings both shear and bulk viscosity into prominence within the set of first-order transport coefficients. Consequently, our approach will involve the use of the minimal set of fundamental gradients $\{\nabla_{\perp \alpha}T,\nabla_{\perp \alpha}\mu, \Theta, \sigma^{\alpha \beta},\Omega^{\alpha \beta}, R^{\alpha \beta \sigma \gamma}\}$.
This set encompasses two first-order transverse vectors, one first-order scalar, one first-order rank-2 TST tensor, one
first-order rank-2 antisymmetric tensor, and the Riemann tensor, a second-order rank-4 tensor with specific symmetry properties. The systematic construction of the hydrodynamic gradient expansion using the irreducible structures (IS) as fundamental blocks is hereby referred to as the IS algorithm.

To minimize potential errors and validate the final results in an independent way, we have implemented the Irreducible-Structure (IS) and Grozdanov-Kaplis (GK) algorithms in two computational codes. These codes utilize the SymPy library,
a Python-written tool designed for symbolic mathematics. Specifically, we employ the Tensor module within SymPy and
integrate the Butler-Portugal canonicalization method \cite{Butler,portugal1998,portugal1999,manssur2002} into the codes.
This method is used for transforming a tensor expression into its canonical form by permuting its index labels with respect
to slot symmetries and dummy-index renaming, and is especially advantageous for handling terms involving the Riemann tensor.

Although the IS algorithm has the benefit of dealing with a small number of combinations of physically motivated quantities,
it has the limitation that some equivalences between the products of $\sigma, \Omega, \Theta$ and their derivatives cannot,
in principle, be uncovered without recourse to their definitions in terms of the transverse derivatives of $u^\alpha$, a
feature inherent to the GK algorithm. An observation that enables the automatic identification of independent structures
in the IS algorithm is that all elements involving derivatives of $\Omega$ can be written as combinations of elements that contain derivatives of $\sigma$ and/or $\Theta$.

After making the necessary adjustments, both codes produce the same quantity of independent scalar, vector, and tensor structures for the hydrodynamic gradient expansion of a specified order. In fact, to evaluate the robustness of the codes,
we elected to extend one order higher in the expansion, thus deriving the fourth-order structures for a charged fluid in
a flat (zero curvature) spacetime. We found perfect agreement between the codes employing the two algorithms, with the
following numbers of independent structures: 106 scalars, 193 vectors, and 244 tensors.

\subsection{First- and second-order constitutive relations}

The gradient expansion of a charged fluid encompasses one set of scalars, one set of TST tensors, and one set of transverse vectors. Importantly, the presence of a conserved baryon charge, in addition to introducing a new fundamental field
(the chemical potential), also augments the gradient expansion with a set of vectors and a set of scalars.

In the lowest-order covariant list, three first-order gradients can be identified: $\nabla_{\perp}^{\alpha}T$, $\nabla_{\perp}^{\alpha}\mu$, and $\nabla_{\perp}^{\alpha}u^{\beta}$. The first two gradients are first-order transverse vectors and hence should appear in $\Upsilon_{\scriptscriptstyle{(1)}}^{\alpha}$. The third gradient is a rank-two tensor
that contributes two independent structures associated with its trace $\Theta=\nabla_\alpha u^\alpha$, and its traceless
part, $\sigma^{\alpha\beta}=\nabla_{\perp}^{\langle \alpha}u^{\beta\rangle}$. By taking these terms into account, the
most general first-order corrections to the conserved currents can be expressed as
\begin{equation}
\begin{split}
\Pi_{\scriptscriptstyle{(1)}}^{\alpha \beta} & = \mathcal{E}_{\scriptscriptstyle{(1)}}u^\alpha u^\beta +
\mathcal{P}_{\scriptscriptstyle{(1)}}\Delta^{\alpha \beta} + 2\mathcal{Q}_{\scriptscriptstyle{(1)}}^{(\alpha} u^{\beta)}
+ \tau_{\scriptscriptstyle{(1)}}^{\alpha \beta},\\
\Upsilon_{\scriptscriptstyle{(1)}}^\alpha &= \mathcal{N}_{\scriptscriptstyle{(1)}}u^\alpha + \mathcal{J}_{\scriptscriptstyle{(1)}}^\alpha,
\end{split}
\end{equation}
where
\begin{equation}
\begin{aligned}
    \mathcal{E}_{\scriptscriptstyle{(1)}} & = \varepsilon_{1}^{\scriptscriptstyle{(1)}}\Theta, & \qquad 
    \mathcal{Q}_{\scriptscriptstyle{(1)}}^{\alpha} & = \theta_{1}^{\scriptscriptstyle{(1)}}\nabla_{\perp}^{\alpha}T
    + \theta_{2}^{\scriptscriptstyle{(1)}}\nabla_{\perp}^{\alpha}\mu,\\
    \mathcal{P}_{\scriptscriptstyle{(1)}} & = \pi_{1}^{\scriptscriptstyle{(1)}}\Theta, & \qquad 
    \mathcal{J}_{\scriptscriptstyle{(1)}}^\alpha & = \kappa_{1}^{\scriptscriptstyle{(1)}}\nabla_{\perp}^\alpha T
  + \kappa_{2}^{\scriptscriptstyle{(1)}} \nabla_{\perp}^{\alpha} \mu,\\
  \mathcal{N}_{\scriptscriptstyle{(1)}} & = \nu_{1}^{\scriptscriptstyle{(1)}} \Theta, & \qquad 
  \tau_{\scriptscriptstyle{(1)}}^{\alpha \beta} & = \eta_{1}^{\scriptscriptstyle{(1)}}\sigma^{\alpha \beta}.
\end{aligned}     
\label{Pi_Up_first_order}
\end{equation}

The set of transport coefficients introduced here, $\{\varepsilon_{1}^{\scriptscriptstyle{(1)}}$, $\pi_{1}^{\scriptscriptstyle{(1)}},\nu_{1}^{\scriptscriptstyle{(1)}}, \theta_{1}^{\scriptscriptstyle{(1)}}, \theta_{2}^{\scriptscriptstyle{(1)}}, \kappa_{1}^{\scriptscriptstyle{(1)}}, \kappa_{2}^{\scriptscriptstyle{(1)}}, \eta_{1}^{\scriptscriptstyle{(1)}}\}$, is closely related to traditional first-order coefficients
$\{\eta,\zeta,\sigma,\chi_{\scriptscriptstyle{\text{T}}}\}$. To establish these connections, we consider
the one-derivative order terms in relations \eqref{rel_transf_para_E}--\eqref{rel_transf_para_Tau}:
\begin{equation}
\begin{gathered}
    \mathcal{E}_{\scriptscriptstyle{(1)}}'(T',u',\mu') + \bar{\delta} \epsilon_{\scriptscriptstyle{(1)}}
 = \mathcal{E}_{\scriptscriptstyle{(1)}}(T,u,\mu), \\
   \mathcal{P}_{\scriptscriptstyle{(1)}}'(T',u',\mu') + \bar{\delta} p_{\scriptscriptstyle{(1)}} 
 = \mathcal{P}_{\scriptscriptstyle{(1)}}(T,u,\mu), \\
  \mathcal{N}_{\scriptscriptstyle{(1)}}'(T',u',\mu') + \bar{\delta} n_{\scriptscriptstyle{(1)}}
 = \mathcal{N}_{\scriptscriptstyle{(1)}}(T,u,\mu),\\
    \mathcal{Q}_{\scriptscriptstyle{(1)}}'^\alpha(T',u',\mu')
 = \mathcal{Q}_{\scriptscriptstyle{(1)}}^\alpha(T,u,\mu)
    -h\bar{\delta} u_{\scriptscriptstyle{(1)}}^\alpha,\\
    \mathcal{J}_{\scriptscriptstyle{(1)}}'^\alpha(T',u',\mu')
 = \mathcal{J}_{\scriptscriptstyle{(1)}}^\alpha(T,u,\mu)
  - n \bar{\delta} u_{\scriptscriptstyle{(1)}}^\alpha,\\
  \mathcal{\tau}_{\scriptscriptstyle{(1)}}'^{\alpha\beta}(T',u',\mu')
 = \mathcal{\tau}_{\scriptscriptstyle{(1)}}'^{\alpha\beta}(T,u,\mu),
\end{gathered}     
\label{rel_first_order_1}
\end{equation}
which can be rewritten in the form
\begin{equation}
\begin{gathered}
\mathcal{E}'_{\scriptscriptstyle{(1)}}(T,u,\mu) = \mathcal{E}_{\scriptscriptstyle{(1)}}(T,u,\mu)
- \epsilon_{,{\scriptscriptstyle{T}}}\bar{\delta} T_{\scriptscriptstyle{(1)}} - \epsilon_{,\mu}\bar{\delta}\mu_{\scriptscriptstyle{(1)}},\\
\mathcal{P}'_{\scriptscriptstyle{(1)}}(T,u,\mu) = \mathcal{P}_{\scriptscriptstyle{(1)}}(T,u,\mu)
- p_{,{\scriptscriptstyle{T}}}\bar{\delta} T_{\scriptscriptstyle{(1)}}
- p_{,\mu}\bar{\delta}\mu_{\scriptscriptstyle{(1)}},\\
\mathcal{N}'_{\scriptscriptstyle{(1)}}(T,u,\mu) = \mathcal{N}_{\scriptscriptstyle{(1)}}(T,u,\mu)
- n_{,{\scriptscriptstyle{T}}}\bar{\delta} T_{\scriptscriptstyle{(1)}}
- n_{,\mu}\bar{\delta}\mu_{\scriptscriptstyle{(1)}},\\
 \mathcal{Q}'^\alpha_{\scriptscriptstyle{(1)}}(T,u,\mu) = \mathcal{Q}_{\scriptscriptstyle{(1)}}^\alpha(T,u,\mu)
-h\bar{\delta} u_{\scriptscriptstyle{(1)}}^\alpha,\\
\mathcal{J}'^\alpha_{\scriptscriptstyle{(1)}}(T,u,\mu) = \mathcal{J}_{\scriptscriptstyle{(1)}}^\alpha(T,u,\mu)
-n\bar{\delta} u_{\scriptscriptstyle{(1)}}^\alpha,\\
\mathcal{\tau}'^{\alpha\beta}_{\scriptscriptstyle{(1)}}(T,u,\mu) =
\mathcal{\tau}_{\scriptscriptstyle{(1)}}^{\alpha\beta}(T,u,\mu),
\end{gathered}     
\label{rel_first_order_2}
\end{equation}
where the comma subscript indicates the partial derivative in relation to the parameter that follows, 
as in $\epsilon_{,{\scriptscriptstyle{T}}} \equiv \left(\partial \epsilon/\partial T\right)_{\mu}$.

In the above equations, the dependence of a given variable on the parameters $(T', u',\mu')$ implies an
expansion of this variable in terms of the gradients of $(T', u',\mu')$. For example,
\begin{equation}
\begin{split}
\mathcal{E}'_{\scriptscriptstyle{(1)}}(T',u',\mu') &= \varepsilon'^{\scriptscriptstyle{(1)}}_{1}\Theta'\\
& = \varepsilon'^{\scriptscriptstyle{(1)}}_{1}\nabla_\alpha \left(A u^\alpha + \bar{\delta} u^\alpha\right)\\
& = \varepsilon'^{\scriptscriptstyle{(1)}}_{1}\Theta + \mathcal{O}(\partial^2),
\end{split}
\end{equation}
whereas $\mathcal{E}'_{\scriptscriptstyle{(1)}}(T,u,\mu) = \varepsilon'^{\scriptscriptstyle{(1)}}_{1}\Theta$ and
$\mathcal{E}_{\scriptscriptstyle{(1)}}(T,u,\mu) = \varepsilon^{\scriptscriptstyle{(1)}}_{1}\Theta$. The substitution of
the gradient expansions of the original and primed variables into the relations given in equation \eqref{rel_first_order_2}, together with the first-order versions of \eqref{expansions_delta}, leads to the following relations among the transport coefficients:
\begin{equation}
\begin{aligned}
\varepsilon'^{\scriptscriptstyle{(1)}}_{1} & = \varepsilon^{\scriptscriptstyle{(1)}}_{1} -
\epsilon_{,{\scriptscriptstyle{T}}} a^{\scriptscriptstyle{(1)}}_{1} - \epsilon_{,\mu} c^{\scriptscriptstyle{(1)}}_{1},
& \quad \theta'^{\scriptscriptstyle{(1)}}_{l} & = \theta^{\scriptscriptstyle{(1)}}_{l} -
h b^{\scriptscriptstyle{(1)}}_{l}, \\
\pi'^{\scriptscriptstyle{(1)}}_{1} & = \pi^{\scriptscriptstyle{(1)}}_{1} -
p_{,{\scriptscriptstyle{T}}} a^{\scriptscriptstyle{(1)}}_{1} - p_{,\mu} c^{\scriptscriptstyle{(1)}}_{1},
& \quad \kappa'^{\scriptscriptstyle{(1)}}_{l} & = \kappa^{\scriptscriptstyle{(1)}}_{l} -
n b^{\scriptscriptstyle{(1)}}_{l},\\
\nu'^{\scriptscriptstyle{(1)}}_{1} & = \nu^{\scriptscriptstyle{(1)}}_{1} -
n_{,{\scriptscriptstyle{T}}} a^{\scriptscriptstyle{(1)}}_{1} - n_{,\mu} c^{\scriptscriptstyle{(1)}}_{1},
& \quad \eta'^{\scriptscriptstyle{(1)}}_{1} & = \eta^{\scriptscriptstyle{(1)}}_{1},
\end{aligned}
\label{rel_coeff_first_order}
\end{equation}
where $l = 1, 2$. With the exception of $\eta^{\scriptscriptstyle{(1)}}_{1}$, all the foregoing coefficients
are frame-dependent, but we can combine them to obtain a set of frame-invariant coefficients,
defined by \cite{Kovtun:2019hdm}:
\begin{equation}
f^{\scriptscriptstyle{(1)}}_{1} \equiv \pi^{\scriptscriptstyle{(1)}}_{1} - \beta_\epsilon \varepsilon^{\scriptscriptstyle{(1)}}_{1}
- \beta_n \nu^{\scriptscriptstyle{(1)}}_{1},
\quad \ell^{\scriptscriptstyle{(1)}}_{l} \equiv \kappa^{\scriptscriptstyle{(1)}}_{l} - \frac{n}{h} \theta^{\scriptscriptstyle{(1)}}_{l},
\label{def_f_and_ell}
\end{equation}
where
\begin{equation}
\beta_\epsilon \equiv \left(\frac{\partial p}{\partial\epsilon}\right)_n,\quad
\beta_n \equiv \left(\frac{\partial p}{\partial n}\right)_{\epsilon}.
\end{equation}

In terms of the traditional transport coefficients of first-order hydrodynamics, the constitutive relations
in a general frame \cite{Kovtun:2012rj} assume the form
\begin{equation}
\begin{gathered}
\tau_{\scriptscriptstyle{(1)}}^{\alpha\beta}=-2\eta\sigma^{\alpha\beta},\qquad
f_{\scriptscriptstyle{(1)}} =-\zeta\Theta,\\
\ell_{\scriptscriptstyle{(1)}}^{\alpha}=\left(\sigma\frac{\mu}{T}+\chi_{\scriptscriptstyle{\text{T}}}\right)
\nabla_{\perp}^{\alpha}T-\sigma\nabla_{\perp}^{\alpha}\mu.
\end{gathered}
\label{1st_order_const_relations}
\end{equation}
A direct comparison of the first relation in \eqref{1st_order_const_relations}
with the expression for $\tau_{\scriptscriptstyle{(1)}}^{\alpha\beta}$ provided
in \eqref{Pi_Up_first_order} yields
\begin{equation}
t_{1}^{\scriptscriptstyle{(1)}}\equiv\eta_{1}^{\scriptscriptstyle{(1)}} = -2\eta\qquad\Longrightarrow\qquad
\eta = -\frac{1}{2}\eta_{1}^{\scriptscriptstyle{(1)}},
\label{shear_viscosity_relation}
\end{equation}
where $\eta$ denotes the shear viscosity.
Furthermore, substituting \eqref{Pi_Up_first_order} into \eqref{def_f_and_ell} and comparing the resulting equations
with the latter two of Eqs.~\eqref{1st_order_const_relations}, we derive
\begin{equation}
\begin{gathered}
\zeta = - \pi_{1}^{\scriptscriptstyle{(1)}} + \beta_\epsilon \varepsilon_{1}^{\scriptscriptstyle{(1)}}
+\beta_n \nu_{1}^{\scriptscriptstyle{(1)}},\\
\sigma\frac{\mu}{T} =  \left(\theta_{1}^{\scriptscriptstyle{(1)}}
- \frac{n}{h}\kappa_{1}^{\scriptscriptstyle{(1)}}\right) - \chi_{\scriptscriptstyle{\text{T}}}
= - \left(\theta_{2}^{\scriptscriptstyle{(1)}}
-\frac{n}{h}\kappa_{2}^{\scriptscriptstyle{(1)}}\right)\frac{\mu}{T},
\end{gathered}
\label{bulk_conductivity_relations}
\end{equation}
where $\zeta$ and $\sigma$ represent the bulk viscosity and the charge conductivity, respectively.

The second law of thermodynamics mandates the positive divergence of the entropy current,
which consequently imposes constraints on the transport coefficients
\cite{Kovtun:2012rj,Muronga:2003ta,Bhattacharya:2011tra}:
\begin{equation}
\eta\geq 0,\qquad \zeta\geq 0,\qquad \sigma\geq 0,\qquad \chi_{\scriptscriptstyle{\text{T}}} = 0.
\label{constraints_second_law}
\end{equation}
In connection with Eqs.~\eqref{shear_viscosity_relation} and \eqref{bulk_conductivity_relations},
these constraints can be reformulated in terms of our notation as
\begin{equation}
\begin{aligned}
\eta_{1}^{\scriptscriptstyle{(1)}} & \leq 0, &\quad
 \theta_{1}^{\scriptscriptstyle{(1)}} & \geq \frac{n}{h}\kappa_{1}^{\scriptscriptstyle{(1)}},\\
\pi_{1}^{\scriptscriptstyle{(1)}} & \leq \beta_\epsilon \varepsilon_{1}^{\scriptscriptstyle{(1)}}
+\beta_n \nu_{1}^{\scriptscriptstyle{(1)}}, & \quad
\theta_{2}^{\scriptscriptstyle{(1)}} & \leq \frac{n}{h}\kappa_{2}^{\scriptscriptstyle{(1)}}.
\end{aligned}
\end{equation}
Thus, the combination of the flexibility in the choice of hydrodynamic frames and the constraints stemming from the
second law of thermodynamics simplifies the constitutive relations by eliminating five transport coefficients.
Consequently, only three transport coefficients remain in the first-order theory.

To derive the subsequent corrections, we first note that the complete list of scalars, vectors, and tensors
of the dynamics of a neutral fluid will also be present in the charged case. This implies that by taking
$n,\mu\to0$, we should be able to replicate previous results concerning the gradient expansion of uncharged
fluids \cite{Romatschke:2009kr,Grozdanov:2015kqa, Diles:2019uft}. Moreover, corrections that involve at least
one gradient of the chemical potential will also be present.

To identify additional structures of the charged fluid, we utilize the computational codes discussed in the preceding subsection. The Grozdanov-Kaplis (GK) algorithm treats the fundamental degrees of freedom, $\{T,\mu, u^{\alpha}, g^{\alpha\beta}\}$, as the starting point. A notable advantage of this procedure lies in its capacity to be implemented
in a computational code that not only generates all elements of the gradient expansion but also eliminates redundancies
within the list. In contrast, the Irreducible Structure (IS) algorithm begins with the fundamental gradients
$\{\nabla_{\perp \alpha}T,\nabla_{\perp \alpha}\mu, \Theta, \sigma^{\alpha\beta},\Omega^{\alpha\beta}, R_{\alpha\beta\sigma\delta}\}$. Recent research indicates the irrelevance of the ordering of transverse
derivatives in gradient expansion \cite{Diles:2020cjy}, which is directly correlated with the redundancy of
$\nabla_{\perp \alpha}\Omega^{\beta \gamma}$. Therefore, $\nabla_{\perp \alpha}\Omega^{\beta \gamma}$ can
be omitted from the gradient expansion, an insight of considerable significance for the computational implementation
of the IS algorithm. A proof supporting this claim is provided in Appendix \ref{equiv_symmetry}.

We observe that the gradient of the chemical potential assumes a role analogous to that of the temperature (or entropy) gradient, owing to their identical index structures. All corrections stemming from the gradient of the chemical potential
can be acquired by sequentially replacing the linear gradients of the temperature. The resulting list should maintain
symmetry under exchange $T\leftrightarrow\mu$. Indeed, two degrees of freedom possessing the same tensor rank will play
identical roles in the construction of their respective gradients. In our specific case, this only occurs with the scalars,
as there are just one vector and one tensor present as zero-order fields. We then find the following independent
second-order scalars:
\begin{equation}
\begin{aligned}
\mathcal{S}^{\scriptscriptstyle{(2)}}_1 &= \nabla_{\perp}^2 T, \qquad &
\mathcal{S}^{\scriptscriptstyle{(2)}}_2 & = \nabla_{\perp}^2 \mu,\\
\mathcal{S}^{\scriptscriptstyle{(2)}}_3 & = \Theta^2,
\qquad & \mathcal{S}^{\scriptscriptstyle{(2)}}_4 & = \sigma^2,\\
\mathcal{S}^{\scriptscriptstyle{(2)}}_5 & = \Omega^2,
\qquad & \mathcal{S}^{\scriptscriptstyle{(2)}}_6 &= (\nabla_{\perp} T)^2, \\
\mathcal{S}^{\scriptscriptstyle{(2)}}_7  & = (\nabla_{\perp} \mu)^2,
\qquad & \mathcal{S}^{\scriptscriptstyle{(2)}}_8  & = \nabla_{\perp \alpha} T \nabla_{\perp}^\alpha \mu, \\
\mathcal{S}^{\scriptscriptstyle{(2)}}_9  &= R,
\qquad & \mathcal{S}^{\scriptscriptstyle{(2)}}_{10}  & = u^\alpha u^\beta R_{\alpha\beta}.
\end{aligned}
\end{equation}
In the above expressions and in the subsequent ones, differential operators act only on the immediate neighbor
to the right. Compared to the case of an uncharged fluid, there exist three additional scalar quantities:
$\mathcal{S}^{\scriptscriptstyle{(2)}}_2$, $\mathcal{S}^{\scriptscriptstyle{(2)}}_7$, and
$\mathcal{S}^{\scriptscriptstyle{(2)}}_8$. These scalars vanish when the chemical potential $\mu$
is constant.

In our search for second-order tensors, we observe that, despite the tensorial index structure, the presence of a baryonic charge plays the same role for tensors as it does for scalars. All second-order tensors present in the uncharged fluid will remain, accompanied by the additional tensors that vanish for constant chemical potential. The additional tensors involving gradients of the chemical potential are derived by taking those with temperature gradients and then exchanging them, one
by one, for gradients of the chemical potential. This is a direct consequence of using the systematic procedure of
Grozdanov-Kaplins while implementing the additional information on the allowed gradients to appear in constitutive
relations. We list below the independent second-order tensors, omitting the free indexes on the left-hand side:
\begin{equation}
\begin{aligned}
\mathcal{T}^{\scriptscriptstyle{(2)}}_1 &= \nabla_{_\perp}^{\langle \alpha} \nabla_{_\perp}^{\beta\rangle} T, \qquad & \mathcal{T}^{\scriptscriptstyle{(2)}}_2 & = \nabla_{_\perp}^{\langle  \alpha} \nabla_{_\perp}^{\beta\rangle} \mu, \\
\mathcal{T}^{\scriptscriptstyle{(2)}}_3 &= \Theta\sigma^{\alpha\beta}, \qquad & \mathcal{T}^{\scriptscriptstyle{(2)}}_4 &= \sigma_{\gamma}^{~\langle \alpha}\sigma^{\beta\rangle \gamma}, \\
\mathcal{T}^{\scriptscriptstyle{(2)}}_5 &= \Omega_{\gamma}^{~\langle \alpha}\Omega^{\beta\rangle \gamma}, \qquad & \mathcal{T}^{\scriptscriptstyle{(2)}}_6 &= \sigma_{\gamma}^{~\langle \alpha}\Omega^{\beta\rangle \gamma},\\
\mathcal{T}^{\scriptscriptstyle{(2)}}_7 &= \nabla_{_\perp}^{\langle \alpha} T\nabla_{_\perp}^{\beta\rangle} T, \qquad & \mathcal{T}^{\scriptscriptstyle{(2)}}_8 &= \nabla_{_\perp}^{\langle  \alpha} \mu\nabla_{_\perp}^{\beta\rangle} \mu, \\
\mathcal{T}^{\scriptscriptstyle{(2)}}_9 &= \nabla_{_\perp}^{\langle \alpha} T\nabla_{_\perp}^{\beta\rangle} \mu, \qquad & \mathcal{T}^{\scriptscriptstyle{(2)}}_{10} & = R^{\expva{\alpha\beta}}, \\
\mathcal{T}^{\scriptscriptstyle{(2)}}_{11} & = u_\gamma u_\delta R^{\gamma\expva{\alpha\beta}\delta}.
\end{aligned}
\end{equation}

Charged fluids also require vector corrections to the constitutive relations. Whether working in a general frame or in a specific one like the Eckart frame, these corrections manifest in the energy-momentum tensor through a non-vanishing heat current that contributes to the longitudinal projection along the local velocity. In the Landau frame case, where it is
imposed Eqs.~\eqref{Landau}, vector corrections appear in the matter current, including the effects of dissipation in the continuity equations. In both frames, these corrections lead to the same number of transport coefficients, since the frames
are equivalents. Second-order vectors have been first presented in \cite{Romatschke:2009kr} and are also discussed in the appendix of \cite{Diles:2019uft} for uncharged fluids:
\begin{equation}
\begin{aligned}
\mathcal{V}^{\scriptscriptstyle{(2)}}_1 &= \nabla_{\perp}^\alpha \Theta, \qquad & \mathcal{V}^{\scriptscriptstyle{(2)}}_2 & = \Delta^\alpha_{~\beta}\nabla_{\perp \gamma}\sigma^{\gamma\beta}, \\
\mathcal{V}^{\scriptscriptstyle{(2)}}_3 &= \Theta\nabla_{\perp }^\alpha T, \qquad & \mathcal{V}^{\scriptscriptstyle{(2)}}_4 &=  \Theta\nabla_{\perp }^\alpha \mu, \\
\mathcal{V}^{\scriptscriptstyle{(2)}}_5 &=  \sigma^{\alpha}_{~\beta}\nabla_{\perp}^\beta T, \qquad & \mathcal{V}^{\scriptscriptstyle{(2)}}_6 &= \sigma^{\alpha}_{~\beta}\nabla_{\perp}^\beta \mu, \\
\mathcal{V}^{\scriptscriptstyle{(2)}}_7 &= \Omega^{\alpha}_{~\beta}\nabla_{\perp}^\beta T, \qquad & \mathcal{V}^{\scriptscriptstyle{(2)}}_8 &= \Omega^{\alpha}_{~\beta}\nabla_{\perp}^\beta \mu, \\
\mathcal{V}^{\scriptscriptstyle{(2)}}_9 &= \Delta^{\alpha\beta}u^\gamma R_{\beta\gamma}.
\end{aligned}
\end{equation}

Finally, we identify 30 independent second-order tensorial structures that establish the constitutive relations for
a fluid with a single global charge. Most of these second-order corrections align with those presented by Lahiri in Ref.~\cite{Lahiri:2019lpk}, but there are exceptions. The tensor $\Theta \sigma^{\alpha\beta}$ is not present in that
list, resulting in our list having one additional tensor. Concerning second-order vectors, we find three discrepancies
in Lahiri's list: the vector $\Omega^{\alpha\beta}\nabla_{\perp \beta} T$ is missing; the vector
$\Theta\nabla_{\perp}^\alpha \mu$ appears twice, in $\mathcal{N}_4$ and $\mathcal{N}_5$; and the
vector $R^{\alpha\beta}\nabla_{\perp \alpha} \mu$ is, in fact, a third-order vector. However, our list
of second-order scalars is in complete agreement with that of \cite{Lahiri:2019lpk}.

The following constitutive relations are then obtained for the dissipative parts of the energy-momentum tensor,
\begin{equation}
\begin{split}
\Pi_{\scriptscriptstyle{(2)}}^{\alpha\beta} & = \sum_{j=1}^{10} \varepsilon_{j}^{\scriptscriptstyle{(2)}}\,\mathcal{S}^{\scriptscriptstyle{(2)}}_{j} u^\alpha u^\beta + \sum_{j=1}^{10} \pi_{j}^{\scriptscriptstyle{(2)}}\,\mathcal{S}^{\scriptscriptstyle{(2)}}_{j}\Delta^{\alpha\beta}\\
& + \sum_{j=1}^{9} \theta_{j}^{\scriptscriptstyle{(2)}} (\mathcal{V}_{j}^{\scriptscriptstyle{(2)}})^{(\alpha} u^{\beta)} 
+\sum_{j=1}^{11} \eta_{j}^{\scriptscriptstyle{(2)}}  (\mathcal{T}_{j}^{\scriptscriptstyle{(2)}})^{\alpha\beta},
\end{split}
\end{equation}
and of the current
\begin{equation}
\Upsilon_{\scriptscriptstyle{(2)}}^{\alpha} = \sum_{j=1}^{10} \nu_{j}^{\scriptscriptstyle{(2)}}\,\mathcal{S}_{j}^{\scriptscriptstyle{(2)}}u^\alpha + \sum_{j=1}^{9} \kappa_{j}^{\scriptscriptstyle{(2)}}\,(\mathcal{V}_{j}^{\scriptscriptstyle{(2)}})^{\alpha}.
\end{equation}

Due to the flexibility in choosing the hydrodynamic frame inherent in the relativistic description of a fluid,
the 49 transport coefficients present in the aforementioned constitutive relations are not all independent. In
fact, they can be combined to form 30 frame-invariant quantities. In this work, we limit ourselves to identifying
the set of linear frame-invariant coefficients in the flat spacetime case, as these are the quantities that appear
in the dispersion relations explored in the final part of this paper. The term ``linear'' refers to tensorial
structures (and their associated coefficients) that are of the first order in the amplitudes of the fundamental
hydrodynamic degrees of freedom ${T,\mu, u^{\alpha}}$, regardless of their order in the gradient expansion.

By retaining only the two derivative terms of linear order in relations
\eqref{rel_transf_para_E}--\eqref{rel_transf_para_Tau}, we find the following:
\begin{equation}
\begin{gathered}
\mathcal{E}_{\scriptscriptstyle{(2)}}'(T,u,\mu) + \triangle_1 \mathcal{E}_{\scriptscriptstyle{(1)}}'
+ \bar{\delta} \epsilon_{\scriptscriptstyle{(2)}} = \mathcal{E}_{\scriptscriptstyle{(2)}}(T,u,\mu),\\
\mathcal{P}_{\scriptscriptstyle{(2)}}'(T,u,\mu) + \triangle_1\mathcal{P}_{\scriptscriptstyle{(1)}}'
+ \bar{\delta} p_{\scriptscriptstyle{(2)}} = \mathcal{P}_{\scriptscriptstyle{(2)}}(T,u,\mu),\\
\mathcal{N}_{\scriptscriptstyle{(2)}}'(T,u,\mu) + \triangle_1\mathcal{N}_{\scriptscriptstyle{(1)}}'
+ \bar{\delta} n_{\scriptscriptstyle{(2)}} = \mathcal{N}_{\scriptscriptstyle{(2)}}(T,u,\mu),\\
\mathcal{Q}_{\scriptscriptstyle{(2)}}'^{\alpha}(T,u,\mu) + \triangle_1\mathcal{Q}_{\scriptscriptstyle{(1)}}'^{\alpha}
= \mathcal{Q}_{\scriptscriptstyle{(2)}}^\alpha(T,u,\mu) - h\bar{\delta} u_{\scriptscriptstyle{(2)}}^\alpha,\\
\mathcal{J}_{\scriptscriptstyle{(2)}}'^\alpha(T,u,\mu) + \triangle_1\mathcal{J}_{\scriptscriptstyle{(1)}}'^{\alpha}
= \mathcal{J}_{\scriptscriptstyle{(2)}}^\alpha(T,u,\mu) - n\bar{\delta} u_{\scriptscriptstyle{(2)}}^\alpha,\\
\mathcal{\tau}_{\scriptscriptstyle{(2)}}'^{\alpha\beta}(T,u,\mu) + \triangle_1\tau_{\scriptscriptstyle{(1)}}'^{\alpha\beta}
= \mathcal{\tau}_{\scriptscriptstyle{(2)}}^{\alpha\beta}(T,u,\mu).
\end{gathered}     
\label{rel_second_order_1}
\end{equation}
In these equations, $\triangle_1\mathcal{E}_{\scriptscriptstyle{(1)}}'$,
$\triangle_1 \mathcal{P}_{\scriptscriptstyle{(1)}}'$, ...,
$\triangle_1\tau_{\scriptscriptstyle{(1)}}'^{\alpha\beta}$ denote the second-order terms obtained
from the gradient expansions of $\mathcal{E}_{\scriptscriptstyle{(1)}}'(T',u',\mu')$,
$\mathcal{P}_{\scriptscriptstyle{(1)}}'(T',u',\mu')$, ...,
$\tau_{\scriptscriptstyle{(1)}}'^{\alpha\beta}(T',u',\mu')$:
\begin{equation}
\begin{gathered}
\begin{aligned}
\triangle_1 \mathcal{E}_{\scriptscriptstyle{(1)}}'
&= \varepsilon_1'^{\scriptscriptstyle{(1)}}\triangle_1\left[\nabla_{\perp\alpha}'\left(u^\alpha +
     \bar{\delta} u_{\scriptscriptstyle{(1)}}^\alpha\right) \right]\\
&= \varepsilon_1'^{\scriptscriptstyle{(1)}} \left(\nabla_{\perp \alpha}
+ D u_\alpha\right) \bar{\delta} u_{\scriptscriptstyle{(1)}}^{\alpha}, \\
\end{aligned}\\
\triangle_1 \mathcal{P}_{\scriptscriptstyle{(1)}}' =  \pi_1'^{\scriptscriptstyle{(1)}}
     \left(\nabla_{\perp \alpha}+ D u_\alpha\right)
     \bar{\delta} u_{\scriptscriptstyle{(1)}}^{\alpha},\\
 \triangle_1 \mathcal{N}_{\scriptscriptstyle{(1)}}'= \nu_1'^{\scriptscriptstyle{(1)}}
     \left(\nabla_{\perp \alpha}+ D u_\alpha\right)
     \bar{\delta} u_{\scriptscriptstyle{(1)}}^{\alpha},\\
\begin{aligned}
    \triangle_1 \mathcal{Q}_{\scriptscriptstyle{(1)}}'^\alpha & = \theta_1'^{\scriptscriptstyle{(1)}}
     \left(\nabla_{\perp}^{\alpha}\bar{\delta} T_{\scriptscriptstyle{(1)}} + 2u^{^{\scriptstyle{(\alpha}}}
     \bar{\delta} u_{\scriptscriptstyle{(1)}}^{\beta)}\nabla_{\beta}T\right)\\
     & +\theta_2'^{\scriptscriptstyle{(1)}}
     \left(\nabla_{\perp}^{\alpha}\bar{\delta} \mu_{\scriptscriptstyle{(1)}} + 2u^{^{\scriptstyle{(\alpha}}}
     \bar{\delta} u_{\scriptscriptstyle{(1)}}^{\beta)}\nabla_{\beta}\mu\right),
\end{aligned}\\
\begin{aligned}
\triangle_1 \mathcal{J}_{\scriptscriptstyle{(1)}}'^\alpha & = \kappa_1'^{\scriptscriptstyle{(1)}}
     \left(\nabla_{\perp}^{\alpha}\bar{\delta} T_{\scriptscriptstyle{(1)}} + 2u^{^{\scriptstyle{(\alpha}}}
     \bar{\delta} u_{\scriptscriptstyle{(1)}}^{\beta)}\nabla_{\beta}T\right)\\
& + \kappa_2'^{\scriptscriptstyle{(1)}}
     \left(\nabla_{\perp}^{\alpha}\bar{\delta} \mu_{\scriptscriptstyle{(1)}} + 2u^{^{\scriptstyle{(\alpha}}}
     \bar{\delta} u_{\scriptscriptstyle{(1)}}^{\beta)}\nabla_{\beta}\mu\right),
\end{aligned}\\
\begin{aligned}
  \triangle_1 \tau_{\scriptscriptstyle{(1)}}'^{\alpha\beta} & = \eta_1'^{\scriptscriptstyle{(1)}}
     \Big[\big(\nabla_{\perp}^{\langle\alpha} + D u^{\langle\alpha}\big)\bar{\delta}
     u_{\scriptscriptstyle{(1)}}^{\beta\rangle}\\
&+u^{(\alpha}\bar{\delta} u_{\scriptscriptstyle{(1)}}^{\gamma}
     \Big(\nabla_{\perp\gamma}u^{\beta)}+\nabla_{\perp}^{\beta)}u_{\gamma}
     - \frac{2}{d-1}\Delta^{\beta)}_{~~\gamma}\Theta\bigg)\Big].
\end{aligned}
\end{gathered}     
\label{triangle_terms_expansion}
\end{equation}
In the general case, $\triangle_j\Phi_{\scriptscriptstyle{(i)}}'(T+\bar{\delta} T,
Au+\bar{\delta} u ,\mu+\bar{\delta}\mu)$ represents the $(i+j)$-th term of the expansion of
a given coefficient $\Phi_{\scriptscriptstyle{(i)}}'(T',u',\mu')$ in terms of the transverse
gradients of the unprimed variables $\{T,\mu, u^{\alpha}\}$.

By truncating the terms in \eqref{triangle_terms_expansion} to the linear order in the amplitudes and
using the resulting expressions in \eqref{rel_second_order_1}, we obtain
\begin{equation*}
\begin{gathered}
\begin{aligned}
\mathcal{E}_{\scriptscriptstyle{(2)}}'(T,u,\mu) & = \mathcal{E}_{\scriptscriptstyle{(2)}}(T,u,\mu)
- \epsilon_{,\scriptscriptstyle{T}}\bar{\delta} T_{\scriptscriptstyle{(2)}}\\
& - \epsilon_{,\mu}\bar{\delta}\mu_{\scriptscriptstyle{(2)}}
-\varepsilon_1'^{\scriptscriptstyle{(1)}}\nabla_{\perp \alpha}
     \bar{\delta} u_{\scriptscriptstyle{(1)}}^{\alpha},
\end{aligned}\\
\begin{aligned}
\mathcal{P}_{\scriptscriptstyle{(2)}}'(T,u,\mu) &= \mathcal{P}_{\scriptscriptstyle{(2)}}(T,u,\mu)
- p_{,\scriptscriptstyle{T}}\bar{\delta} T_{\scriptscriptstyle{(2)}}\\
&- p_{,\mu}\bar{\delta}\mu_{\scriptscriptstyle{(2)}}
-\pi'^{\scriptscriptstyle{(1)}}\nabla_{\perp \alpha}
     \bar{\delta} u_{\scriptscriptstyle{(1)}}^{\alpha},
\end{aligned}\\
\begin{aligned}
\mathcal{N}_{\scriptscriptstyle{(2)}}'(T,u,\mu) & = \mathcal{N}_{\scriptscriptstyle{(2)}}(T,u,\mu)
- n_{,\scriptscriptstyle{T}}\bar{\delta} T_{\scriptscriptstyle{(2)}}\\
&- n_{,\mu}\bar{\delta}\mu_{\scriptscriptstyle{(2)}} -\nu_1'^{\scriptscriptstyle{(1)}}\nabla_{\perp \alpha}
\bar{\delta} u_{\scriptscriptstyle{(1)}}^{\alpha},
\end{aligned}
\end{gathered}
\end{equation*}
\begin{equation}
\begin{gathered}
\begin{aligned}
\mathcal{Q}_{\scriptscriptstyle{(2)}}'^{\alpha}(T,u,\mu) & = \mathcal{Q}_{\scriptscriptstyle{(2)}}^\alpha(T,u,\mu)
- h\bar{\delta} u_{\scriptscriptstyle{(2)}}^\alpha\\
&- \theta_1'^{\scriptscriptstyle{(1)}}\nabla_{\perp}^{\alpha}\bar{\delta} T_{\scriptscriptstyle{(1)}}
- \theta_2'^{\scriptscriptstyle{(1)}}
\nabla_{\perp}^{\alpha}\bar{\delta} \mu_{\scriptscriptstyle{(1)}},
\end{aligned}\\
\begin{aligned}
\mathcal{J}_{\scriptscriptstyle{(2)}}'^{\alpha}(T,u,\mu) & = \mathcal{J}_{\scriptscriptstyle{(2)}}^\alpha(T,u,\mu)
    - n\bar{\delta} u_{\scriptscriptstyle{(2)}}^\alpha\\
&- \kappa_1'^{\scriptscriptstyle{(1)}}
    \nabla_{\perp}^{\alpha}\bar{\delta} T_{\scriptscriptstyle{(1)}}
- \kappa_2'^{\scriptscriptstyle{(1)}}
     \nabla_{\perp}^{\alpha}\bar{\delta} \mu_{\scriptscriptstyle{(1)}},
\end{aligned}\\
\mathcal{\tau}_{\scriptscriptstyle{(2)}}'^{\alpha\beta}(T,u,\mu) = 
  \mathcal{\tau}_{\scriptscriptstyle{(2)}}'^{\alpha\beta}(T,u,\mu) - \eta_1'^{\scriptscriptstyle{(1)}}
     \nabla_{\perp}^{\langle\alpha} \bar{\delta} u_{\scriptscriptstyle{(1)}}^{\beta\rangle}.
\end{gathered}     
\label{rel_second_order_2}
\end{equation}
Substituting the gradient expansions of both the original and primed variables into \eqref{rel_second_order_2},
together with the second-order versions of \eqref{expansions_delta}, we establish the relations among
the transport coefficients that persist after the linearization process:
\begin{equation}
\begin{aligned}
\varepsilon'^{\scriptscriptstyle{(2)}}_{l} & = \varepsilon^{\scriptscriptstyle{(2)}}_{l} -
\epsilon_{,\scriptscriptstyle{T}} a^{\scriptscriptstyle{(2)}}_{l} - \epsilon_{,\mu} c^{\scriptscriptstyle{(2)}}_{l} -
\varepsilon_1'^{\scriptscriptstyle{(1)}}b_l^{\scriptscriptstyle{(1)}}, \\
\pi'^{\scriptscriptstyle{(2)}}_{l} & = \pi^{\scriptscriptstyle{(2)}}_{l} -
p_{,\scriptscriptstyle{T}} a^{\scriptscriptstyle{(2)}}_{l} - p_{,\mu} c^{\scriptscriptstyle{(2)}}_{l} -
\pi_1'^{\scriptscriptstyle{(1)}}b_l^{\scriptscriptstyle{(1)}}, \\
\nu'^{\scriptscriptstyle{(2)}}_{l} & = \nu^{\scriptscriptstyle{(2)}}_{l} -
n_{,\scriptscriptstyle{T}} a^{\scriptscriptstyle{(2)}}_{l} - n_{,\mu} c^{\scriptscriptstyle{(2)}}_{l} -
\nu_1'^{\scriptscriptstyle{(1)}}b_l^{\scriptscriptstyle{(1)}}, \\
\theta'^{\scriptscriptstyle{(2)}}_{l} & = \theta^{\scriptscriptstyle{(2)}}_{l} -
h b_l^{\scriptscriptstyle{(2)}} - (\theta'^{\scriptscriptstyle{(1)}}_{1} a^{\scriptscriptstyle{(1)}}_{1} +
\theta'^{\scriptscriptstyle{(1)}}_{2} c^{\scriptscriptstyle{(1)}}_{1})\delta_{l}^1, \\
\kappa'^{\scriptscriptstyle{(2)}}_{l} & = \kappa^{\scriptscriptstyle{(2)}}_{l} -
n b_l^{\scriptscriptstyle{(2)}} -  (\kappa'^{\scriptscriptstyle{(1)}}_{1} a^{\scriptscriptstyle{(1)}}_{1} -
\kappa'^{\scriptscriptstyle{(1)}}_{2} c^{\scriptscriptstyle{(1)}}_{1})\delta_{l}^1, \\
\eta'^{\scriptscriptstyle{(2)}}_{l} & = \eta^{\scriptscriptstyle{(2)}}_{l} -
\eta^{\scriptscriptstyle{(1)}}_{1} b_l^{\scriptscriptstyle{(1)}}, 
\end{aligned}
\end{equation}
where $l = 1, 2$ and $\delta_{l}^{j}$ represents the Kronecker delta.

The foregoing second-order coefficients depend on the hydrodynamic frame chosen; however, they can be combined
in a manner analogous to that of the first-order case, thereby yielding the subsequent equations:
\begin{equation}
\begin{gathered}
\pi'^{\scriptscriptstyle{(2)}}_{l} - \beta_\epsilon \varepsilon'^{\scriptscriptstyle{(2)}}_{l}
- \beta_n \nu'^{\scriptscriptstyle{(2)}}_{l}  = \pi^{\scriptscriptstyle{(2)}}_{l} - \beta_\epsilon \varepsilon^{\scriptscriptstyle{(2)}}_{l}
- \beta_n \nu^{\scriptscriptstyle{(2)}}_{l} - f_1^{\scriptscriptstyle{(1)}}b_{l}^{\scriptscriptstyle{(1)}},\\
\kappa'^{\scriptscriptstyle{(2)}}_{l} - \frac{n}{h} \theta'^{\scriptscriptstyle{(2)}}_{l} =
\kappa^{\scriptscriptstyle{(2)}}_{l} - \frac{n}{h} \theta^{\scriptscriptstyle{(2)}}_{l}
- \left(a_{1}^{\scriptscriptstyle{(1)}} \ell^{\scriptscriptstyle{(1)}}_{1}
+ c_{1}^{\scriptscriptstyle{(1)}}\ell^{\scriptscriptstyle{(1)}}_{2}\right)\delta_{l}^1,\\
\eta'^{\scriptscriptstyle{(2)}}_{l} = \eta^{\scriptscriptstyle{(2)}}_{l} - t^{\scriptscriptstyle{(1)}}_{1} b_{l}^{\scriptscriptstyle{(1)}}.
\end{gathered}
\label{rel_coef_second_order}
\end{equation}
The exact functional forms of the coefficients $a_{1}^{\scriptscriptstyle{(1)}}$ and
$c_{1}^{\scriptscriptstyle{(1)}}$ depend on the specific equations selected from
\eqref{rel_coeff_first_order}. These coefficients can be expressed in terms of one of
the following sets of transport coefficients, along with their primed counterparts:
$\{\varepsilon_{1}^{\scriptscriptstyle{(1)}}, \pi_{1}^{\scriptscriptstyle{(1)}}\}$,
$\{\varepsilon_{1}^{\scriptscriptstyle{(1)}}, \nu_{1}^{\scriptscriptstyle{(1)}}\}$ or
$\{\pi_{1}^{\scriptscriptstyle{(1)}}, \nu_{1}^{\scriptscriptstyle{(1)}}\}$.
For example, it is feasible to determine $a_{1}^{\scriptscriptstyle{(1)}}$ and
$c_{1}^{\scriptscriptstyle{(1)}}$ in terms of $\{\varepsilon_{1}^{\scriptscriptstyle{(1)}}, \pi_{1}^{\scriptscriptstyle{(1)}}\}$ and $\{\varepsilon'^{\scriptscriptstyle{(1)}}_{1},
\pi'^{\scriptscriptstyle{(1)}}_{1}\}$ using the subsequent system of equations:
\begin{equation}
\begin{aligned}
\epsilon_{,{\scriptscriptstyle{T}}} a^{\scriptscriptstyle{(1)}}_{1} + \epsilon_{,\mu} c^{\scriptscriptstyle{(1)}}_{1}
& = \varepsilon^{\scriptscriptstyle{(1)}}_{1} - \varepsilon'^{\scriptscriptstyle{(1)}}_{1},\\
p_{,{\scriptscriptstyle{T}}} a^{\scriptscriptstyle{(1)}}_{1} + p_{,\mu} c^{\scriptscriptstyle{(1)}}_{1}
& = \pi^{\scriptscriptstyle{(1)}}_{1} - \pi'^{\scriptscriptstyle{(1)}}_{1}.
\end{aligned}
\end{equation}
Indeed, by solving the above equations for $a_{1}^{\scriptscriptstyle{(1)}}$ and
$c_{1}^{\scriptscriptstyle{(1)}}$, the following expressions are yielded:
\begin{equation}
\begin{split}
a_{1}^{\scriptscriptstyle{(1)}}  & =
\frac{1}{\beta_n (\epsilon_{,\mu} n_{,\scriptscriptstyle{T}}-\epsilon_{,\scriptscriptstyle{T}} n_{,\mu})}
\Big[\epsilon_{,\mu} \left({\pi_{1}^{\scriptscriptstyle{(1)}}} - \pi_{1}'^{\scriptscriptstyle{(1)}}\right)\\
& - \left({\beta_{\epsilon}} {\epsilon_{,\mu}} + {\beta_n} {n_{,\mu}}\right)\varepsilon_{1}^{\scriptscriptstyle{(1)}}
+\left({\beta_{\epsilon}} {\epsilon_{,\mu}} + {\beta_n} {n_{,\mu}}\right)\varepsilon_{1}'^{\scriptscriptstyle{(1)}}\Big],
\end{split}
\label{rel_a11}
\end{equation}
\begin{equation}
\begin{split}
c_{1}^{\scriptscriptstyle{(1)}} & =
\frac{1}{\beta_n (\epsilon_{,\scriptscriptstyle{T}} n_{,\mu} - \epsilon_{,\mu} n_{,\scriptscriptstyle{T}})}
\Big[\epsilon_{,\scriptscriptstyle{T}} {\left({\pi_{1}^{\scriptscriptstyle{(1)}}} - {\pi_{1}'^{\scriptscriptstyle{(1)}}}\right)}\\
&- {\left({\beta_{\epsilon}} {\epsilon_{,\scriptscriptstyle{T}}} + {\beta_n} {n_{,\scriptscriptstyle{T}}}\right)} {\varepsilon_{1}^{\scriptscriptstyle{(1)}}} + {\left({\beta_{\epsilon}} {\epsilon_{,\scriptscriptstyle{T}}} + {\beta_n} {n_{,\scriptscriptstyle{T}}}\right)} {\varepsilon_{1}'^{\scriptscriptstyle{(1)}}}\Big].
\end{split}
\label{rel_c11}
\end{equation}
Similarly, $b_{l}^{\scriptscriptstyle{(1)}}$ can be obtained from $\theta'^{\scriptscriptstyle{(1)}}_{l} = \theta^{\scriptscriptstyle{(1)}}_{l} - h b^{\scriptscriptstyle{(1)}}_{l}$ or $\kappa'^{\scriptscriptstyle{(1)}}_{l} = \kappa^{\scriptscriptstyle{(1)}}_{l} - n b^{\scriptscriptstyle{(1)}}_{l}$, resulting in
\begin{equation}
b^{\scriptscriptstyle{(1)}}_{l} = \frac{\theta^{\scriptscriptstyle{(1)}}_{l} - \theta'^{\scriptscriptstyle{(1)}}_{l}}{h}
\qquad\mbox{or}\qquad
b^{\scriptscriptstyle{(1)}}_{l} = \frac{\kappa^{\scriptscriptstyle{(1)}}_{l} - \kappa'^{\scriptscriptstyle{(1)}}_{l}}{n},
\label{expressoes_bl1}
\end{equation}
where $l = 1, 2$ for all the above equations with the index $l$. Substituting the expressions
\eqref{rel_a11}-\eqref{rel_c11} and the first expression of \eqref{expressoes_bl1} in equations
\eqref{rel_coef_second_order}, we obtain the following set of frame-invariant coefficients:
\begin{equation}
\begin{gathered}
f^{\scriptscriptstyle{(2)}}_{l} \equiv \pi^{\scriptscriptstyle{(2)}}_{l} - \beta_\epsilon \varepsilon^{\scriptscriptstyle{(2)}}_{l}
- \beta_n \nu^{\scriptscriptstyle{(2)}}_{l} - f_1^{\scriptscriptstyle{(1)}}[b_{l}^{\scriptscriptstyle{(1)}}]_{(\theta)},\\
\ell^{\scriptscriptstyle{(2)}}_{l} \equiv \kappa^{\scriptscriptstyle{(2)}}_{l} - \frac{n}{h} \theta^{\scriptscriptstyle{(2)}}_{l}
-\left\{[a_{1}^{\scriptscriptstyle{(1)}}]_{(\varepsilon, \pi)}\,\ell^{\scriptscriptstyle{(1)}}_{1}
+ [c_{1}^{\scriptscriptstyle{(1)}}]_{(\varepsilon, \pi)}\,\ell^{\scriptscriptstyle{(1)}}_{2}\right\}\delta_{l}^1,\\
t^{\scriptscriptstyle{(2)}}_{l} \equiv \eta^{\scriptscriptstyle{(2)}}_{l} - t^{\scriptscriptstyle{(1)}}_{1}[b_{l}^{\scriptscriptstyle{(1)}}]_{(\theta)},\\
\end{gathered}
\label{2order_frame_invariants}
\end{equation}
where $[a_{1}^{\scriptscriptstyle{(1)}}]_{(\varepsilon, \pi)}$, $[c_{1}^{\scriptscriptstyle{(1)}}]_{(\varepsilon,\pi)}$ and $[b_{l}^{\scriptscriptstyle{(1)}}]_{(\theta)}$ refer to the respective parts of $a_{1}^{\scriptscriptstyle{(1)}}$,
$c_{1}^{\scriptscriptstyle{(1)}}$ and $b_{l}^{\scriptscriptstyle{(1)}}$ that depend on $(\varepsilon, \pi)$ and $(\theta)$.
These terms are given by
\begin{equation}
\begin{aligned}
\left [ a_{1}^{\scriptscriptstyle{(1)}} \right ]_{(\varepsilon, \pi)}
& = \bigg(\frac{\partial T}{\partial\epsilon}\bigg)_p \varepsilon_1^{\scriptscriptstyle{(1)}} + \bigg(\frac{\partial T}{\partial p}\bigg)_\epsilon\pi_1^{\scriptscriptstyle{(1)}},\\
\left [ c_{1}^{\scriptscriptstyle{(1)}} \right ]_{(\varepsilon, \pi)} & = \bigg(\frac{\partial \mu}{\partial\epsilon}\bigg)_p\varepsilon_1^{\scriptscriptstyle{(1)}} + \bigg(
\frac{\partial \mu}{\partial p}\bigg)_\epsilon\pi_1^{\scriptscriptstyle{(1)}},\\
\left [ b_{l}^{\scriptscriptstyle{(1)}} \right ]_{(\theta)} & = \frac{1}{h}\theta_l^{\scriptscriptstyle{(1)}},
\qquad\mbox{for}\quad l = 1,\,2,
\end{aligned}
\label{rel_abc_with_coefficients}
\end{equation}
where the partial derivatives in the foregoing equations can be expressed as
\begin{equation}
\begin{gathered}
\bigg(\frac{\partial T}{\partial\epsilon}\bigg)_p =  \bigg(\frac{\partial T}{\partial\epsilon}\bigg)_n
-\frac{\beta_\epsilon}{\beta_n}\bigg(\frac{\partial T}{\partial n}\bigg)_\epsilon
=\frac{\beta_{\epsilon}\epsilon_{,\mu} + \beta_n n_{,\mu}}{\beta_n (\epsilon_{,\scriptscriptstyle{T}} n_{,\mu} - \epsilon_{,\mu} n_{,\scriptscriptstyle{T}})},\\
\bigg(\frac{\partial T}{\partial p}\bigg)_\epsilon = \frac{T_{,n}}{\beta_n} =
\frac{\epsilon_{,\mu}}{\beta_n (\epsilon_{,\mu} n_{,\scriptscriptstyle{T}} - \epsilon_{,\scriptscriptstyle{T}} n_{,\mu})},\\
\bigg(\frac{\partial \mu}{\partial\epsilon}\bigg)_p = \bigg(\frac{\partial \mu}{\partial\epsilon}\bigg)_n
-\frac{\beta_\epsilon}{\beta_n}\bigg(\frac{\partial \mu}{\partial n}\bigg)_\epsilon
=\frac{\beta_{\epsilon}\epsilon_{,\scriptscriptstyle{T}} + \beta_n n_{,\scriptscriptstyle{T}}}{\beta_n (\epsilon_{,\mu} n_{,\scriptscriptstyle{T}} - \epsilon_{,\scriptscriptstyle{T}} n_{,\mu})},\\
\bigg(\frac{\partial \mu}{\partial p}\bigg)_\epsilon = \frac{\mu_{,n}}{\beta_n} =
\frac{\epsilon_{,\scriptscriptstyle{T}}}{\beta_n (\epsilon_{,\scriptscriptstyle{T}} n_{,\mu} - \epsilon_{,\mu} n_{,\scriptscriptstyle{T}})}.
\end{gathered}
\label{partial_derivatives}
\end{equation}

\subsection{Third-order charged-fluid structures}

Third-order gradients furnish the first corrections to even dispersion relations, exemplified by the shear channel
in a static flow. This corresponds to the lowest order in which a longitudinal derivative consistently appears in
the expansion, as the Riemann tensor is not subjected to constraints imposed by ideal fluid equations. For charged
relativistic fluids, the third-order gradient expansion remains incomplete. In the case of a neutral (uncharged)
fluid, this has been rigorously established in \cite{Grozdanov:2015kqa}, hereafter referred to as GK, and further
updated in \cite{Diles:2019uft}.
Third-order corrections have also been explored in the framework of kinetic
theory, and their results are consistent with the gradient expansion approach \cite{Denicol:2010xn, Jaiswal:2012qm, Jaiswal:2013vta, Grozdanov:2015kqa, Brito:2021iqr}.
We present in the following the results pertaining to independent
third-order corrections in the constitutive relations of a charged fluid.

In relation to the third-order scalars, we obtain the following list of structures for one conserved charge:
\begin{equation}
\begin{aligned}
\mathcal{S}^{\scriptscriptstyle{(3)}}_1 &= \nabla_{\perp}^{2}\Theta,\quad
& \mathcal{S}^{\scriptscriptstyle{(3)}}_2 &= \Theta \nabla_{\perp}^2 T,\\
\mathcal{S}^{\scriptscriptstyle{(3)}}_3 &= \Theta \nabla_{\perp}^2 \mu,\quad 
& \mathcal{S}^{\scriptscriptstyle{(3)}}_4 &= \sigma_{\alpha\beta}\nabla_{\perp}^{\alpha}\nabla_{\perp}^{\beta}T,\\
\mathcal{S}^{\scriptscriptstyle{(3)}}_5 &= \sigma_{\alpha\beta}\nabla_{\perp}^{\alpha}\nabla_{\perp}^{\beta}\mu,\quad
& \mathcal{S}^{\scriptscriptstyle{(3)}}_6 &= \nabla_{\perp}^{\alpha}T\nabla_{\perp\alpha}\Theta,\\
\mathcal{S}^{\scriptscriptstyle{(3)}}_{7} &= \nabla_{\perp}^{\alpha}\mu\nabla_{\perp\alpha}\Theta,\quad
& \mathcal{S}^{\scriptscriptstyle{(3)}}_{8} &= \nabla_{\perp}^{\alpha}T\nabla_{\perp}^{\beta}\sigma_{\alpha\beta},\\
\mathcal{S}^{\scriptscriptstyle{(3)}}_{9} &= \nabla_{\perp}^{\alpha}\mu\nabla_{\perp}^{\beta}\sigma_{\alpha\beta},\quad
& \mathcal{S}^{\scriptscriptstyle{(3)}}_{10} &= \Theta^{3},\\
\mathcal{S}^{\scriptscriptstyle{(3)}}_{11} &= \sigma^{\alpha\beta}\sigma_{\alpha}^{\;\gamma}\sigma_{\beta\gamma},\quad
& \mathcal{S}^{\scriptscriptstyle{(3)}}_{12} &= \Theta\sigma^{2},\\
\mathcal{S}^{\scriptscriptstyle{(3)}}_{13} &= \Theta\Omega^{2},\quad 
& \mathcal{S}^{\scriptscriptstyle{(3)}}_{14} & = \sigma^{\alpha\beta}\Omega_{\alpha}^{\;\gamma}\Omega_{\beta\gamma},\\
\mathcal{S}^{\scriptscriptstyle{(3)}}_{15} &= \Theta(\nabla_{\perp}T)^2,\quad 
& \mathcal{S}^{\scriptscriptstyle{(3)}}_{16} &=  \Theta(\nabla_{\perp}\mu)^2,\\
\mathcal{S}^{\scriptscriptstyle{(3)}}_{17} &= \Theta\nabla_{\perp\alpha}T\nabla_{\perp}^{\alpha}\mu,\quad
& \mathcal{S}^{\scriptscriptstyle{(3)}}_{18} &= \sigma_{\alpha\beta}\nabla_{\perp}^{\alpha}T\nabla_{\perp}^{\beta}T,\\
\mathcal{S}^{\scriptscriptstyle{(3)}}_{19} &= \sigma_{\alpha\beta}\nabla_{\perp}^{\alpha}\mu\nabla_{\perp}^{\beta}\mu,\quad
& \mathcal{S}^{\scriptscriptstyle{(3)}}_{20} &= \sigma_{\alpha\beta}\nabla_{\perp}^{\alpha}T\nabla_{\perp}^{\beta}\mu,\\
\mathcal{S}^{\scriptscriptstyle{(3)}}_{21} &= \Omega_{\alpha\beta}\nabla_{\perp}^{\alpha}T\nabla_{\perp}^{\beta}\mu,\quad
& \mathcal{S}^{\scriptscriptstyle{(3)}}_{22} &= \Theta R,\\
\mathcal{S}^{\scriptscriptstyle{(3)}}_{23} &= \sigma^{\alpha\beta}R_{\alpha\beta},\quad
& \mathcal{S}^{\scriptscriptstyle{(3)}}_{24} &= D R,\\
\mathcal{S}^{\scriptscriptstyle{(3)}}_{25} &= u^{\alpha}\nabla_{\perp}^{\beta}T R_{\alpha\beta},\quad
& \mathcal{S}^{\scriptscriptstyle{(3)}}_{26} &= u^{\alpha}\nabla_{\perp}^{\beta}\mu R_{\alpha\beta},\\
\mathcal{S}^{\scriptscriptstyle{(3)}}_{27} &= \Theta u^{\alpha}u^{\beta}R_{\alpha\beta},\quad
& \mathcal{S}^{\scriptscriptstyle{(3)}}_{28} &= \sigma^{\alpha\beta}u^{\gamma}u^{\delta}R{}_{\alpha\gamma\beta\delta},\\
\mathcal{S}^{\scriptscriptstyle{(3)}}_{29} &= u^{\alpha}u^{\beta}DR_{\alpha\beta}.
\end{aligned}
\end{equation}
We note that the scalars $\mathcal{S}^{\scriptscriptstyle{(3)}}_{15}$ and $\mathcal{S}^{\scriptscriptstyle{(3)}}_{18}$ incorporate products of two temperature gradients. Each of these yields two novel scalars for a charged fluid,
corresponding to the two independent substitutions that can be made by replacing the temperature with the chemical
potential. This shows the importance of working with irreducible representations; for each irreducible product of
temperature gradients, there exists a unique and straightforward way for obtaining the corresponding irreducible
gradients of the chemical potential. Furthermore, the presence of two independent scalar gradients leads to the
emergence of terms such as $\mathcal{S}^{\scriptscriptstyle{(3)}}_{16}$, which are absent in the uncharged case
where only a single scalar gradient exists. Symmetry is another key point here: the product of gradients
from the same scalar field remains symmetric upon permutation. Conversely, the direct product of gradients from
two distinct scalars does not possess a definite symmetry, thereby allowing for a non-vanishing anti-symmetric
part being realized in the non-vanishing scalar $\mathcal{S}^{\scriptscriptstyle{(3)}}_{21}$.

To derive the third-order tensors of a charged fluid, we employ the same strategy that is based on prior knowledge
of the gradient expansion in uncharged fluids. The comprehensive list of tensors is obtained by considering all
possible combinations of temperature and chemical potential gradients for each third-order tensor present in the
gradient expansion of the uncharged fluid. The outcomes are ascertained through the computational implementation of
both the Grozdanov-Kaplis (GK) and the Irreducible-Structure (IS) algorithms. This approach yields the subsequent
list of tensors in the case of a system with one conserved charge:
\begin{equation}
\begin{aligned}
\mathcal{T}^{\scriptscriptstyle{(3)}}_1 &= \nabla_{\perp}^{\langle\alpha} \nabla_{\perp}^{\beta\rangle}\Theta,
& \mathcal{T}^{\scriptscriptstyle{(3)}}_2 &= \nabla_{\perp}^{2}\sigma^{\alpha\beta},\\
\mathcal{T}^{\scriptscriptstyle{(3)}}_3 &= \Theta\nabla_{\perp}^{\langle\alpha} \nabla_{\perp}^{\beta\rangle}T,
& \mathcal{T}^{\scriptscriptstyle{(3)}}_4 &= \Theta\nabla_{\perp}^{\langle\alpha} \nabla_{\perp}^{\beta\rangle}\mu,\\
\mathcal{T}^{\scriptscriptstyle{(3)}}_5 &= \sigma^{\alpha\beta}\nabla_{\perp}^{2}T,
& \mathcal{T}^{\scriptscriptstyle{(3)}}_6 &= \sigma^{\alpha\beta}\nabla_{\perp}^{2}\mu,\\
\mathcal{T}^{\scriptscriptstyle{(3)}}_7 &= \sigma_{\gamma}^{\;\langle\alpha}\nabla_{\perp}^{\beta\rangle}\nabla_{\perp}^{\gamma}T,
& \mathcal{T}^{\scriptscriptstyle{(3)}}_8 &= \sigma_{\gamma}^{\;\langle\alpha}\nabla_{\perp}^{\beta\rangle}\nabla_{\perp}^{\gamma}\mu,\\
\mathcal{T}^{\scriptscriptstyle{(3)}}_9 &= \Omega_{\gamma}^{\;\langle\alpha}\nabla_{\perp}^{\beta\rangle}\nabla_{\perp}^{\gamma}T,
& \mathcal{T}^{\scriptscriptstyle{(3)}}_{10} &= \Omega_{\gamma}^{\;\langle\alpha}\nabla_{\perp}^{\beta\rangle}\nabla_{\perp}^{\gamma}\mu,\\
\mathcal{T}^{\scriptscriptstyle{(3)}}_{11} &= \nabla_{\perp}^{\langle\alpha}T\nabla_{\perp}^{\beta\rangle}\Theta,
& \mathcal{T}^{\scriptscriptstyle{(3)}}_{12} &= \nabla_{\perp}^{\langle\alpha}\mu\nabla_{\perp}^{\beta\rangle}\Theta,\\
\mathcal{T}^{\scriptscriptstyle{(3)}}_{13} &= \nabla_{\perp}^{\gamma}T\nabla_{\perp\gamma}\sigma^{\alpha\beta},
& \mathcal{T}^{\scriptscriptstyle{(3)}}_{14} &= \nabla_{\perp}^{\gamma}\mu\nabla_{\perp\gamma}\sigma^{\alpha\beta},\\
\mathcal{T}^{\scriptscriptstyle{(3)}}_{15} &= \nabla_{\perp}^{\gamma}\sigma_{\gamma}^{\;\langle\alpha}\nabla_{\perp}^{\beta\rangle}T,
& \mathcal{T}^{\scriptscriptstyle{(3)}}_{16} &= \nabla_{\perp}^{\gamma}\sigma_{\gamma}^{\;\langle\alpha}\nabla_{\perp}^{\beta\rangle}\mu,\\
\mathcal{T}^{\scriptscriptstyle{(3)}}_{17} &= \nabla_{\perp}^{\gamma}T\nabla_{\perp}^{\langle\alpha}\sigma^{\beta\rangle}_{\;\;\gamma},
& \mathcal{T}^{\scriptscriptstyle{(3)}}_{18} &= \nabla_{\perp}^{\gamma}\mu\nabla_{\perp}^{\langle\alpha}\sigma^{\beta\rangle}_{\;\;\gamma},\\
\mathcal{T}^{\scriptscriptstyle{(3)}}_{19} &= \Theta\nabla_{\perp}^{\langle\alpha}T\nabla_{\perp}^{\beta\rangle}T,
& \mathcal{T}^{\scriptscriptstyle{(3)}}_{20} &= \Theta\nabla_{\perp}^{\langle\alpha}\mu\nabla_{\perp}^{\beta\rangle}\mu,\\
\mathcal{T}^{\scriptscriptstyle{(3)}}_{21} &= \Theta\nabla_{\perp}^{\langle\alpha}T\nabla_{\perp}^{\beta\rangle}\mu,
& \mathcal{T}^{\scriptscriptstyle{(3)}}_{22} &= \sigma^{\alpha\beta}(\nabla_{\perp}T)^2,\\
\mathcal{T}^{\scriptscriptstyle{(3)}}_{23} &= \sigma^{\alpha\beta}(\nabla_{\perp}\mu)^2,
& \mathcal{T}^{\scriptscriptstyle{(3)}}_{24} &= \sigma^{\alpha\beta}\nabla_{\perp}^{\gamma}\mu\nabla_{\perp\gamma}T,\\
\mathcal{T}^{\scriptscriptstyle{(3)}}_{25} &= \sigma_{\gamma}^{\;\langle\alpha}\nabla_{\perp}^{\beta\rangle}T\nabla_{\perp}^{\gamma}T,
& \mathcal{T}^{\scriptscriptstyle{(3)}}_{26} &= \sigma_{\gamma}^{\;\langle\alpha}\nabla_{\perp}^{\beta\rangle}\mu\nabla_{\perp}^{\gamma}\mu,\\
\mathcal{T}^{\scriptscriptstyle{(3)}}_{27} &= \sigma_{\gamma}^{\;\langle\alpha}\nabla_{\perp}^{\beta\rangle}T\nabla_{\perp}^{\gamma}\mu,
& \mathcal{T}^{\scriptscriptstyle{(3)}}_{28} &= \sigma_{\gamma}^{\;\langle\alpha}\nabla_{\perp}^{\beta\rangle}\mu\nabla_{\perp}^{\gamma}T,\\
\mathcal{T}^{\scriptscriptstyle{(3)}}_{29} &= \Omega_{\gamma}^{\;\langle\alpha}\nabla_{\perp}^{\beta\rangle}T\nabla_{\perp}^{\gamma}T,
& \mathcal{T}^{\scriptscriptstyle{(3)}}_{30} &= \Omega_{\gamma}^{\;\langle\alpha}\nabla_{\perp}^{\beta\rangle}\mu\nabla_{\perp}^{\gamma}\mu,\\
\mathcal{T}^{\scriptscriptstyle{(3)}}_{31} &= \Omega_{\gamma}^{\;\langle\alpha}\nabla_{\perp}^{\beta\rangle}T\nabla_{\perp}^{\gamma}\mu,
& \mathcal{T}^{\scriptscriptstyle{(3)}}_{32} &= \Omega_{\gamma}^{\;\langle\alpha}\nabla_{\perp}^{\beta\rangle}\mu\nabla_{\perp}^{\gamma}T,\\
\mathcal{T}^{\scriptscriptstyle{(3)}}_{33} &= \sigma^{\alpha\beta} \sigma^{2},
& \mathcal{T}^{\scriptscriptstyle{(3)}}_{34} &= \sigma_{\gamma}^{\;\;\langle\alpha}\sigma^{\beta\rangle\delta}\sigma_{\delta}^{\;\,\gamma},\\
\mathcal{T}^{\scriptscriptstyle{(3)}}_{35} &= \sigma^{\alpha\beta}\Theta^{2},
& \mathcal{T}^{\scriptscriptstyle{(3)}}_{36} &= \sigma_{\gamma}^{\;\;\langle\alpha}\sigma^{\beta\rangle\gamma}\Theta,\\
\mathcal{T}^{\scriptscriptstyle{(3)}}_{37} &= \Omega_{\gamma}^{\;\;\langle\alpha}\Omega^{\beta\rangle\gamma}\Theta,
& \mathcal{T}^{\scriptscriptstyle{(3)}}_{38} &= \sigma_{\gamma}^{\;\;\langle\alpha}\Omega^{\beta\rangle\delta}\sigma_{\delta}^{\;\,\gamma},\\
\mathcal{T}^{\scriptscriptstyle{(3)}}_{39} &= \sigma^{\alpha\beta}\Omega^{2},
& \mathcal{T}^{\scriptscriptstyle{(3)}}_{40} &= \sigma_{\gamma}^{\;\;\langle\alpha}\Omega^{\beta\rangle\delta}\Omega_{\delta}^{\;\,\gamma},\\
\mathcal{T}^{\scriptscriptstyle{(3)}}_{41} &= \Omega_{\gamma}^{\;\;\langle\alpha}\Omega^{\beta\rangle\delta}\sigma_{\delta}^{\;\,\gamma},
& \mathcal{T}^{\scriptscriptstyle{(3)}}_{42} &= \sigma_{\gamma}^{\;\;\langle\alpha}\Omega^{\beta\rangle\gamma}\Theta,\\
\mathcal{T}^{\scriptscriptstyle{(3)}}_{43} &= \Theta R^{\langle\alpha\beta\rangle},
& \mathcal{T}^{\scriptscriptstyle{(3)}}_{44} &= \sigma^{\alpha\beta}R,\\
\mathcal{T}^{\scriptscriptstyle{(3)}}_{45} &= \sigma_{\gamma}^{\;\;\langle\alpha}R^{\beta\rangle\gamma},
& \mathcal{T}^{\scriptscriptstyle{(3)}}_{46} &= \Omega_{\gamma}^{\;\;\langle\alpha}R^{\beta\rangle\gamma},\\
\mathcal{T}^{\scriptscriptstyle{(3)}}_{47} &= \sigma_{\gamma\delta}R^{\gamma\langle\alpha\beta\rangle\delta},
& \mathcal{T}^{\scriptscriptstyle{(3)}}_{48} &= DR^{\langle\alpha\beta\rangle},\\
\mathcal{T}^{\scriptscriptstyle{(3)}}_{49} &= u_{\gamma}\nabla_{\perp}^{\langle\alpha}R^{\beta\rangle\gamma},
& \mathcal{T}^{\scriptscriptstyle{(3)}}_{50} &= u_{\gamma}\nabla_{\delta}R^{\gamma\langle\alpha\beta\rangle\delta},\\
\mathcal{T}^{\scriptscriptstyle{(3)}}_{51} &= u_{\gamma}R^{\gamma\langle\alpha}\nabla_{\perp}^{\beta\rangle}T,
& \mathcal{T}^{\scriptscriptstyle{(3)}}_{52} &= u_{\gamma}R^{\gamma\langle\alpha}\nabla_{\perp}^{\beta\rangle}\mu,\\
\mathcal{T}^{\scriptscriptstyle{(3)}}_{53} &= u_{\gamma}R^{\gamma\langle\alpha\beta\rangle}_{\qquad\delta}\nabla_{\perp}^{\delta}T,
& \mathcal{T}^{\scriptscriptstyle{(3)}}_{54} &= u_{\gamma}R^{\gamma\langle\alpha\beta\rangle}_{\qquad\delta}\nabla_{\perp}^{\delta}\mu,\\
\mathcal{T}^{\scriptscriptstyle{(3)}}_{55} &= u_{\gamma}u_{\delta}DR^{\gamma\langle\alpha\beta\rangle\delta},
& \mathcal{T}^{\scriptscriptstyle{(3)}}_{56} &= \Theta u_{\gamma}u_{\delta}R^{\gamma\langle\alpha\beta\rangle\delta},\\
\mathcal{T}^{\scriptscriptstyle{(3)}}_{57} &= \sigma^{\alpha\beta}u_{\gamma}u_{\delta}R^{\gamma\delta},
& \mathcal{T}^{\scriptscriptstyle{(3)}}_{58} &= u_{\gamma}u_{\delta}\sigma_{\eta}^{\;\langle\alpha}R^{\beta\rangle\gamma\delta\eta},\\
\mathcal{T}^{\scriptscriptstyle{(3)}}_{59} &= u_{\gamma}u_{\delta}\Omega_{\eta}^{\;\langle\alpha}R^{\beta\rangle\gamma\delta\eta}.
\end{aligned}
\end{equation}

If the Landau frame is chosen, no vector corrections appear in the energy-momentum tensor. For a fluid with one conserved charge, there are 88 new transport coefficients associated with third-order corrections in $T^{\alpha\beta}$. This substantial number of transport coefficients requires simplifications to solve real-world problems; let us examine some special cases. Imposing that spacetime is flat, i.e. $\nabla_\alpha g_{\beta\gamma} = 0$, is of particular interest on the scales of elementary particle processes. In flat spacetime, 21 third-order scalars and 42 third-order tensors survive, requiring 63 transport coefficients. Note that by construction, for a constant chemical potential, we obtain the gradient expansion of
the uncharged fluid.

For a charged fluid in the Landau frame, higher-order corrections appear in the conserved current in the form of transverse vectors. Although there are no vector corrections in the gradient expansion of a neutral fluid in the Landau frame,
the possible third-order transverse vectors that can be constructed for an uncharged fluid were obtained in \cite{Grozdanov:2015kqa} and updated in \cite{Diles:2019uft}. We use this prior knowledge, together with the
computational implementation of both the GK and IS algorithms, to compile the following list of third-order
transverse vectors for the charged fluid:
\begin{equation}
\begin{aligned}
\mathcal{V}^{\scriptscriptstyle{(3)}}_1 &= \nabla_{\perp}^{\alpha}\nabla_{\perp}^{2}T,
& \mathcal{V}^{\scriptscriptstyle{(3)}}_2 &= \nabla_{\perp}^{\alpha}\nabla_{\perp}^{2}\mu,\\
\mathcal{V}^{\scriptscriptstyle{(3)}}_3 &= \nabla_{\perp}^{\alpha}\Theta^{2},
& \mathcal{V}^{\scriptscriptstyle{(3)}}_{4} &= \nabla_{\perp}^{\alpha}\sigma^{2},\\
\mathcal{V}^{\scriptscriptstyle{(3)}}_{5} &= \Theta\nabla_{\perp}^{\beta}\sigma^{\alpha}_{\;\;\beta},
& \mathcal{V}^{\scriptscriptstyle{(3)}}_{6} &= \sigma^{\alpha}_{\;\;\beta}\nabla_{\perp}^{\beta}\Theta,\\
\mathcal{V}^{\scriptscriptstyle{(3)}}_{7} &= \Omega^{\alpha}_{\;\;\beta}\nabla_{\perp}^{\beta}\Theta,
& \mathcal{V}^{\scriptscriptstyle{(3)}}_{8} &=
\sigma^{\alpha}_{\;\;\beta}\nabla_{\perp}^{\gamma}\sigma^{\beta}_{\;\;\gamma},\\
\mathcal{V}^{\scriptscriptstyle{(3)}}_{9} &= \sigma_{\beta\gamma}\nabla_{\perp}^{\beta}\sigma^{\alpha\gamma},
& \mathcal{V}^{\scriptscriptstyle{(3)}}_{10} &=
\Omega^{\alpha}_{\;\;\beta}\nabla_{\perp}^{\gamma}\sigma^{\beta}_{\;\;\gamma},\\
\mathcal{V}^{\scriptscriptstyle{(3)}}_{11} &= \Omega_{\beta\gamma}\nabla_{\perp}^{\beta}\sigma^{\alpha\gamma},
& \mathcal{V}^{\scriptscriptstyle{(3)}}_{12} &= \nabla_{\perp}^{\alpha}T\nabla_{\perp}^{2}T,\\
\mathcal{V}^{\scriptscriptstyle{(3)}}_{13} &= \nabla_{\perp}^{\alpha}\mu\nabla_{\perp}^{2}\mu,
& \mathcal{V}^{\scriptscriptstyle{(3)}}_{14} &= \nabla_{\perp}^{\alpha}T\nabla_{\perp}^{2}\mu,\\
\mathcal{V}^{\scriptscriptstyle{(3)}}_{15} &= \nabla_{\perp}^{\alpha}\mu\nabla_{\perp}^{2}T,
& \mathcal{V}^{\scriptscriptstyle{(3)}}_{16} &=
\nabla_{\perp}^{\alpha}\nabla_{\perp}^{\beta}T\nabla_{\perp\beta}T,\\
\mathcal{V}^{\scriptscriptstyle{(3)}}_{17} &= \nabla_{\perp}^{\alpha}\nabla_{\perp}^{\beta}\mu\nabla_{\perp\beta}\mu,
& \mathcal{V}^{\scriptscriptstyle{(3)}}_{18} &= \nabla_{\perp}^{\alpha}\nabla_{\perp}^{\beta}T\nabla_{\perp\beta}\mu,\\
\mathcal{V}^{\scriptscriptstyle{(3)}}_{19} &= \nabla_{\perp}^{\alpha}\nabla_{\perp}^{\beta}\mu\nabla_{\perp\beta}T,
& \mathcal{V}^{\scriptscriptstyle{(3)}}_{20} &= \Theta^{2}\nabla_{\perp}^{\alpha}T,\\
\mathcal{V}^{\scriptscriptstyle{(3)}}_{21} &= \Theta^{2}\nabla_{\perp}^{\alpha}\mu,
& \mathcal{V}^{\scriptscriptstyle{(3)}}_{22} &= \sigma^{2}\nabla_{\perp}^{\alpha}T,\\
\mathcal{V}^{\scriptscriptstyle{(3)}}_{23} &= \sigma^{2}\nabla_{\perp}^{\alpha}\mu,
& \mathcal{V}^{\scriptscriptstyle{(3)}}_{24} &= \Omega^{2}\nabla_{\perp}^{\alpha}T,\\
\mathcal{V}^{\scriptscriptstyle{(3)}}_{25} &= \Omega^{2}\nabla_{\perp}^{\alpha}\mu,
& \mathcal{V}^{\scriptscriptstyle{(3)}}_{26} &= \Theta\sigma^{\alpha}_{\;\;\beta}\nabla_{\perp}^{\beta}T,\\
\mathcal{V}^{\scriptscriptstyle{(3)}}_{27} &= \Theta\sigma^{\alpha}_{\;\;\beta}\nabla_{\perp}^{\beta}\mu,
& \mathcal{V}^{\scriptscriptstyle{(3)}}_{28} &= \Theta\Omega^{\alpha}_{\;\;\beta}\nabla_{\perp}^{\beta}T,\\
\mathcal{V}^{\scriptscriptstyle{(3)}}_{29} &= \Theta\Omega^{\alpha}_{\;\;\beta}\nabla_{\perp}^{\beta}\mu,
& \mathcal{V}^{\scriptscriptstyle{(3)}}_{30} &= \sigma^{\alpha\beta}\sigma_{\beta\gamma}\nabla_{\perp}^{\gamma}T,\\
\mathcal{V}^{\scriptscriptstyle{(3)}}_{31} &= \sigma^{\alpha\beta}\sigma_{\beta\gamma}\nabla_{\perp}^{\gamma}\mu,
& \mathcal{V}^{\scriptscriptstyle{(3)}}_{32} &= \Omega^{\alpha\beta}\Omega_{\beta\gamma}\nabla_{\perp}^{\gamma}T,\\
\mathcal{V}^{\scriptscriptstyle{(3)}}_{33} &= \Omega^{\alpha\beta}\Omega_{\beta\gamma}\nabla_{\perp}^{\gamma}\mu,
& \mathcal{V}^{\scriptscriptstyle{(3)}}_{34} &= \sigma^{\alpha\beta}\Omega_{\beta\gamma}\nabla_{\perp}^{\gamma}T,\\
\mathcal{V}^{\scriptscriptstyle{(3)}}_{35} &= \sigma^{\alpha\beta}\Omega_{\beta\gamma}\nabla_{\perp}^{\gamma}\mu,
& \mathcal{V}^{\scriptscriptstyle{(3)}}_{36} &= \Omega^{\alpha\beta}\sigma_{\beta\gamma}\nabla_{\perp}^{\gamma}T,\\
\mathcal{V}^{\scriptscriptstyle{(3)}}_{37} &= \Omega^{\alpha\beta}\sigma_{\beta\gamma}\nabla_{\perp}^{\gamma}\mu,
& \mathcal{V}^{\scriptscriptstyle{(3)}}_{38} &= \nabla_{\perp}^{\alpha}T(\nabla_{\perp}T)^2,\\
\mathcal{V}^{\scriptscriptstyle{(3)}}_{39} &= \nabla_{\perp}^{\alpha}\mu(\nabla_{\perp}\mu)^2,
& \mathcal{V}^{\scriptscriptstyle{(3)}}_{40} &= \nabla_{\perp}^{\alpha}T(\nabla_{\perp}\mu)^2,\\
\mathcal{V}^{\scriptscriptstyle{(3)}}_{41} &= \nabla_{\perp}^{\alpha}\mu(\nabla_{\perp}T)^2,
& \mathcal{V}^{\scriptscriptstyle{(3)}}_{42} &= \nabla_{\perp}^{\alpha}T\nabla_{\perp\beta}T\nabla_{\perp}^{\beta}\mu,\\
\mathcal{V}^{\scriptscriptstyle{(3)}}_{43} &= \nabla_{\perp}^{\alpha}\mu\nabla_{\perp\beta}T\nabla_{\perp}^{\beta}\mu,
& \mathcal{V}^{\scriptscriptstyle{(3)}}_{44} &= \nabla_{\perp}^{\alpha}R,\\
\mathcal{V}^{\scriptscriptstyle{(3)}}_{45} &= R\nabla_{\perp}^{\alpha}T,
& \mathcal{V}^{\scriptscriptstyle{(3)}}_{46} &= R\nabla_{\perp}^{\alpha}\mu,\\
\mathcal{V}^{\scriptscriptstyle{(3)}}_{47} &= \Delta^{\alpha\beta}R_{\beta\gamma}\nabla_{\perp}^{\gamma}T,
& \mathcal{V}^{\scriptscriptstyle{(3)}}_{48} &= \Delta^{\alpha\beta}R_{\beta\gamma}\nabla_{\perp}^{\gamma}\mu,\\
\mathcal{V}^{\scriptscriptstyle{(3)}}_{49} &= \Theta\Delta^{\alpha\beta}u^{\gamma}R_{\beta\gamma},
& \mathcal{V}^{\scriptscriptstyle{(3)}}_{50} &= \sigma^{\alpha\beta}u^{\gamma}R_{\beta\gamma},\\
\mathcal{V}^{\scriptscriptstyle{(3)}}_{51} &= \Omega^{\alpha\beta}u^{\gamma}R_{\beta\gamma},
& \mathcal{V}^{\scriptscriptstyle{(3)}}_{52} &=
\Delta^{\alpha\beta}\sigma^{\gamma\delta}u^{\eta}R_{\beta\gamma\delta\eta},\\
\mathcal{V}^{\scriptscriptstyle{(3)}}_{53} &= \Delta^{\alpha\beta}\Omega^{\gamma\delta}u^{\eta}R_{\beta\gamma\delta\eta},
& \mathcal{V}^{\scriptscriptstyle{(3)}}_{54} &= \Delta^{\alpha\beta}u^{\gamma}u^{\delta}\nabla^{\eta}R_{\beta\gamma\delta\eta},\\
\mathcal{V}^{\scriptscriptstyle{(3)}}_{55} &= u^{\beta}u^{\gamma}R_{\beta\gamma}\nabla_{\perp}^{\alpha}T,
& \mathcal{V}^{\scriptscriptstyle{(3)}}_{56} &=
\Delta^{\alpha\beta} u^{\gamma}u^{\delta} R_{\beta\gamma\delta\eta} \nabla_{\perp}^{\eta}T, \\
\mathcal{V}^{\scriptscriptstyle{(3)}}_{57} &= u^{\beta}u^{\gamma}R_{\beta\gamma}\nabla_{\perp}^{\alpha}\mu,
& \mathcal{V}^{\scriptscriptstyle{(3)}}_{58} &=
\Delta^{\alpha\beta}u^{\gamma}u^{\delta}R_{\beta\gamma\delta\eta} \nabla_{\perp}^{\eta}\mu,\\
\mathcal{V}^{\scriptscriptstyle{(3)}}_{59} &=
u^{\beta}u^{\gamma}\nabla_{\perp}^{\alpha}R_{\beta\gamma}.
\end{aligned}
\end{equation}

Based on the foregoing tensorial structures, we obtain the following constitutive relations
for the energy-momentum tensor
\begin{equation}
\begin{split}
\Pi_{\scriptscriptstyle{(3)}}^{\alpha\beta} & = \sum_{j=1}^{29} \varepsilon_{j}^{\scriptscriptstyle{(3)}}\mathcal{S}^{\scriptscriptstyle{(3)}}_{j} u^\alpha u^\beta
+ \sum_{i=j}^{29} \pi_{j}^{\scriptscriptstyle{(3)}}\mathcal{S}^{\scriptscriptstyle{(3)}}_{j}\Delta^{\alpha\beta}\\
&+ \sum_{j=1}^{59} \theta_{j}^{\scriptscriptstyle{(3)}}(\mathcal{V}^{\scriptscriptstyle{(3)}}_{j})^{(\alpha} u^{\beta)} 
+\sum_{j=1}^{59} \eta_{j}^{\scriptscriptstyle{(3)}} (\mathcal{T}^{\scriptscriptstyle{(3)}}_{j})^{\alpha\beta},
\end{split}
\end{equation}
and for the current
\begin{equation}
\Upsilon_{\scriptscriptstyle{(3)}}^{\alpha} = \sum_{j=1}^{29} \nu_{j}^{\scriptscriptstyle{(3)}}\mathcal{S}^{\scriptscriptstyle{(3)}}_{j}u^\alpha + \sum_{j=1}^{59} \kappa_{j}^{\scriptscriptstyle{(3)}}(\mathcal{V}^{\scriptscriptstyle{(3)}}_{j})^{\alpha}.
\end{equation}
Therefore, the most general third-order correction to the energy-momentum tensor of a fluid with one conserved charge
involves 176 transport coefficients. Of these, 127 are relevant in flat spacetime. The third-order correction to the
matter current contains 88 transport coefficients, with 64 being present in the flat spacetime case.

It is worth noting that not all of the coefficients of the constitutive relations mentioned above are independent. Subsequently, we demonstrate that they can be combined in a set of 147 frame-invariant quantities. Our primary focus
is on identifying the linear frame-invariant coefficients relevant in flat space-time. These are the coefficients that
manifest themselves in the dispersion relations of sound waves, diffusion, and shear modes. By solely considering the
three derivative terms of linear order in relations \eqref{rel_transf_para_E}--\eqref{rel_transf_para_Tau},
we find the following:
\begin{equation}
\begin{gathered}
    \mathcal{E}_{\scriptscriptstyle{(3)}}'(T,u,\mu) + \triangle_2
    \mathcal{E}_{\scriptscriptstyle{(1)}}' + \triangle_1
    \mathcal{E}_{\scriptscriptstyle{(2)}}' + \bar{\delta} \epsilon_{\scriptscriptstyle{(3)}} 
     = \mathcal{E}_{\scriptscriptstyle{(3)}}(T,u,\mu), \\
   \mathcal{P}_{\scriptscriptstyle{(3)}}'(T,u,\mu) + \triangle_2
 \mathcal{P}_{\scriptscriptstyle{(1)}}' + \triangle_1
 \mathcal{P}_{\scriptscriptstyle{(2)}}' +  \bar{\delta} p_{\scriptscriptstyle{(3)}} 
 = \mathcal{P}_{\scriptscriptstyle{(3)}}(T,u,\mu),\\
  \mathcal{N}_{\scriptscriptstyle{(3)}}'(T,u,\mu) + \triangle_2
    \mathcal{N}_{\scriptscriptstyle{(1)}}' + \triangle_1
    \mathcal{N}_{\scriptscriptstyle{(2)}}' + \bar{\delta} n_{\scriptscriptstyle{(3)}} 
    = \mathcal{N}_{\scriptscriptstyle{(3)}}(T,u,\mu),\\
    \mathcal{Q}_{\scriptscriptstyle{(3)}}'^{\alpha}(T,u,\mu) + \triangle_2
    \mathcal{Q}_{\scriptscriptstyle{(1)}}'^{\alpha} + \triangle_1
    \mathcal{Q}_{\scriptscriptstyle{(2)}}'^{\alpha}  = \mathcal{Q}_{\scriptscriptstyle{(3)}}^\alpha(T,u,\mu)
    -h\bar{\delta} u_{\scriptscriptstyle{(3)}}^\alpha,\\
    \mathcal{J}_{\scriptscriptstyle{(3)}}'^\alpha(T,u,\mu) + \triangle_2
    \mathcal{J}_{\scriptscriptstyle{(1)}}'^{\alpha} + \triangle_1
    \mathcal{J}_{\scriptscriptstyle{(2)}}'^{\alpha} = \mathcal{J}_{\scriptscriptstyle{(3)}}^\alpha(T,u,\mu)
  - n \bar{\delta} u_{\scriptscriptstyle{(3)}}^\alpha,\\
  \mathcal{\tau}_{\scriptscriptstyle{(3)}}'^{\alpha\beta}(T,u,\mu) + \triangle_2
    \tau_{\scriptscriptstyle{(1)}}'^{\alpha\beta} + \triangle_1
    \tau_{\scriptscriptstyle{(2)}}'^{\alpha\beta} = \mathcal{\tau}_{\scriptscriptstyle{(3)}}'^{\alpha\beta}(T,u,\mu).
\end{gathered}     
\label{rel_third_order_1}
\end{equation}
In these equations, $\triangle_2\mathcal{E}_{\scriptscriptstyle{(1)}}'$,
$\triangle_2 \mathcal{P}_{\scriptscriptstyle{(1)}}'$, ...,
$\triangle_2\tau_{\scriptscriptstyle{(1)}}'^{\alpha\beta}$ denote the third-order terms
originated from the gradient expansions of $\mathcal{E}_{\scriptscriptstyle{(1)}}'(T',u',\mu')$,
$\mathcal{P}_{\scriptscriptstyle{(1)}}'(T',u',\mu')$, ...,
$\tau_{\scriptscriptstyle{(1)}}'^{\alpha\beta}(T',u',\mu')$:
\begin{equation}
\begin{gathered}
\triangle_2 \mathcal{E}_{\scriptscriptstyle{(1)}}' = \varepsilon_1'^{\scriptscriptstyle{(1)}} \nabla_{\perp \alpha}
     \bar{\delta} u_{\scriptscriptstyle{(2)}}^{\alpha} + \cdots,\\
\triangle_2 \mathcal{P}_{\scriptscriptstyle{(1)}}' =  \pi_1'^{\scriptscriptstyle{(1)}}
     \nabla_{\perp \alpha}\bar{\delta} u_{\scriptscriptstyle{(2)}}^{\alpha} + \cdots,\\
\triangle_2 \mathcal{N}_{\scriptscriptstyle{(1)}}' = \nu_1'^{\scriptscriptstyle{(1)}}
     \nabla_{\perp \alpha}\bar{\delta} u_{\scriptscriptstyle{(1)}}^{\alpha}+ \cdots,\\
\triangle_2 \mathcal{Q}_{\scriptscriptstyle{(1)}}'^\alpha = \theta_1'^{\scriptscriptstyle{(1)}}\nabla_{\perp}^{\alpha}\bar{\delta} T_{\scriptscriptstyle{(2)}} +\theta_2'^{\scriptscriptstyle{(1)}}
     \nabla_{\perp}^{\alpha}\bar{\delta} \mu_{\scriptscriptstyle{(2)}} + \cdots,\\
\triangle_2 \mathcal{J}_{\scriptscriptstyle{(1)}}'^\alpha = \kappa_1'^{\scriptscriptstyle{(1)}}
    \nabla_{\perp}^{\alpha}\bar{\delta} T_{\scriptscriptstyle{(2)}}
     +\kappa_2'^{\scriptscriptstyle{(1)}}
     \nabla_{\perp}^{\alpha}\bar{\delta} \mu_{\scriptscriptstyle{(2)}}+ \cdots,\\
    \triangle_2 \tau_{\scriptscriptstyle{(1)}}'^{\alpha} = \eta_1'^{\scriptscriptstyle{(1)}}
     \nabla_{\perp}^{\langle\alpha}\bar{\delta} u_{\scriptscriptstyle{(2)}}^{\beta\rangle} + \cdots\,.
\end{gathered}     
\label{Delta2_expansion_3rd}
\end{equation}
In a similar way, $\triangle_1\mathcal{E}_{\scriptscriptstyle{(2)}}'$,
$\triangle_1 \mathcal{P}_{\scriptscriptstyle{(2)}}'$, ...,
$\triangle_1\tau_{\scriptscriptstyle{(2)}}'^{\alpha\beta}$ represent the third-order terms obtained
from the gradient expansions of $\mathcal{E}_{\scriptscriptstyle{(2)}}'(T',u',\mu')$,
$\mathcal{P}_{\scriptscriptstyle{(2)}}'(T',u',\mu')$, ...,
$\tau_{\scriptscriptstyle{(2)}}'^{\alpha\beta}(T',u',\mu')$:
\begin{equation}
\begin{gathered}
    \triangle_1 \mathcal{E}_{\scriptscriptstyle{(2)}}' = \varepsilon_1'^{\scriptscriptstyle{(2)}} \nabla_{\perp}^2
     \bar{\delta} T_{\scriptscriptstyle{(1)}} + \varepsilon_2'^{\scriptscriptstyle{(2)}} \nabla_{\perp}^2
     \bar{\delta} \mu_{\scriptscriptstyle{(1)}} + \cdots,\\  
   \triangle_2 \mathcal{P}_{\scriptscriptstyle{(1)}}' =  \pi_1'^{\scriptscriptstyle{(2)}} \nabla_{\perp}^2
     \bar{\delta} T_{\scriptscriptstyle{(1)}} + \pi_2'^{\scriptscriptstyle{(2)}} \nabla_{\perp}^2
     \bar{\delta} \mu_{\scriptscriptstyle{(1)}} + \cdots,\\
 \triangle_2 \mathcal{N}_{\scriptscriptstyle{(1)}}'  = \nu_1'^{\scriptscriptstyle{(2)}} \nabla_{\perp}^2
     \bar{\delta} T_{\scriptscriptstyle{(1)}} + \nu_2'^{\scriptscriptstyle{(2)}} \nabla_{\perp}^2
     \bar{\delta} \mu_{\scriptscriptstyle{(1)}} + \cdots,\\
\begin{aligned}
     \triangle_2 \mathcal{Q}_{\scriptscriptstyle{(1)}}'^\alpha = \theta_1'^{\scriptscriptstyle{(2)}}
    & \nabla_{\perp}^\alpha \nabla_{\perp \beta}\bar{\delta} u_{\scriptscriptstyle{(1)}}^{\beta}\\
& + \theta_2'^{\scriptscriptstyle{(2)}}\Delta^{\alpha}_{~\beta}\nabla_{\perp \gamma}
    \nabla_{\perp}^{\langle \gamma}\bar{\delta} u_{\scriptscriptstyle{(1)}}^{\beta\rangle} + \cdots,
\end{aligned}\\
\begin{aligned}
   \triangle_2 \mathcal{J}_{\scriptscriptstyle{(1)}}'^\alpha = \kappa_1'^{\scriptscriptstyle{(2)}}
    & \nabla_{\perp}^\alpha \nabla_{\perp \beta}\bar{\delta} u_{\scriptscriptstyle{(1)}}^{\beta}\\
& + \kappa_2'^{\scriptscriptstyle{(2)}}\Delta^{\alpha}_{~\beta}\nabla_{\perp \gamma}
     \nabla_{\perp}^{\langle \gamma}\bar{\delta} u_{\scriptscriptstyle{(1)}}^{\beta\rangle} + \cdots,
\end{aligned}\\
\triangle_2 \tau_{\scriptscriptstyle{(1)}}'^{\alpha} = \eta_1'^{\scriptscriptstyle{(2)}}
     \nabla_{_\perp}^{\langle \alpha} \nabla_{_\perp}^{\beta\rangle} \bar{\delta} T_{\scriptscriptstyle{(1)}}
     + \eta_2'^{\scriptscriptstyle{(2)}} \nabla_{_\perp}^{\langle  \alpha} \nabla_{_\perp}^{\beta\rangle}
\bar{\delta} \mu_{\scriptscriptstyle{(1)}} + \cdots.
\end{gathered}     
\label{Delta1_expansion_3rd}
\end{equation}

The ellipses in \eqref{Delta2_expansion_3rd} and \eqref{Delta1_expansion_3rd} indicate
omitted nonlinear terms in the amplitudes. Since we are interested in the frame-invariant coefficients that
persist in the linear regime, these terms can be ignored. Thus, after substitution of \eqref{Delta2_expansion_3rd}
and \eqref{Delta1_expansion_3rd} in \eqref{rel_third_order_1}, the linearized resulting equations are given by
\begin{equation}
\begin{aligned}
    \mathcal{E}_{\scriptscriptstyle{(3)}}'(T,u,\mu) & = \mathcal{E}_{\scriptscriptstyle{(3)}}(T,u,\mu)
            - \epsilon_{,\scriptscriptstyle{T}}\delta T_{\scriptscriptstyle{(3)}}\\
            & - \epsilon_{,\mu}\delta \mu_{\scriptscriptstyle{(3)}}
            - \varepsilon_1'^{\scriptscriptstyle{(1)}}\nabla_{\perp \alpha}\delta u_{\scriptscriptstyle{(2)}}^{\alpha}\\
            &- \varepsilon_1'^{\scriptscriptstyle{(2)}} \nabla_{\perp}^2\delta T_{\scriptscriptstyle{(1)}}
            - \varepsilon_2'^{\scriptscriptstyle{(2)}} \nabla_{\perp}^2\delta \mu_{\scriptscriptstyle{(1)}},
\end{aligned}
\end{equation}
\begin{equation}
\begin{aligned}
    \mathcal{P}_{\scriptscriptstyle{(3)}}'(T,u,\mu) & = \mathcal{P}_{\scriptscriptstyle{(3)}}(T,u,\mu)
            - p_{,\scriptscriptstyle{T}}\delta T_{\scriptscriptstyle{(3)}}\\
            &- p_{,\mu}\delta \mu_{\scriptscriptstyle{(3)}}
            - \pi_1'^{\scriptscriptstyle{(1)}}\nabla_{\perp \alpha}\delta u_{\scriptscriptstyle{(2)}}^{\alpha}\\
            &- \pi_1'^{\scriptscriptstyle{(2)}} \nabla_{\perp}^2\delta T_{\scriptscriptstyle{(1)}}
            - \pi_2'^{\scriptscriptstyle{(2)}} \nabla_{\perp}^2\delta \mu_{\scriptscriptstyle{(1)}},
\end{aligned}
\end{equation}
\begin{equation}
\begin{aligned}
    \mathcal{N}_{\scriptscriptstyle{(3)}}' (T,u,\mu) & = \mathcal{N}_{\scriptscriptstyle{(3)}}(T,u,\mu)
            - n_{,\scriptscriptstyle{T}}\delta T_{\scriptscriptstyle{(3)}}\\
            &- n_{,\mu}\delta \mu_{\scriptscriptstyle{(3)}}
            - \nu_1'^{\scriptscriptstyle{(1)}}\nabla_{\perp \alpha}\delta u_{\scriptscriptstyle{(2)}}^{\alpha}\\
            & - \nu_1'^{\scriptscriptstyle{(2)}} \nabla_{\perp}^2\delta T_{\scriptscriptstyle{(1)}}
            - \nu_2'^{\scriptscriptstyle{(2)}} \nabla_{\perp}^2\delta \mu_{\scriptscriptstyle{(1)}},
\end{aligned}
\end{equation}
\begin{equation}
\begin{aligned}
    \mathcal{Q}_{\scriptscriptstyle{(3)}}'^{\alpha} (T,u,\mu) & = \mathcal{Q}_{\scriptscriptstyle{(3)}}^\alpha(T,u,\mu)
            - h\delta u_{\scriptscriptstyle{(3)}}^\alpha\\
            & - \theta_1'^{\scriptscriptstyle{(1)}}\nabla_{\perp}^{\alpha}\delta T_{\scriptscriptstyle{(2)}}
            - \theta_2'^{\scriptscriptstyle{(1)}}\nabla_{\perp}^{\alpha}\delta \mu_{\scriptscriptstyle{(2)}}\\
            & - \theta_1'^{\scriptscriptstyle{(2)}}\nabla_{\perp}^\alpha \nabla_{\perp\beta}
              \delta u_{\scriptscriptstyle{(1)}}^{\beta}
            - \theta_2'^{\scriptscriptstyle{(2)}}\Delta^{\alpha}_{~\beta}\nabla_{\perp \gamma}
              \nabla_{\perp}^{\langle \gamma}\delta u_{\scriptscriptstyle{(1)}}^{\beta\rangle},
\end{aligned}
\end{equation}
\begin{equation}
\begin{aligned}
    \mathcal{J}_{\scriptscriptstyle{(3)}}'^{\alpha}(T,u,\mu) & = \mathcal{J}_{\scriptscriptstyle{(3)}}^\alpha(T,u,\mu)
            - n\delta u_{\scriptscriptstyle{(3)}}^\alpha\\
            &- \kappa_1'^{\scriptscriptstyle{(1)}}\nabla_{\perp}^{\alpha}\delta T_{\scriptscriptstyle{(2)}}
            - \kappa_2'^{\scriptscriptstyle{(1)}}\nabla_{\perp}^{\alpha}\delta \mu_{\scriptscriptstyle{(2)}}\\
            &- \kappa_1'^{\scriptscriptstyle{(2)}}\nabla_{\perp}^\alpha \nabla_{\perp \beta}
              \delta u_{\scriptscriptstyle{(1)}}^{\beta}
            - \kappa_2'^{\scriptscriptstyle{(2)}}\Delta^{\alpha}_{~\beta}\nabla_{\perp \gamma}
              \nabla_{\perp}^{\langle \gamma}\delta u_{\scriptscriptstyle{(1)}}^{\beta\rangle},
\end{aligned}
\end{equation}
\begin{equation}
\begin{aligned}
    \mathcal{\tau}_{\scriptscriptstyle{(3)}}'^{\alpha\beta}(T,u,\mu) & =
            \mathcal{\tau}_{\scriptscriptstyle{(3)}}'^{\alpha\beta}(T,u,\mu)
            - \eta_1'^{\scriptscriptstyle{(1)}}\nabla_{\perp}^{\langle\alpha}\delta
            u_{\scriptscriptstyle{(2)}}^{\beta\rangle}\\
            & - \eta_1'^{\scriptscriptstyle{(2)}} \nabla_{_\perp}^{\langle \alpha} \nabla_{_\perp}^{\beta\rangle}
                \delta T_{\scriptscriptstyle{(1)}}
            - \eta_2'^{\scriptscriptstyle{(2)}} \nabla_{_\perp}^{\langle  \alpha} \nabla_{_\perp}^{\beta\rangle}
                \delta \mu_{\scriptscriptstyle{(1)}}.
\end{aligned}
\end{equation}     
Substituting the gradient expansions of the original and primed variables into the above equations,
together with the third-order versions of \eqref{expansions_delta}, we obtain the following relations among the transport coefficients that survive the linearization process:
\begin{equation}
\begin{gathered}
\begin{aligned}
\varepsilon'^{\scriptscriptstyle{(3)}}_{1} & = \varepsilon^{\scriptscriptstyle{(3)}}_{1}
        - a_{1}^{\scriptscriptstyle{(3)}} \epsilon_{,\scriptscriptstyle{T}}
        - c_{1}^{\scriptscriptstyle{(3)}}\epsilon_{,\mu}\\
        &- \Big(b_{1}^{\scriptscriptstyle{(2)}}
        + \frac{d-2}{d-1}b_{2}^{\scriptscriptstyle{(2)}}\Big) \varepsilon'^{\scriptscriptstyle{(1)}}_{1}
        - {a_{1}^{\scriptscriptstyle{(1)}}} \varepsilon'^{\scriptscriptstyle{(2)}}_{1}
        - {c_{1}^{\scriptscriptstyle{(1)}}} \varepsilon'^{\scriptscriptstyle{(2)}}_{2},
\end{aligned}\\
\begin{aligned}
\pi'^{\scriptscriptstyle{(3)}}_{1} & = \pi^{\scriptscriptstyle{(3)}}_{1}
        - a_{1}^{\scriptscriptstyle{(3)}} p_{,\scriptscriptstyle{T}}
        - c_{1}^{\scriptscriptstyle{(3)}} p_{,\mu} \\
        &- \Big(b_{1}^{\scriptscriptstyle{(2)}}
        + \frac{d-2}{d-1}b_{2}^{\scriptscriptstyle{(2)}}\Big) \pi'^{\scriptscriptstyle{(1)}}_{1}
        - {a_{1}^{\scriptscriptstyle{(1)}}} \varepsilon'^{\scriptscriptstyle{(2)}}_{1}
        - {c_{1}^{\scriptscriptstyle{(1)}}} \varepsilon'^{\scriptscriptstyle{(2)}}_{2},
\end{aligned}\\
\begin{aligned}
\nu'^{\scriptscriptstyle{(3)}}_{1} & = \nu^{\scriptscriptstyle{(3)}}_{1}
        - a_{1}^{\scriptscriptstyle{(3)}} n_{,\scriptscriptstyle{T}}
        - c_{1}^{\scriptscriptstyle{(3)}} n_{,\mu}\\
        &- \Big(b_{1}^{\scriptscriptstyle{(2)}}
        + \frac{d-2}{d-1}b_{2}^{\scriptscriptstyle{(2)}}\Big) \nu'^{\scriptscriptstyle{(1)}}_{1}
        - a_{1}^{\scriptscriptstyle{(1)}} \nu'^{\scriptscriptstyle{(2)}}_{1}
        - c_{1}^{\scriptscriptstyle{(1)}} \nu'^{\scriptscriptstyle{(2)}}_{2},
\end{aligned}\\
\begin{aligned}
\theta'^{\scriptscriptstyle{(3)}}_{l} = \theta^{\scriptscriptstyle{(3)}}_{l}
        &- h b_{l}^{\scriptscriptstyle{(3)}} - a_{l}^{\scriptscriptstyle{(2)}} \theta'^{\scriptscriptstyle{(1)}}_{1}\\
        & - c_{l}^{\scriptscriptstyle{(2)}} \theta'^{\scriptscriptstyle{(1)}}_{2}
        - b_{l}^{\scriptscriptstyle{(1)}}\Big(\theta'^{\scriptscriptstyle{(2)}}_{1}
        + \frac{d-2}{d-1}\theta'^{\scriptscriptstyle{(2)}}_{2}\Big),
\end{aligned}\\
\begin{aligned}
\kappa'^{\scriptscriptstyle{(3)}}_{l} = \kappa^{\scriptscriptstyle{(3)}}_{l}
        &- n b_{l}^{\scriptscriptstyle{(3)}} - a_{l}^{\scriptscriptstyle{(2)}} \kappa'^{\scriptscriptstyle{(1)}}_{l}\\
        &- c_{l}^{\scriptscriptstyle{(2)}} \kappa'^{\scriptscriptstyle{(1)}}_{2}
        - b_{l}^{\scriptscriptstyle{(1)}}\Big(\kappa'^{\scriptscriptstyle{(2)}}_{1}
        + \frac{d-2}{d-1}\kappa'^{\scriptscriptstyle{(2)}}_{2}\Big),
\end{aligned}\\
\begin{aligned}
\eta'^{\scriptscriptstyle{(3)}}_{l} & = \eta^{\scriptscriptstyle{(3)}}_{l}
        - \Big[a_{1}^{\scriptscriptstyle{(1)}} \eta'^{\scriptscriptstyle{(2)}}_{1}
        + c_{1}^{\scriptscriptstyle{(1)}} \eta'^{\scriptscriptstyle{(2)}}_{2}\\
        &+ \Big(b_{1}^{\scriptscriptstyle{(2)}}
        + \frac{d-3}{2(d-1)}b_{2}^{\scriptscriptstyle{(2)}}\Big)\eta'^{\scriptscriptstyle{(1)}}_{1}\Big]\delta_{l}^{1}
        - \frac{1}{2} b_{2}^{\scriptscriptstyle{(2)}} \eta'^{\scriptscriptstyle{(1)}}_{1} \delta_{l}^{2},
\end{aligned}
\end{gathered}
\end{equation}
where $l = 1, 2$ and $\delta_l^j$ represents the Kronecker delta.

The foregoing third-order coefficients can now be combined in a manner analogous to that of
the first- and second-order cases, which end to yielding the following equations:
\begin{equation}
\begin{gathered}
\begin{aligned}
    f^{\scriptscriptstyle{(3)}}_{1} & \equiv \pi^{\scriptscriptstyle{(3)}}_{1}
            - \beta_\epsilon \varepsilon^{\scriptscriptstyle{(3)}}_{1}
            - \beta_n \nu^{\scriptscriptstyle{(3)}}_{1}
            - \big([a_{1}^{\scriptscriptstyle{(1)}}]_{(\varepsilon, \pi)} f_{1}^{\scriptscriptstyle{(2)}}\\
            &+ [c_{1}^{\scriptscriptstyle{(1)}}]_{(\varepsilon,\pi)} f_{2}^{\scriptscriptstyle{(2)}}\big)
            - \frac{f_{1}^{\scriptscriptstyle{(1)}}}{h} \Big(\theta_{1}^{\scriptscriptstyle{(2)}}
            + \frac{d-2}{d-1}\theta_{2}^{\scriptscriptstyle{(2)}}\Big),
\end{aligned}\\
\begin{aligned}
    \ell^{\scriptscriptstyle{(3)}}_{l} & \equiv \kappa^{\scriptscriptstyle{(3)}}_{l}
            - \frac{n}{h} \theta^{\scriptscriptstyle{(3)}}_{l} 
            - \big([a_{l}^{\scriptscriptstyle{(2)}}]_{(\varepsilon, \pi)}\,\ell^{\scriptscriptstyle{(1)}}_{1}\\
            & + [c_{l}^{\scriptscriptstyle{(2)}}]_{(\varepsilon, \pi)}\,\ell^{\scriptscriptstyle{(1)}}_{2}\big)
            - [b_{l}^{\scriptscriptstyle{(1)}}]_{(\theta)}
            \Big(\ell_{1}^{\scriptscriptstyle{(2)}} + \frac{d-2}{d-1}\ell_{2}^{\scriptscriptstyle{(2)}}\Big),
\end{aligned}\\
\begin{aligned}
    t^{\scriptscriptstyle{(3)}}_{l} &\equiv \eta_{l}^{\scriptscriptstyle{(3)}}
            - \Big\{[a_{1}^{\scriptscriptstyle{(1)}}]_{(\varepsilon, \pi)}\,t^{\scriptscriptstyle{(2)}}_{1}
            + [c_{1}^{\scriptscriptstyle{(1)}}]_{(\varepsilon, \pi)}\,t^{\scriptscriptstyle{(2)}}_{2}\\
            &+ \frac{t_{1}^{\scriptscriptstyle{(1)}}}{h}\Big(\theta_{1}^{\scriptscriptstyle{(2)}}
            + \frac{d-3}{2(d-1)}\theta_{2}^{\scriptscriptstyle{(2)}}\Big)\Big\} \delta_{l}^{1}
            - \frac{t_{1}^{\scriptscriptstyle{(1)}}}{2h}\theta_{2}^{\scriptscriptstyle{(2)}} \delta_{l}^{2},
\end{aligned}
\end{gathered}
\label{frame_invariants_3order}
\end{equation}
where the terms $[a_{l}^{\scriptscriptstyle{(2)}}]_{(\varepsilon, \pi)}$ and $[c_{l}^{\scriptscriptstyle{(2)}}]_{(\varepsilon,\pi)}$ refer to the respective parts of
$a_{l}^{\scriptscriptstyle{(2)}}$ and $c_{l}^{\scriptscriptstyle{(2)}}$ that depend on $(\varepsilon, \pi)$,
given by
\begin{equation}
\begin{aligned}
\left [ a_{l}^{\scriptscriptstyle{(2)}} \right ]_{(\varepsilon, \pi)}
& = \bigg(\frac{\partial T}{\partial\epsilon}\bigg)_p \varepsilon_l^{\scriptscriptstyle{(2)}}
+ \bigg(\frac{\partial T}{\partial p}\bigg)_\epsilon\pi_l^{\scriptscriptstyle{(2)}},\\
\left [ c_{l}^{\scriptscriptstyle{(2)}} \right ]_{(\varepsilon, \pi)} & = \bigg(\frac{\partial \mu}{\partial\epsilon}\bigg)_p\varepsilon_l^{\scriptscriptstyle{(2)}} + \bigg(
\frac{\partial \mu}{\partial p}\bigg)_\epsilon\pi_l^{\scriptscriptstyle{(2)}}.
\end{aligned}
\label{rel_abc_with_coefficients_v2}
\end{equation}
The expressions for $[a_{1}^{\scriptscriptstyle{(1)}}]_{(\varepsilon, \pi)}$, $[c_{1}^{\scriptscriptstyle{(1)}}]_{(\varepsilon,\pi)}$, $[b_{l}^{\scriptscriptstyle{(1)}}]_{(\theta)}$,
and for the above partial derivatives are presented, respectively, in \eqref{rel_abc_with_coefficients}
and \eqref{partial_derivatives}.

\section{Conformal-invariant charged fluids}
\label{conformal-fluid}

A conformal field theory with massless fermions in the fundamental representation gives rise to conformal hydrodynamics
with conserved matter currents. Our aim here is to extend the systematic approach used in the previous section to this
case, where conformal symmetry holds. We refer to local conformal symmetry, or Weyl invariance, as the invariance of
the theory under scaling transformations $g_{\alpha \beta}\to \tilde{g}_{\alpha\beta} = e^{2\phi}g_{\alpha\beta}$.
In the context of hydrodynamics, the existence of a smooth congruence characterizing the flow requires its tangent
vector to transform according to $u^\beta\to\tilde{u}^\beta=e^{-\phi}u^\beta$, keeping its norm invariant.

Here, we are fundamentally interested in describing fluid dynamics in a general curved spacetime. We treat the metric
tensor as a hydrodynamic degree of freedom, allowing the metric derivatives to be translated into the Riemann tensor
within the gradient expansion. It is crucial to note that for a Weyl-invariant fluid, one should \textit{a priori}
include these terms in the gradient expansion since a local Weyl scaling can map a flat space-time to a curved one. Additionally, we adhere to the algebraic structure of Riemannian geometry, demanding the energy-momentum tensor and
the matter currents to be contravariant tensors with rank two and one, respectively, with respect to diffeomorphism transformations. The possibility of Weyl scaling introduces further constraints and requires that both the
energy-momentum tensor and the gauge currents be tensor densities. These transform under Weyl scaling as
\begin{equation}
T^{\alpha\beta}\to e^{-w_{\scriptscriptstyle{T}}}T^{\alpha\beta},\qquad
J^\alpha\to e^{-w_{\scriptscriptstyle{J}}}J^\alpha,
\end{equation}
where $w_{\scriptscriptstyle{T}}=d+2$ and $w_{\scriptscriptstyle{J}}=d$ represent their respective scaling dimensions.

Following the important lesson from \cite{Loganayagam:2008is}, we use the Weyl covariant derivative to identify the
allowed terms in the gradient expansion. We implement the covariant derivative using the minimal coupling prescription $\partial_\mu \phi \to (\partial_\mu + w \mathcal{A_\mu})\phi$, which requires a Weyl connection \cite{Diles:2017azi}.
In this manner, we ensure that all gradients we build transform as covariant tensor densities under Weyl scaling,
preserving the scaling dimension of the zeroth-order degree of freedom. For each correction in the derivative
expansion, we can fit the appropriate scaling dimension by factoring powers of temperature (or entropy).

We denote the Weyl-covariant derivative of an arbitrary rank tensor $\psi$ by $\mathcal{D}_\alpha \psi$. Let
$\phi$ be a scalar density of conformal weight $w\phi$ and $l^\alpha$ be a vector density of conformal weight
$w_l$. Their respective Weyl covariant derivatives can be expressed as
\begin{equation}
\begin{aligned}
    \mathcal{D}_\alpha\phi & = \partial_\alpha\phi + w_\phi\mathcal{A}_\alpha \phi,\\
    \mathcal{D}_\beta l^\alpha & = \nabla_\beta l^\alpha + w_l\mathcal{A}_\beta l^\alpha
    + \left( l^\alpha\mathcal{A}_\beta + \delta^\alpha_\beta l^\sigma\mathcal{A}_\sigma - l_\beta\mathcal{A}^\alpha \right).
\end{aligned}
\label{weylderiv}
\end{equation}
In the realm of hydrodynamics, the Weyl connection can be ascertained either by ensuring that the covariant
derivative of the velocity field is transverse and traceless or by observing that certain first-order derivative
combinations transform as a connection under Weyl scaling \cite{Loganayagam:2008is}. Both approaches lead to the
same Weyl connection for a fluid with a tangent vector field $u^\alpha$:
\begin{equation}\label{WeylConnection}
     \mathcal{A}_\alpha = u^\beta\nabla_\beta u_\alpha - \frac{\Theta}{d-1}u_\alpha.
\end{equation}

The concept of a Weyl connection, as defined in Ref.~\cite{Fukuma:2012ws}, aligns perfectly with the result of
employing the minimal coupling prescription \cite{Weyl:1918ib, Diles:2017azi} in the Christoffel symbol by substituting $\partial_\gamma g_{\alpha\beta} $ with $\partial_\gamma g_{\alpha\beta} + 2\mathcal{A}_\gamma g_{\alpha\beta}$.
Here, we refer to the Weyl connection as the vector field $\mathcal{A}_\alpha $. It is crucial to emphasize that
once a Weyl connection is established, the Weyl covariant derivative is always precisely defined through minimal
coupling. By design, Weyl covariance is thus guaranteed for any gradient order. This step is pivotal in obtaining
the gradient expansion of a conformal fluid. Our ultimate goal is to formulate both the energy-momentum tensor
and the matter current up to the third order in the gradient expansion.

The program for formulating the gradient expansion in a general frame has not been explicitly carried out
using the structure of a Weyl geometry $(\mathcal{M},g,\mathcal{A})$. For a discussion on Weyl geometry, see
\cite{hall:92, Delhom:2019yeo}. As we work in a general frame, we include all scalar corrections in the gradient
expansions of the conformal charged fluid. The energy-momentum tensor and vector current retain the general forms
given in Eq.~\eqref{general_TJ}, just as in the ordinary nonconformal case. Weyl symmetry constrains the
energy-momentum tensor by requiring it to be traceless, and thus 
\begin{equation}
\mathcal{P} = \frac{\mathcal{E}}{d-1}. \label{traceless}
\end{equation}
The equilibrium equation of state for the conformal fluid is the expected one, $p=\epsilon/(d-1)$, and the
transport coefficients of pressure corrections are proportional to the transport coefficients for energy corrections. 
In the Landau frame, we impose $u_\beta T^{\alpha \beta} = -\epsilon u^\alpha$, so that $\mathcal{E}=0$ and, consequently,
in the conformal case, $\mathcal{P}=0$. This leads to the energy-momentum tensor of the conformal fluid:
\begin{equation}
     T^{\alpha\beta}_{\text{ideal}} = \epsilon u^\alpha u^\beta +
     \frac{\epsilon}{d-1}\Delta^{\alpha\beta} + \tau^{\alpha \beta}.
     \label{confideal}
\end{equation}
The dissipative corrections for an uncharged conformal fluid in the Landau frame are restricted to be
of the tensor type only.

The matter current of an ideal conformal fluid matches the ordinary one, given by $J_{\text{ideal}}^\alpha = n u^\alpha$.
It should be noted that both the energy-momentum tensor and the matter current of an ideal fluid possess the correct
weight under Weyl scaling of $d+2$ and $d$, respectively.

The equations for the ideal conformal fluid are derived by applying the Weyl covariant derivative to the divergence-free
energy-momentum tensor and the matter current:
\begin{equation}
 \mathcal{D}_\alpha   T^{\alpha\beta}_{\text{ideal}} = 0,
 \qquad\qquad
 \mathcal{D}_\alpha J^\alpha_{\text{ideal}} = 0.
 \label{eomconformal}
\end{equation}
For this conformal scenario, we continue to consider the gradient expansion on-shell. Using Eqs.~\eqref{eomconformal},
it follows that both the longitudinal and transverse derivatives of the energy density are of higher order in gradients
and thus vanish at zero order. This excludes $\mathcal{D}_\alpha \epsilon$ from the list of fundamental gradients allowed
in the gradient expansion, or equivalently, $\mathcal{D}_\alpha T$. Within the gradient expansion of the conformal charged fluid, $\mathcal{D}^\alpha\mu$ emerges as an additional fundamental derivative compared to the uncharged case. As we will discuss later, this addition significantly expands the full nonlinear second- and third-order viscous corrections in the constitutive relations.

Curvature structures warrant special attention in our formulation based on the Weyl covariant derivative. Within this formulation, the Weyl covariant versions of the curvature tensor incorporate gradients of the velocity field present
in the Weyl connection; the conformal Riemann tensor is defined by the commutation of the Weyl covariant derivatives.
It is also important to note that the conformal Riemann tensor encompasses linear gradients of the velocity field, which are essential for describing propagating waves.

In addition to the geometric curvature, there is another tensor that appears in the commutation of the Weyl covariant derivatives. The gauge curvature, denoted $\mathcal{F}$, is defined by the exterior derivative of the Weyl connection: $\mathcal{F} = d\mathcal{A},~\mathcal{F}_{\alpha\beta} =\partial_\alpha\mathcal{A}_\beta-\partial_\beta\mathcal{A}_\alpha$.

In the gradient expansion, we encounter the conformal Riemann tensor $\mathcal{R}_{\alpha\beta\gamma\delta}$,
as well as the gauge curvature for the Weyl connection, $\mathcal{F}_{\alpha \beta}$. It is worth noting that
in a Weyl covariant theory, one cannot disregard geometric curvature, since a flat metric can be mapped into a
curved one through a Weyl scaling. The list of the lowest-order gradients compatible with conformal symmetry
is thus represented as $\{\mathcal{D}_\alpha\mu, \sigma_{\alpha\beta},\Omega_{\alpha\beta}, \mathcal{F}_{\alpha\beta},\mathcal{R}_{\alpha\beta\gamma\delta}\}$. The chemical potential introduces a
first-order vector into the list of conformal gradients, which is absent in the case of an uncharged
conformal fluid. Consequently, we cannot adopt the approach of substituting temperature with the chemical
potential, as described in the previous section for a non-conformal fluid; temperature gradients are always
of higher order and can be omitted in the gradient expansion. In this work, we will exclude temperature
gradients from the gradient expansion. We also note that $\mathcal{D}^\alpha\mu \sim \mathcal{D}^\alpha (\mu/T)$
at any gradient expansion level. Thus, we can replace $\mu$ with $\alpha\equiv \mu/T$ in the gradients.

\subsection{First- and second-order constitutive relations}

Conformal invariance imposes a traceless condition on the energy-momentum tensor. When applying this to the constitutive relation of the ideal fluid, the resulting equation of state is $p(\epsilon)=\epsilon/(d-1)$. It is important to note that
we are working in a general hydrodynamic frame, which means that scalar corrections for the energy-momentum tensor of the conformal fluid are allowed. However, conformal symmetry ensures that the transport coefficients related to these
scalar corrections are zero. In this paper, we will not dive into the topic of Weyl anomaly. As highlighted in \cite{Grozdanov:2015kqa}, the Weyl anomaly is proportional to $R^2$, making it pertinent only for corrections of
the fourth order or beyond.

In our approach, since we are not limiting ourselves to a specific hydrodynamic frame, the corrections to the ideal conformal fluid incorporate two unconstrained scalars, two transverse vectors, and one TST tensor. Corrections of first order are based solely on first-order gradients, which do not involve any curvature tensor. It is noteworthy that there is no first-order scalar in an on-shell description of a conformal charged fluid. However, there is a first-order vector that is transverse, represented as $\mathcal{D}^\alpha \mu$. This is the only structure that comes up in the constitutive relation for the matter current. Similarly, there is also just one TST tensor: the shear viscosity $\sigma^{\alpha\beta}$, which shows up in the
constitutive relation of the energy-momentum tensor. So we obtain the following corrections
to $T^{\alpha\beta}$ and $J^\alpha$:
\begin{equation}
\begin{aligned}
\Pi^{\alpha\beta}_{\scriptscriptstyle{(1)}} & = \widetilde{\theta}_1^{\scriptscriptstyle{(1)}}[(\mathcal{D}^\alpha\mu)
u^\beta +(\mathcal{D}^\beta\mu) u^\alpha] + \widetilde{\eta}_1^{\,\scriptscriptstyle{(1)}} \sigma^{\alpha\beta},\\
\Upsilon^{\alpha}_{\scriptscriptstyle{(1)}} & = \widetilde{\kappa}_1^{\scriptscriptstyle{(1)}} \mathcal{D}^\alpha\mu.
\end{aligned}
\end{equation}

The second-order contributions to the constitutive relations are constructed using the systematic algorithm we
previously outlined for ordinary (nonconformal) fluids. This time, however, we employ a different set of fundamental
gradients. As expected, the constitutive relations for the conformal charged fluid include all the transport coefficients
found in the uncharged fluid, in addition to those tied to the chemical potential gradients. Accordingly, we organized
the list of corrections so that all gradients related to the chemical potential are positioned at the end.

For the matter current, we obtain two vectors from the chemical potential gradient: $\sigma^{\alpha\beta}\mathcal{D}_\beta\mu$ and $\Omega^{\alpha\beta}\mathcal{D}_{\beta}\mu$. From these we find the second-order contributions to the matter current:
\begin{equation}
    \Upsilon^{\alpha}_{\scriptscriptstyle{(2)}} = \sum_{j=1}^{6} \widetilde{\nu}_{j}^{\scriptscriptstyle{(2)}} \mathfrak{S}^{\scriptscriptstyle{(2)}}_{j} u^\alpha +  \sum_{j=1}^{5} \widetilde{\kappa}_{j}^{\scriptscriptstyle{(2)}} (\mathcal{U}^{\scriptscriptstyle{(2)}}_{j})^{\alpha},
\end{equation}
where
\begin{equation}
\begin{aligned}
\mathcal{U}^{\scriptscriptstyle{(2)}}_{1} &=\Delta^{\alpha\beta}\mathcal{D}^{\gamma}\sigma_{\beta\gamma},\quad
&\mathcal{U}^{\scriptscriptstyle{(2)}}_{2} &  = \Delta^{\alpha\beta}u^{\gamma}\mathcal{R}_{\beta\gamma},\\
\mathcal{U}^{\scriptscriptstyle{(2)}}_{3} &=u_\beta\mathcal{F}^{\alpha\beta},\quad
&\mathcal{U}^{\scriptscriptstyle{(2)}}_{4} &= \sigma^{\alpha\beta}\mathcal{D}_\beta\mu,\\
\mathcal{U}^{\scriptscriptstyle{(2)}}_{5} &= \Omega^{\alpha\beta}\mathcal{D}_\beta\mu,
\end{aligned}
\label{vector_conforme_2nd_order}
\end{equation} 
and 
\begin{equation}
\begin{aligned}
\mathfrak{S}^{\scriptscriptstyle{(2)}}_{1} & = \sigma^{2}, \qquad
&\mathfrak{S}^{\scriptscriptstyle{(2)}}_{2} & = \Omega^{2},\\
\mathfrak{S}^{\scriptscriptstyle{(2)}}_{3} &= \mathcal{R},\qquad
&\mathfrak{S}^{\scriptscriptstyle{(2)}}_{4} &=  u^\alpha u^\beta \mathcal{R}_{\alpha\beta},\\
\mathfrak{S}^{\scriptscriptstyle{(2)}}_{5} &= \mathcal{D}^2\mu,\qquad
&\mathfrak{S}^{\scriptscriptstyle{(2)}}_{6} &= \mathcal{D}_\alpha\mu\mathcal{D}^\alpha\mu,
\end{aligned}
\label{scalar_conforme_2nd_order}
\end{equation}
where the free indices on the left-hand side of \eqref{vector_conforme_2nd_order} have been omitted.
In Ref.~\cite{Diles:2019uft}, four vectors were presented, including $\mathcal{D}_\beta \Omega^{\alpha\beta}$.
We remove this vector from our list because we find that it is equivalent to $\mathcal{D}_\beta\sigma^{\alpha\beta}$,
due to the irrelevance of the order of Weyl covariant derivatives.

For the tensor structures, we have two new second-order transport coefficients that arise from the chemical potential gradients. These coefficients are associated with $\mathcal{D}^{\langle\alpha}\mathcal{D}^{\beta\rangle}\mu$ and $\mathcal{D}^{\langle\alpha}\mu\,\mathcal{D}^{\beta\rangle}\mu$. The complete list combines both the existing
tensors from the ``uncharged-case list'' and these new additions. The general expression for the second-order
energy-momentum tensor corrections is given by
\begin{equation}
\begin{split}
    \Pi_{\scriptscriptstyle{(2)}}^{\alpha\beta} & = \sum_{j=1}^{6}\widetilde{\varepsilon}_{j}^{\,\scriptscriptstyle{(2)}}\mathfrak{S}^{\scriptscriptstyle{(2)}}_{j}
    u^\alpha u^\beta + \sum_{j=1}^{6}\widetilde{\pi}_{j}^{\scriptscriptstyle{(2)}}\mathfrak{S}^{\scriptscriptstyle{(2)}}_{j}
    \Delta^{\alpha \beta}\\
    & + \sum_{j=1}^{5} \widetilde{\theta}_{j}^{\scriptscriptstyle{(2)}} (\mathcal{U}^{\scriptscriptstyle{(2)}}_{j})^{(\alpha} u^{\beta)} + \sum_{j=1}^{7} \widetilde{\eta}_{j}^{\,\scriptscriptstyle{(2)}} (\mathfrak{T}^{\scriptscriptstyle{(2)}}_{j})^{\alpha\beta},
\end{split}
\end{equation}
where the second-order conformal TST gradients are
\begin{equation}
\begin{aligned}
\mathfrak{T}^{\scriptscriptstyle{(2)}}_1 &= \sigma_{\gamma}^{~\langle \alpha}\sigma^{\beta\rangle \gamma},\quad
&\mathfrak{T}^{\scriptscriptstyle{(2)}}_2 & = \sigma_{\gamma}^{~\langle \alpha}\Omega^{\beta\rangle \gamma},\\
\mathfrak{T}^{\scriptscriptstyle{(2)}}_3 &= \Omega_\gamma^{~\langle\alpha}\Omega^{\beta\rangle\gamma},\quad
&\mathfrak{T}^{\scriptscriptstyle{(2)}}_4 &= \mathcal{R}^{\expva{\alpha\beta}},\\
\mathfrak{T}^{\scriptscriptstyle{(2)}}_5 &= u_\gamma u_\delta \mathcal{R}^{\gamma\expva{\alpha\beta}\delta},\quad  
&\mathfrak{T}^{\scriptscriptstyle{(2)}}_6 &= \mathcal{D}^{\langle \alpha} \mu\mathcal{D}^{\beta\rangle} \mu,\\
\mathfrak{T}^{\scriptscriptstyle{(2)}}_7 &= \mathcal{D}^{\langle \alpha} \mathcal{D}^{\beta\rangle} \mu,
\end{aligned}
\label{tensor_conforme_2nd_order}
\end{equation}
and, again, the free indices on the left-hand side of \eqref{tensor_conforme_2nd_order} have been omitted.

As a result of equation \eqref{traceless}, we find that
\begin{equation}
\widetilde{\pi}_{j}^{\scriptscriptstyle{(n)}} =\frac{\widetilde{\varepsilon}_{j}^{\,\scriptscriptstyle{(n)}}}{d-1},
\end{equation}
where $n=2,\,3,\ldots$ represents the order of the gradient expansion. It is worth noting that $u^\nu\mathcal{D}_\nu \sigma^{\alpha\beta}$ is absent from the previous list; it is equivalent to $\mathfrak{T}^{\scriptscriptstyle{(2)}}_5$.
Since $\mathcal{R}$ represents the conformal Riemann tensor, it carries information about fluid flow and is retained in linearized equations that model wave propagation. Conversely, the choice to omit longitudinal derivatives of $u, \mu$ via equations of motion proves beneficial for systematically developing the gradient expansion. We opt to retain the conformal Ricci tensor over its equivalent longitudinal velocity gradient for the sake of this systematization.

The list of second-order conformal tensors obtained here aligns with the one found in appendix A of \cite{Banerjee:2008th},
except for some redundancies that we have eliminated. However, our lists of conformal vectors and scalars differ.
The present list of conformal vectors includes two additional structures, while the list of conformal scalars includes
one more structure. In Ref.~\cite{Banerjee:2008th}, the conformal Ricci scalar is included, but the projection of the
conformal Ricci tensor onto the velocities is missing. Similarly, projections of the conformal Ricci and Weyl gauge
curvatures onto the velocity are absent in \cite{Banerjee:2008th}. The systematic procedure described in \cite{Grozdanov:2015kqa} proved to be effective in identifying these missing second-order corrections.

\subsection{Third-order corrections for conformal fluids}

We go ahead to apply the procedure for obtaining higher-order constitutive relations for a charged conformal fluid
in a general frame. In this context, we present a complete list of third-order gradients and enumerate the total
number of transport coefficients.

To identify the third-order transverse vectors, it is helpful to rely on existing information from both lower-order
gradients of the charged fluid and third-order gradients of the uncharged conformal fluid. We use the former to find
vectors related to gradients of the chemical potential, and the latter to complete this list, ultimately yielding the
full set of third-order corrections. These third-order vectors are constructed by multiplying a second-order scalar
or by combining a second-order tensor with the derivative term $\mathcal{D}^\alpha\mu$. The contributions to the
matter current at the third order can be expressed as
\begin{equation}
     \Upsilon^{\alpha}_{\scriptscriptstyle{(3)}} =  \sum_{j=1}^{13}\widetilde{\nu}_{j}^{\scriptscriptstyle{(3)}} \mathfrak{S}^{\scriptscriptstyle{(3)}}_{j} u^\alpha + \sum_{j=1}^{24}\widetilde{\kappa}_{j}^{\scriptscriptstyle{(3)}} (\mathcal{U}^{\scriptscriptstyle{(3)}}_{j})^{\alpha}.
\end{equation}
The complete list of third-order conformal vectors is presented as follows:
\begin{equation}
\begin{aligned}
\mathcal{U}^{\scriptscriptstyle{(3)}}_1 &=\mathcal{D}^\alpha \sigma^2,\quad
&\mathcal{U}^{\scriptscriptstyle{(3)}}_2 &= \sigma^{\alpha}_{~\beta}\mathcal{D}_\gamma\sigma^{\beta\gamma},\\
\mathcal{U}^{\scriptscriptstyle{(3)}}_3 &= \Omega^{\alpha}_{~\beta}\mathcal{D}_\gamma\sigma^{\beta\gamma},\quad
&\mathcal{U}^{\scriptscriptstyle{(3)}}_4 &= \Delta^\alpha_{~\beta}\mathcal{D}_{\gamma} \mathcal{F}^{\beta\gamma},\\
\mathcal{U}^{\scriptscriptstyle{(3)}}_5 &= \Delta^\alpha_{~\beta}\mathcal{D}_{\gamma} \mathcal{R}^{\beta\gamma},\quad
&\mathcal{U}^{\scriptscriptstyle{(3)}}_6 &= \sigma^{\alpha\beta}u^{\gamma}\mathcal{R}_{\beta\gamma},\\
\mathcal{U}^{\scriptscriptstyle{(3)}}_7 &= \Omega^{\alpha\beta}u^{\gamma}\mathcal{R}_{\beta\gamma},\quad
&\mathcal{U}^{\scriptscriptstyle{(3)}}_8 & = \Delta^{\alpha\beta}\sigma^{\gamma\delta}u^{\eta} \mathcal{R}_{\beta\gamma\delta\eta},\\
\mathcal{U}^{\scriptscriptstyle{(3)}}_9 &= \Delta^{\alpha\beta}\Omega^{\gamma\delta}u^{\eta} \mathcal{R}_{\beta\gamma\delta\eta},\quad
&\mathcal{U}^{\scriptscriptstyle{(3)}}_{10} & = \Delta^{\alpha\beta}u^{\gamma} u^{\delta} \mathcal{D}^{\eta}\mathcal{R}_{\beta\gamma\delta\eta},\\
\mathcal{U}^{\scriptscriptstyle{(3)}}_{11} &= \sigma^2 \,\mathcal{D}^\alpha\mu,\quad  
&\mathcal{U}^{\scriptscriptstyle{(3)}}_{12} &=  \Omega^2 \,\mathcal{D}^\alpha\mu, \\
\mathcal{U}^{\scriptscriptstyle{(3)}}_{13} &= \mathcal{D}^\alpha\mathcal{D}^{2}\mu,\quad
&\mathcal{U}^{\scriptscriptstyle{(3)}}_{14} &  = \sigma^{\alpha\beta}\sigma_{\beta\gamma}\mathcal{D}^{\gamma}\mu,\\
\mathcal{U}^{\scriptscriptstyle{(3)}}_{15} &= \sigma^{\alpha\beta}\Omega_{\beta\gamma}\mathcal{D}^{\gamma}\mu,\quad
&\mathcal{U}^{\scriptscriptstyle{(3)}}_{16} &= \Omega^{\alpha\beta}\Omega_{\beta\gamma}\mathcal{D}^{\gamma}\mu,\\
\mathcal{U}^{\scriptscriptstyle{(3)}}_{17} &= \mathcal{D}^{\alpha}\mu\,\mathcal{D}^2\mu,\quad
&\mathcal{U}^{\scriptscriptstyle{(3)}}_{18} &= \mathcal{D}^\alpha\mathcal{D}^\beta\mu\,\mathcal{D}_\beta\mu,\\
\mathcal{U}^{\scriptscriptstyle{(3)}}_{19} &=   \mathcal{D}^\alpha\mu\,\mathcal{D}^\beta\mu\,\mathcal{D}_\beta\mu,\quad
&\mathcal{U}^{\scriptscriptstyle{(3)}}_{20} &= \mathcal{R}\,\mathcal{D}^\alpha \mu,\\
\mathcal{U}^{\scriptscriptstyle{(3)}}_{21} &= \Delta^{\alpha\beta}\mathcal{F}_{\beta\gamma}\mathcal{D}^{\gamma}\mu,\quad
&\mathcal{U}^{\scriptscriptstyle{(3)}}_{22} &= \Delta^{\alpha\beta}\mathcal{R}_{\beta\gamma}\mathcal{D}^{\gamma}\mu,\\
\mathcal{U}^{\scriptscriptstyle{(3)}}_{23} &= u^\beta u^\gamma\mathcal{R}_{\beta \gamma} \mathcal{D}^\alpha \mu,\quad
&\mathcal{U}^{\scriptscriptstyle{(3)}}_{24} &= \Delta^{\alpha\beta}u^{\gamma} u^{\delta}\mathcal{R}_{\beta\gamma\delta\eta}\mathcal{D}^{\eta}\mu.
\end{aligned} 
\end{equation}    
Below, we enumerate the corresponding third-order conformal scalars:
\begin{equation}
\begin{aligned}
\mathfrak{S}^{\scriptscriptstyle{(3)}}_1 &= \mathcal{D}_\alpha\mathcal{D}_\beta\sigma^{\alpha\beta},\quad
&\mathfrak{S}^{\scriptscriptstyle{(3)}}_2 &= \sigma_{\alpha\beta}\sigma^{\beta}_{~\gamma}\sigma^{\gamma\alpha},\\
\mathfrak{S}^{\scriptscriptstyle{(3)}}_3 &= \sigma_{\alpha\beta}\Omega^{\beta}_{~\gamma}\Omega^{\gamma\alpha},\quad
&\mathfrak{S}^{\scriptscriptstyle{(3)}}_4 &= u^\alpha\mathcal{D}_\alpha\mathcal{R},\\
\mathfrak{S}^{\scriptscriptstyle{(3)}}_5 &= \sigma_{\alpha\beta}\mathcal{R}^{\alpha\beta},\quad
&\mathfrak{S}^{\scriptscriptstyle{(3)}}_6 &= \Omega_{\alpha\beta}\mathcal{F}^{\alpha\beta},\\
\mathfrak{S}^{\scriptscriptstyle{(3)}}_7 &= u^\alpha\sigma^{\beta\gamma}u^\delta\mathcal{R}_{\alpha\beta\gamma\delta},\quad
&\mathfrak{S}^{\scriptscriptstyle{(3)}}_8 &= u^\alpha u^\beta u^\gamma \mathcal{D}_\alpha \mathcal{R}_{\beta\gamma},\\
\mathfrak{S}^{\scriptscriptstyle{(3)}}_9 &= \sigma^{\alpha\beta} \mathcal{D}_\alpha  \mathcal{D}_\beta\mu,\quad
&\mathfrak{S}^{\scriptscriptstyle{(3)}}_{10} &= \mathcal{D}_\alpha\mu\, \mathcal{D}_\beta\sigma^{\alpha\beta},\\
\mathfrak{S}^{\scriptscriptstyle{(3)}}_{11} &= \sigma^{\alpha\beta} \mathcal{D}_\alpha\mu\,  \mathcal{D}_\beta\mu,\quad 
&\mathfrak{S}^{\scriptscriptstyle{(3)}}_{12} &= \mathcal{F}^{\alpha\beta}u_\alpha\mathcal{D}_\beta\mu,\\
\mathfrak{S}^{\scriptscriptstyle{(3)}}_{13} &= \mathcal{R}^{\alpha\beta}u_\alpha\mathcal{D}_\beta\mu.
\end{aligned}
\end{equation}

Of the 24 third-order vectors, 14 incorporate gradients of the chemical potential, which constitute the majority of the
related transport coefficients. In fact, the presence of a conserved charge gives rise to the 14 third-order tensorial structures. Matter in the fundamental representation of the underlying microscopic theory not only introduces an additional degree of freedom, the chemical potential, but also requires the matter current to have its own gradient expansion. It should be noted that the particular third-order vector $\mathcal{U}^{\scriptscriptstyle{(3)}}_4=J_{\scriptscriptstyle{\mathcal{A}}}$
corresponds precisely to the charge current of the Weyl connection \cite{Diles:2017azi}.

We now consider the third-order TST conformal tensors appearing in the gradient expansion. Extra tensors come from combining
a second-order vector with $\mathcal{D}^\alpha\mu$ and the Weyl covariant derivative, as well as from multiplying second-order scalars with $\sigma^{\alpha\beta}$. Although it might seem possible to combine second-order tensors with first-order scalars, Weyl symmetry does not allow for any first-order scalars. Note also that even in a general hydrodynamic frame, our gradient expansion is on-shell, meaning that we rule out the equivalences that arise from the equations of motion
of the ideal fluid. The third-order correction to the tensor $T^{\alpha\beta}$ then takes the form:
\begin{equation}
\begin{split}
    \Pi_{\scriptscriptstyle{(3)}}^{\alpha\beta} & = \sum_{j=1}^{13}\widetilde{\varepsilon}_{j}^{\,\scriptscriptstyle{(3)}}
    \mathfrak{S}^{\scriptscriptstyle{(3)}}_{j}u^\alpha u^\beta
    + \sum_{j=1}^{13}\widetilde{\pi}_{j}^{\scriptscriptstyle{(3)}}\mathfrak{S}^{\scriptscriptstyle{(3)}}_{j}
    \Delta^{\alpha \beta}\\
    &+ \sum_{i=1}^{24} \widetilde{\theta}_{j}^{\scriptscriptstyle{(3)}}
    (\mathcal{U}^{\scriptscriptstyle{(3)}}_{j})^{(\alpha} u^{\beta)}
    + \sum_{i=1}^{32} \widetilde{\eta}_{j}^{\,\scriptscriptstyle{(3)}}
    (\mathfrak{T}^{\scriptscriptstyle{(3)}}_{j})^{\alpha\beta},
\end{split}
\end{equation}
where the above third-order tensor structures are defined as
\begin{equation}
\begin{aligned}
\mathfrak{T}^{\scriptscriptstyle{(3)}}_1 &= \mathcal{D}^2\sigma^{\alpha\beta},\quad
&\mathfrak{T}^{\scriptscriptstyle{(3)}}_2 &= \mathcal{D}_\gamma\mathcal{D}^{\langle\alpha}\sigma^{\beta\rangle\gamma},\\
\mathfrak{T}^{\scriptscriptstyle{(3)}}_3 &= \sigma^2 \sigma^{\alpha\beta},\quad
&\mathfrak{T}^{\scriptscriptstyle{(3)}}_4 &= \Omega^2 \sigma^{\alpha\beta},\\
\mathfrak{T}^{\scriptscriptstyle{(3)}}_5 & = \sigma_{\gamma\delta}\sigma^{\gamma\langle \alpha}\sigma^{\beta\rangle \delta},\quad
&\mathfrak{T}^{\scriptscriptstyle{(3)}}_6 &= \sigma_{\gamma\delta}\Omega^{\gamma\langle \alpha}\sigma^{\beta\rangle \delta},\\
\mathfrak{T}^{\scriptscriptstyle{(3)}}_7 &= \sigma_{\gamma\delta}\Omega^{\gamma\langle \alpha}\Omega^{\beta\rangle \delta},\quad
&\mathfrak{T}^{\scriptscriptstyle{(3)}}_8 &= \Omega_{\gamma\delta}\sigma^{\gamma\langle \alpha}\Omega^{\beta\rangle \delta},\\
\mathfrak{T}^{\scriptscriptstyle{(3)}}_9 &= \mathcal{R}\sigma^{\alpha\beta},\quad
&\mathfrak{T}^{\scriptscriptstyle{(3)}}_{10}&= \mathcal{F}_\gamma^{~\langle \alpha}\sigma^{\beta\rangle \gamma},\\
\mathfrak{T}^{\scriptscriptstyle{(3)}}_{11} &= \mathcal{R}_\gamma^{~\langle \alpha}\sigma^{\beta\rangle\gamma},\quad
&\mathfrak{T}^{\scriptscriptstyle{(3)}}_{12} &= \mathcal{F}_\gamma^{~\langle \alpha}\Omega^{\beta\rangle \gamma},\\
\mathfrak{T}^{\scriptscriptstyle{(3)}}_{13} &= \mathcal{R}_\gamma^{~\langle \alpha}\Omega^{\beta\rangle \gamma},\quad
&\mathfrak{T}^{\scriptscriptstyle{(3)}}_{14} &= \sigma_{\gamma\delta}\mathcal{R}^{\gamma\expva{\alpha\beta}\delta},\\
\mathfrak{T}^{\scriptscriptstyle{(3)}}_{15} &= u_\gamma\mathcal{D}^{\langle\alpha}\mathcal{F}^{\beta\rangle\gamma},\quad
&\mathfrak{T}^{\scriptscriptstyle{(3)}}_{16} &= u_\gamma\mathcal{D}^{\langle \alpha} \mathcal{R}^{\beta\rangle\gamma},\\
\mathfrak{T}^{\scriptscriptstyle{(3)}}_{17} &= u_\gamma\mathcal{D}_\delta\mathcal{R}^{\gamma\expva{\alpha\beta}\delta},\quad
&\mathfrak{T}^{\scriptscriptstyle{(3)}}_{18} &= \sigma^{\alpha\beta} u^\gamma u^\delta\mathcal{R}_{\gamma\delta},\\
\mathfrak{T}^{\scriptscriptstyle{(3)}}_{19} &= u^\gamma u^\delta \sigma^{\eta\langle\alpha}\mathcal{R}^{\beta\rangle}_{~~\gamma\delta\eta},\quad
&\mathfrak{T}^{\scriptscriptstyle{(3)}}_{20} &= u^\gamma u^\delta \Omega^{\eta\langle \alpha}\mathcal{R}^{\beta\rangle}_{~~\gamma\delta\eta},\\
\mathfrak{T}^{\scriptscriptstyle{(3)}}_{21} &=\sigma^{\alpha\beta}\mathcal{D}^2\mu,\quad
&\mathfrak{T}^{\scriptscriptstyle{(3)}}_{22} &= \mathcal{D}^\gamma\sigma^{\alpha\beta}\mathcal{D}_\gamma \mu,\\
\mathfrak{T}^{\scriptscriptstyle{(3)}}_{23} &= \mathcal{D}_\gamma \sigma^{\gamma\langle\alpha}\mathcal{D}^{\beta\rangle}\mu,\quad
&\mathfrak{T}^{\scriptscriptstyle{(3)}}_{24} &= \mathcal{D}^{\langle\alpha}\sigma^{\beta\rangle\gamma}\mathcal{D}_\gamma \mu,\\
\mathfrak{T}^{\scriptscriptstyle{(3)}}_{25} &= \sigma^{\gamma\langle\alpha}\mathcal{D}^{\beta\rangle}\mathcal{D}_\gamma\mu,\quad
&\mathfrak{T}^{\scriptscriptstyle{(3)}}_{26} &= \Omega^{\gamma\langle\alpha}\mathcal{D}^{\beta\rangle}\mathcal{D}_\gamma\mu,\\
\mathfrak{T}^{\scriptscriptstyle{(3)}}_{27} &= \sigma^{\alpha\beta}\mathcal{D}^\gamma\mu\,\mathcal{D}_\gamma\mu,\quad
&\mathfrak{T}^{\scriptscriptstyle{(3)}}_{28} &= \sigma^{\gamma\langle\alpha}\mathcal{D}^{\beta\rangle}\mu\,\mathcal{D}_\gamma\mu,\\
\mathfrak{T}^{\scriptscriptstyle{(3)}}_{29} &= \Omega^{\gamma\langle\alpha}\mathcal{D}^{\beta\rangle}\mu\,\mathcal{D}_\gamma\mu,\quad
&\mathfrak{T}^{\scriptscriptstyle{(3)}}_{30} &= u_\gamma\mathcal{F}^{\gamma\langle\alpha}\mathcal{D}^{\beta\rangle}\mu,\\
\mathfrak{T}^{\scriptscriptstyle{(3)}}_{31} &= u_\gamma\mathcal{R}^{\gamma\langle\alpha}\mathcal{D}^{\beta\rangle}\mu,\quad
&\mathfrak{T}^{\scriptscriptstyle{(3)}}_{32} &= u_\gamma\mathcal{R}^{\gamma\expva{\alpha\beta}\delta}\mathcal{D}_\delta \mu.
\end{aligned}
\end{equation}

The above list contains 32 third-order TST conformal tensors. Of these, 20 are related to the transport coefficients
of an uncharged fluid, and 12 correspond to the variable chemical potential associated with the conserved charge.
It is important to note that none of the 12 tensors associated with chemical potential gradients remains after
linearization, which means that they do not influence wave propagation in the fluid.

\section{Linearized dispersion relations}
\label{dispersion_relations}

In order to determine the dispersion relations of the waves propagating in a charged, non-conformal fluid,
we initially consider the decomposition of the energy-momentum tensor and the current into an ideal part
and a dissipative component:
\begin{align}
T^{\alpha\beta} & = T^{\alpha\beta}_{\text{ideal}} + \Pi^{\alpha\beta},\\
J^{\alpha} & = J^{\alpha}_{\text{ideal}} + \Upsilon^{\alpha},
\end{align}
where $T^{\alpha\beta}_{\text{ideal}}$ and $J^{\alpha}_{\text{ideal}}$ are defined, respectively,
by equations \eqref{Tideal} and \eqref{Jideal}. The dissipative terms under consideration
extend up to the third order in the gradient expansion for both $\Pi^{\alpha\beta}$ and $\Upsilon^\alpha$:
\begin{align}
\Pi^{\alpha\beta} & = \Pi^{\alpha\beta}_{\scriptscriptstyle{(1)}} + \Pi^{\alpha\beta}_{\scriptscriptstyle{(2)}} +
\Pi^{\alpha\beta}_{\scriptscriptstyle{(3)}},\\
\Upsilon^{\alpha} & = \Upsilon^{\alpha}_{\scriptscriptstyle{(1)}} + \Upsilon^{\alpha}_{\scriptscriptstyle{(2)}}
+ \Upsilon^{\alpha}_{\scriptscriptstyle{(3)}},
\end{align}
where $\Pi^{\alpha\beta}_{\scriptscriptstyle{(i)}}$ and $\Upsilon^{\alpha}_{\scriptscriptstyle{(i)}}$
denote terms of $i$-th order in gradients.

The equations of motion are derived by taking the divergence of $T^{\alpha\beta}$
and $J^\alpha$:
\begin{gather}
\nabla_\alpha T^{\alpha\beta} = \nabla_\alpha T_{\text{ideal}}^{\alpha\beta} + \nabla_\alpha \Pi^{\alpha\beta} = 0, \label{divergence1}\\ 
\nabla_\alpha J^{\alpha} = \nabla_\alpha J_{\text{ideal}}^{\alpha} + \nabla_\alpha \Upsilon^{\alpha} = 0.
\label{divergence2}
\end{gather}
Upon projecting equation \eqref{divergence1} in the directions parallel and perpendicular to $u^\mu$
and expanding the terms in equation \eqref{divergence2}, we obtain:
\begin{gather}
u_\beta \nabla_\alpha T^{\alpha\beta} = - D\epsilon - h\Theta - \Pi^{\alpha\beta}\nabla_\alpha u_\beta = 0,
\label{paralelo1}\\
\Delta^{\alpha}_{\;\;\beta}\nabla_\gamma T^{\gamma\beta} =  h Du^\alpha + \nabla_{\perp}^{\alpha}p
+ \Delta^{\alpha}_{\;\;\beta}\nabla_\gamma \Pi^{\gamma\beta} = 0,
\label{perpendicular1}\\
\nabla_\alpha J^{\alpha} = D n + n\Theta + \nabla_\alpha \Upsilon^{\alpha} = 0.
\label{conservacao_corrente1}
\end{gather}

In the Landau frame, there is no energy flow in the local rest frame of the fluid, which
means $\mathcal{Q}^{\mu}=0$. The defining properties of this frame are complemented by
$\mathcal{E} = \mathcal{N} = 0$. By setting the various coefficients $\varepsilon_j^{\scriptscriptstyle{(i)}}$,
$\theta_j^{\scriptscriptstyle{(i)}}$, and $\nu_j^{\scriptscriptstyle{(i)}}$ to zero, the linear dissipative
terms present in equations \eqref{paralelo1}--\eqref{conservacao_corrente1} are reduced to
\begin{equation}
\begin{split}
\Pi^{\alpha\beta} = & \left[\Delta^{\alpha\beta}\left(\pi_1^{\scriptscriptstyle{(1)}} +\pi_1^{\scriptscriptstyle{(3)}}\nabla_{\perp}^{2}\right)+
\eta_1^{\scriptscriptstyle{(3)}}\nabla_{\perp}^{\langle\alpha}\nabla_{\perp}^{\beta\rangle}\right]\Theta\\
&+ \Delta^{\alpha}_{\;\;\gamma}\left(\eta_1^{\scriptscriptstyle{(1)}} + \eta_2^{\scriptscriptstyle{(3)}}\nabla_{\perp}^{2}\right)\sigma^{\gamma\beta}\\
&+ \left(\pi_1^{\scriptscriptstyle{(2)}}\Delta^{\alpha\beta}\nabla_{\perp}^{2} +\eta_1^{\scriptscriptstyle{(2)}}\nabla_{\perp}^{\langle\alpha}\nabla_{\perp}^{\beta\rangle}\right)T\\
&+ \left(\pi_2^{\scriptscriptstyle{(2)}}\Delta^{\alpha\beta}\nabla_{\perp}^{2} +\eta_2^{\scriptscriptstyle{(2)}}\nabla_{\perp}^{\langle\alpha}\nabla_{\perp}^{\beta\rangle}\right)\mu\,,
\end{split}
\label{expansion_Pi}
\end{equation}
\begin{equation}
\begin{split}
\Upsilon^{\alpha} & = \left(\kappa_1^{\scriptscriptstyle{(1)}}
    + \kappa_1^{\scriptscriptstyle{(3)}}\nabla_{\perp}^{2}\right)\nabla_{\perp}^{\alpha}T\\
    & + \left(\kappa_2^{\scriptscriptstyle{(1)}}
    + \kappa_2^{\scriptscriptstyle{(3)}}\nabla_{\perp}^{2}\right)\nabla_{\perp}^{\alpha}\mu\\
    & +\kappa_1^{\scriptscriptstyle{(2)}} \nabla_{\perp}^{\alpha}\Theta
    + \kappa_2^{\scriptscriptstyle{(2)}} \Delta^{\alpha}_{\;\;\beta}\nabla_{\perp\gamma}\sigma^{\gamma\beta}.
\end{split}
\label{expansion_Upsilon}
\end{equation}

Equations specific to a given hydrodynamic frame, such as those presented above, can be expressed
in terms of frame-invariant coefficients. For example, using the equations \eqref{def_f_and_ell}
and \eqref{shear_viscosity_relation}, it is evident that
\begin{equation}
\begin{aligned}
\pi_1^{\scriptscriptstyle{(1)}} & = f_1^{\scriptscriptstyle{(1)}} = - \zeta, \\
\kappa_1^{\scriptscriptstyle{(1)}} & = \ell_{1}^{\scriptscriptstyle{(1)}} = \sigma\frac{\mu}{T},\\
\kappa_2^{\scriptscriptstyle{(1)}} & = \ell_{2}^{\scriptscriptstyle{(1)}} = -\sigma, \\
\eta_1^{\scriptscriptstyle{(1)}} & = t_{1}^{\scriptscriptstyle{(1)}} = -2\eta.
\end{aligned}
\label{rel_coef_transp_1st_order}
\end{equation}
Similarly, based on equations \eqref{2order_frame_invariants} and \eqref{frame_invariants_3order},
the following relations can be established for second- and third-order transport coefficients (with $l = 1, 2$):
\begin{equation}
\begin{gathered}
\pi^{\scriptscriptstyle{(2)}}_{l} = f^{\scriptscriptstyle{(2)}}_{l},\quad\;\;
\eta^{\scriptscriptstyle{(2)}}_{l} = t^{\scriptscriptstyle{(2)}}_{l},\\
\kappa^{\scriptscriptstyle{(2)}}_{l} = \ell^{\scriptscriptstyle{(2)}}_{l}
    + \frac{1}{\beta_n}\left(T_{,n}\ell^{\scriptscriptstyle{(1)}}_{1}
    + \mu_{,n}\ell^{\scriptscriptstyle{(1)}}_{2}\right) f_1^{\scriptscriptstyle{(1)}}\delta_{l}^1,\\
\pi^{\scriptscriptstyle{(3)}}_{1} = f^{\scriptscriptstyle{(3)}}_{1}
    + \frac{1}{\beta_n}\left(T_{,n} f_{1}^{\scriptscriptstyle{(2)}}
    + \mu_{,n} f_{2}^{\scriptscriptstyle{(2)}}\right) f_1^{\scriptscriptstyle{(1)}},\\
\kappa^{\scriptscriptstyle{(3)}}_{l} = \ell^{\scriptscriptstyle{(3)}}_{l} 
    + \frac{1}{\beta_n}\left(T_{,n}\ell^{\scriptscriptstyle{(1)}}_{1}
    + \mu_{,n}\ell^{\scriptscriptstyle{(1)}}_{2}\right) f_l^{\scriptscriptstyle{(2)}},\\
\eta_{l}^{\scriptscriptstyle{(3)}} = t^{\scriptscriptstyle{(3)}}_{l}
    + \frac{1}{\beta_n}\left(T_{,n} t^{\scriptscriptstyle{(2)}}_{1}
    + \mu_{,n} t^{\scriptscriptstyle{(2)}}_{2}\right) f_1^{\scriptscriptstyle{(1)}}\delta_{l}^1.
\end{gathered}
\label{rel_coef_2nd_3rd_order}
\end{equation}

For a flat $d$-dimensional spacetime, the metric is $g_{\alpha\beta} = \eta_{\alpha\beta} =
\mbox{diag}(-1, + 1, \dots, +1)$. In the linearized regime of hydrodynamics, we define
\begin{equation}
\nabla_{\perp\alpha}=\Delta^{\;\;\beta}_{\alpha}\partial_\beta =
u_\alpha u^{\beta}\partial_{\beta} + \partial_\alpha \equiv \partial_{\perp\alpha}
\end{equation}
and introduce the first-order perturbations in amplitude:
\begin{equation}
\begin{aligned}
\epsilon & \rightarrow \epsilon + \delta\epsilon,& \qquad T & \rightarrow T + \delta T,\\
p & \rightarrow p + \delta p,& \qquad \mu & \rightarrow \mu + \delta \mu,\\
n & \rightarrow n + \delta n.
\end{aligned}
\label{linear_perturbations}
\end{equation}
Consequently, equations \eqref{paralelo1}--\eqref{conservacao_corrente1} reduce to
\begin{gather}
u^\alpha\partial_\alpha\delta\epsilon + h\partial_{\perp\alpha}\delta u^\alpha = 0, \\
h u^\beta\partial_\beta\delta u^\alpha + \partial_{\perp}^{\alpha}\delta p +
\Delta^{\alpha}_{\;\;\beta}\partial_{\gamma}\delta\Pi^{\gamma\beta} = 0, \\
u^\alpha\partial_\alpha\delta n + n\partial_{\perp\alpha}\delta u^\alpha +\partial\delta u^\alpha = 0.
\end{gather}
Upon taking the Fourier transform of these perturbations as
$\delta\Psi \rightarrow \delta\Psi e^{ik_{\alpha}x^{\alpha}}$, we obtain
\begin{gather}
k_{\parallel}\delta\epsilon + h k_{\perp\alpha}\delta u^{\alpha} = 0,\label{paralelo2}\\
h k_{\parallel}\delta u^{\alpha} + k_{\perp}^{\alpha}\delta p +\Delta^{\alpha}_{\;\;\beta}
k_\gamma \delta\Pi^{\gamma\beta} = 0, \label{perpendicular2}\\
k_{\parallel}\delta n + n k_{\perp\alpha}\delta u^{\alpha} + k_{\alpha}\delta\Upsilon^{\alpha} = 0,
\label{conservacao_corrente2}
\end{gather}
where $k_{\parallel}= u^{\alpha}k_{\alpha}$ and $k_{\perp}^{\alpha}=\Delta^{\alpha\beta}k_{\beta}$,
and the perturbations in $\Pi^{\alpha\beta}$ and $\Upsilon^{\alpha}$ take the form
\begin{equation}
\begin{aligned}
\delta \Pi^{\alpha\beta} & = i\left[\Delta^{\alpha\beta}\left(\pi_1^{\scriptscriptstyle{(1)}}
            - \frac{\eta_1^{\scriptscriptstyle{(1)}}}{d-1}\right)
            - \eta_1^{\scriptscriptstyle{(3)}} k_{\perp}^{\alpha}k_{\perp}^{\beta}\right.\\
            &\left. - \Delta^{\alpha\beta} \left(\pi_1^{\scriptscriptstyle{(3)}}
            - \frac{\eta_1^{\scriptscriptstyle{(3)}} + \eta_2^{\scriptscriptstyle{(3)}}}{d-1}\right)
            k_{\perp}^{2}\right] k_{\perp\gamma}\delta u^{\gamma}\\
            & -\left[\Delta^{\alpha\beta}\left(\pi_1^{\scriptscriptstyle{(2)}}
            - \frac{\eta_1^{\scriptscriptstyle{(2)}}}{d-1}\right) k_{\perp}^{2}
            + \eta_1^{\scriptscriptstyle{(2)}} k_{\perp}^{\alpha}k_{\perp}^{\beta}\right]\delta T\\
            &- \left[\Delta^{\alpha\beta}\left(\pi_2^{\scriptscriptstyle{(2)}}
            - \frac{\eta_2^{\scriptscriptstyle{(2)}}}{d-1}\right) k_{\perp}^{2}
            + \eta_2^{\scriptscriptstyle{(2)}} k_{\perp}^{\alpha}k_{\perp}^{\beta}\right]\delta\mu \\
& + \frac{i}{2}\left(\eta_1^{\scriptscriptstyle{(1)}} - \eta_2^{\scriptscriptstyle{(3)}}k_{\perp}^{2}\right)\left(k_{\perp}^{\alpha}\delta u^{\beta} + k_{\perp}^{\beta}\delta u^{\alpha}\right),
\end{aligned}
\label{perturba_Pi}
\end{equation}
\begin{equation}
\begin{aligned}
\delta\Upsilon^{\alpha} & = i k_{\perp}^{\alpha}\left(\kappa_1^{\scriptscriptstyle{(1)}}
    - \kappa_1^{\scriptscriptstyle{(3)}} k_{\perp}^{2}\right)\delta T\\
    &+ i k_{\perp}^{\alpha}\left(\kappa_2^{\scriptscriptstyle{(1)}}
    - \kappa_2^{\scriptscriptstyle{(3)}} k_{\perp}^{2}\right)\delta\mu
    - \frac{1}{2}\kappa_2^{\scriptscriptstyle{(2)}} k_{\perp}^{2}\delta u^{\alpha}\\
    &- k_{\perp}^{\alpha}\left(\kappa_1^{\scriptscriptstyle{(2)}}
    + \frac{d-3}{2(d-1)}\kappa_2^{\scriptscriptstyle{(2)}}\right) k_{\perp\gamma}\delta u^{\gamma}.
\end{aligned}
\end{equation}
In these equations, we retain the original transport coefficients $\pi_j^{\scriptscriptstyle{(i)}}$,
$\kappa_j^{\scriptscriptstyle{(i)}}$, and $\eta_j^{\scriptscriptstyle{(i)}}$ for simplicity. However,
these coefficients should be regarded as functions of the frame-invariant quantities $f_j^{\scriptscriptstyle{(i)}}$,
$\ell_j^{\scriptscriptstyle{(i)}}$, and $t_j^{\scriptscriptstyle{(i)}}$ as prescribed by
Eqs.~\eqref{rel_coef_transp_1st_order} and \eqref{rel_coef_2nd_3rd_order}.

As usual in the literature, we will decompose $\delta p$, $\delta T$ and $\delta\mu$ in terms of the
fluctuations of the densities, $\delta\epsilon$ and $\delta n$, which leads to
\begin{align}
\delta p = & \left(\frac{\partial p}{\partial\epsilon}\right)_n \delta\epsilon
                + \left(\frac{\partial p}{\partial n}\right)_{\epsilon} \delta n
                = \beta_\epsilon \delta\epsilon + \beta_n \delta n,\label{decompoe_dp}\\
\delta T = & \left(\frac{\partial T}{\partial\epsilon}\right)_n \delta\epsilon
                + \left(\frac{\partial T}{\partial n}\right)_{\epsilon} \delta n
                =  T_{,\epsilon} \delta\epsilon + T_{,n} \delta n, \label{decompoe_dT}\\
\delta\mu = & \left(\frac{\partial \mu}{\partial\epsilon}\right)_n \delta\epsilon
                + \left(\frac{\partial \mu}{\partial n}\right)_{\epsilon} \delta n
                = \mu_{,\epsilon} \delta\epsilon + \mu_{,n} \delta n.
\label{decompoe_dMu}
\end{align}
Substituting expressions \eqref{perturba_Pi}--\eqref{decompoe_dMu} into Eqs.~\eqref{perpendicular2}
and \eqref{conservacao_corrente2}, we find
\begin{equation}
\begin{aligned}
\left[2hk_{\parallel} \right. & + \left. i(t_1^{\scriptscriptstyle{(1)}} 
     - t_2^{\scriptscriptstyle{(3)}}k_{\perp}^{2})k_{\perp}^{2}\right]\delta u^\alpha\\
    & + 2k_{\perp}^{\alpha}(\beta_\epsilon\delta\epsilon + \beta_n\delta n)\\
    & - 2 k_{\perp}^{\alpha}k_{\perp}^{2}\left(\psi_\epsilon^{\scriptscriptstyle{(2)}} \delta\epsilon
    + \psi_n^{\scriptscriptstyle{(2)}} \delta n\right)\\
    & - 2ihk_{\perp}^{\alpha}\left(\gamma_s^{\scriptscriptstyle{(1)}}
    + \chi_s^{\scriptscriptstyle{(3)}} k_{\perp}^{2}\right)k_{\perp\beta}\delta u^\beta = 0,
\end{aligned}
\label{perpendicular3_transformed}
\end{equation}
\begin{equation}
\begin{aligned}
\left[k_{\parallel} \right. & - \left. ik_{\perp}^{2}\left(\sigma\alpha_n^{\scriptscriptstyle{(1)}}
            + \phi_n^{\scriptscriptstyle{(3)}}k_{\perp}^{2}\right)\right]\delta n\\
            &- ik_{\perp}^{2}\left(\sigma\alpha_\epsilon^{\scriptscriptstyle{(1)}}
            + \phi_\epsilon^{\scriptscriptstyle{(3)}} k_{\perp}^{2}\right)\delta\epsilon\\
            & + \left(n - h\varsigma_s^{\scriptscriptstyle{(2)}} k_{\perp}^{2}\right)k_{\perp\alpha}\delta u^\alpha = 0,
\end{aligned}
\label{conservacao_corrente3_transformed_corrected}
\end{equation}
where the new quantities introduced above are defined by the following expressions: 
\begin{equation}
\begin{gathered}
\alpha_a^{\scriptscriptstyle{(1)}} = - \frac{1}{\sigma}\left(T_{,a}\ell_1^{\scriptscriptstyle{(1)}}
                +\mu_{,a} \ell_2^{\scriptscriptstyle{(1)}}\right)
                = \mu_{,a} -\frac{\mu}{T}T_{,a},\\
\gamma_s^{\scriptscriptstyle{(1)}} = - \frac{1}{h}\left(f_1^{\scriptscriptstyle{(1)}}
                                     + \frac{d-2}{d-1}t_1^{\scriptscriptstyle{(1)}}\right), \\
\psi_a^{\scriptscriptstyle{(2)}} = T_{,a}\left(f_1^{\scriptscriptstyle{(2)}}
                                    + \frac{d-2}{d-1}t_1^{\scriptscriptstyle{(2)}}\right)
                                    + \mu_{,a}\left(f_2^{\scriptscriptstyle{(2)}}
                                    + \frac{d-2}{d-1}t_2^{\scriptscriptstyle{(2)}}\right),\\
\varsigma_s^{\scriptscriptstyle{(2)}} = \frac{1}{h}\left(\ell^{\scriptscriptstyle{(2)}}_{1}
        - \frac{\sigma}{\beta_n}\alpha_n^{\scriptscriptstyle{(1)}} f_1^{\scriptscriptstyle{(1)}}
        + \frac{d-2}{d-1}\ell^{\scriptscriptstyle{(2)}}_{2}\right),\\
\phi_a^{\scriptscriptstyle{(3)}} = T_{,a}\left(\ell^{\scriptscriptstyle{(3)}}_{1} 
     - \frac{\sigma}{\beta_n}\alpha_n^{\scriptscriptstyle{(1)}} f_1^{\scriptscriptstyle{(2)}}\right)
     + \mu_{,a}\left(\ell^{\scriptscriptstyle{(3)}}_{2} 
     - \frac{\sigma}{\beta_n}\alpha_n^{\scriptscriptstyle{(1)}} f_2^{\scriptscriptstyle{(2)}}\right),\\
\begin{aligned}
\chi_{s}^{\scriptscriptstyle{(3)}} & = \frac{1}{h} \left[ f^{\scriptscriptstyle{(3)}}_{1}
    + \frac{1}{\beta_n}\left(T_{,n} f_{1}^{\scriptscriptstyle{(2)}}
    + \mu_{,n} f_{2}^{\scriptscriptstyle{(2)}}\right) f_1^{\scriptscriptstyle{(1)}}\right.\\
    & \left. + \frac{d-2}{d-1}\bigg(t^{\scriptscriptstyle{(3)}}_{1}
    + \frac{1}{\beta_n}\left(T_{,n} t^{\scriptscriptstyle{(2)}}_{1}
    + \mu_{,n} t^{\scriptscriptstyle{(2)}}_{2}\right) f_1^{\scriptscriptstyle{(1)}}
    + t^{\scriptscriptstyle{(3)}}_{2}\bigg)\right],
\end{aligned}
\end{gathered}
\label{new_coeficients}
\end{equation}
with $a$ assuming the values $\epsilon$ and $n$.

Equations \eqref{paralelo2} and \eqref{conservacao_corrente3_transformed_corrected} comprise two scalar equations
that can be used to eliminate $\delta\epsilon$ and $\delta n$, expressing them as functions of the perturbations
in fluid velocity. Upon substituting these expressions for the density fluctuations into the vector equation \eqref{perpendicular3_transformed} and projecting it along the direction transverse to $k_{\perp}^{\alpha}$, we obtain
\begin{equation}
\left[2hk_{\parallel} + i(t_1^{\scriptscriptstyle{(1)}} 
            - t_2^{\scriptscriptstyle{(3)}}k_{\perp}^{2})k_{\perp}^{2}\right]
            \delta u_{\scriptscriptstyle{T}}^{\alpha} = 0,
\label{eq_transversal1}
\end{equation}
where
\begin{equation}
\delta u_{\scriptscriptstyle{T}}^{\alpha} =\left(\Delta^{\alpha}_{\;\;\beta}-
\frac{k_{\perp}^{\alpha}k_{\perp\beta}}{k_{\perp}^{2}}\right)\delta u^\beta.
\end{equation}
A similar procedure, albeit with a projection along the direction longitudinal to
$k_{\perp}^{\alpha}$, yields an equation of the form
\begin{equation}
\mathscr{F}(k_{\parallel}, k_{\perp}^{2}) \delta u_{\scriptscriptstyle{L}}^{\alpha} = 0,
\label{eq_longitudinal1}
\end{equation}
where $\mathscr{F}(k_{\parallel},k_{\perp}^{2})$ is a function of the wavenumber components and the
fluid transport coefficients. The longitudinal projection of the fluid velocity perturbation is given by
\begin{equation}
\delta u_{\scriptscriptstyle{L}}^{\alpha} =
\frac{\left(k_{\perp\beta}\delta u^{\beta}\right)}{k_{\perp}^{2}}k_{\perp}^{\alpha}.
\end{equation}

From the vanishing of the term enclosed in square brackets in Eq.~\eqref{eq_transversal1},
as well as that of the function $\mathscr{F}(k_{\parallel}, k_{\perp}^{2})$, we deduce the
equations governing the transverse and longitudinal hydrodynamic modes, respectively.
To find the dispersion relations, we adopt the rest frame of the fluid, expressed as
\begin{equation}
u^\alpha = (1,\, \vec{0})\;\Rightarrow \; k_{\parallel} = u^{0}k_{0} \equiv -\omega
\;\;\mbox{and}\;\; k_{\perp}^{2}=k^{i}k_{i}\equiv k^2.
\end{equation}
Consequently, the algebraic equations characterizing the hydrodynamic modes assume the forms
\begin{equation}
- 2 h\omega + i(t_1^{\scriptscriptstyle{(1)}} - t_2^{\scriptscriptstyle{(3)}} k^2)k^2 = 0,
\label{eq_transversal2}
\end{equation}
\begin{equation}
\begin{gathered}
    \omega^3 - \omega k^2\left[v_{s}^2 - k^2\Big(\psi_\epsilon^{\scriptscriptstyle{(2)}}
    + \frac{n}{h}\psi_n^{\scriptscriptstyle{(2)}}
    - \sigma\alpha_n^{\scriptscriptstyle{(1)}}\gamma_s^{\scriptscriptstyle{(1)}}
    + \beta_n\varsigma_s^{\scriptscriptstyle{(2)}} \Big)\right]\\
    + i \omega^2 k^2 \left[\gamma_s^{\scriptscriptstyle{(1)}}
    + \sigma\alpha_n^{\scriptscriptstyle{(1)}} + k^2 (\chi_s^{\scriptscriptstyle{(3)}} + \phi_n^{\scriptscriptstyle{(3)}})\right]
    - ik^4 \sigma\Big[\alpha_n^{\scriptscriptstyle{(1)}}(\beta_\epsilon\\
    - k^2 \psi_\epsilon^{\scriptscriptstyle{(2)}}) - \alpha_\epsilon^{\scriptscriptstyle{(1)}}
    (\beta_n - k^2 \psi_n^{\scriptscriptstyle{(2)}}) -
    \frac{k^2}{\sigma}(\beta_n\phi_\epsilon^{\scriptscriptstyle{(3)}}
    - \beta_\epsilon\phi_n^{\scriptscriptstyle{(3)}})\Big] = 0,
\end{gathered}
\label{eq_longitudinal2}
\end{equation}
where the square of the speed of sound $v_{s}$ is defined as
\begin{equation}
v_{s}^2 \equiv \left(\frac{\partial p}{\partial\epsilon}\right)_{s/n}
= \beta_\epsilon + \frac{n}{h}\beta_n.
\end{equation} 

The solution of equation \eqref{eq_transversal2} gives rise to the shear mode, which is
associated with the diffusion of transverse momentum in the fluid. This mode is described by
\begin{equation}
\omega_{\mbox{\scriptsize{shear}}}(k) = i\frac{t_1^{\scriptscriptstyle{(1)}}}{2h}k^2
                                        - i\frac{t_2^{\scriptscriptstyle{(3)}}}{2h}k^4.
\label{shear_dispersion_relation}
\end{equation}
Meanwhile, the solutions of equation \eqref{eq_longitudinal2} yield the dispersion relations
for both the charge diffusion mode and the sound wave mode, expressed as
\begin{equation}
\omega_{\mbox{\scriptsize{diffu}}}(k) = -i\mathscr{D}k^2 +
\frac{i}{v_s^2}\left(\mathscr{Y} - v_{s}^{2}\phi_n^{\scriptscriptstyle{(3)}}\right) k^4,
\label{diffusion_dispersion_relation}
\end{equation}
and
\begin{equation}
\begin{split}
\omega_{\mbox{\scriptsize{sound}}}^{\scriptscriptstyle{(\pm)}}(k) & = \pm v_s k -\frac{i}{2}\Gamma k^2\\
        & \mp \frac{1}{8v_s}\left(\Gamma^2 - 4\mathscr{X}\right) k^3
        - \frac{i}{2v_{s}^{2}}\left(\mathscr{Y} + v_{s}^{2}\chi_s^{\scriptscriptstyle{(3)}}\right) k^4,
\end{split}
\label{soundwave_dispersion_relation}
\end{equation}
where the quantities introduced in the foregoing relations are given by
\begin{equation}
\begin{gathered}
\mathscr{D} = \frac{\sigma}{v_{s}^{2}}(\alpha_n^{\scriptscriptstyle{(1)}}\beta_\epsilon
            - \alpha_\epsilon^{\scriptscriptstyle{(1)}} \beta_n),
\Gamma = \gamma_s^{\scriptscriptstyle{(1)}}
            + \frac{\sigma\beta_n}{v_{s}^{2}}\left[\alpha_\epsilon^{\scriptscriptstyle{(1)}}
            + \frac{n}{h}\alpha_n^{\scriptscriptstyle{(1)}}\right],\\
\mathscr{X} = (\mathscr{D} - \gamma_s^{\scriptscriptstyle{(1)}})
                (\mathscr{D} - \sigma\alpha_n^{\scriptscriptstyle{(1)}})
                - \left[\psi_\epsilon^{\scriptscriptstyle{(2)}}
                + \frac{n}{h}\psi_n^{\scriptscriptstyle{(2)}}\right]
                - \beta_n\varsigma_s^{\scriptscriptstyle{(2)}},\\
\mathscr{Y} = \mathscr{D}\mathscr{X}
        + \sigma(\alpha_n^{\scriptscriptstyle{(1)}}\psi_\epsilon^{\scriptscriptstyle{(2)}}
        - \alpha_\epsilon^{\scriptscriptstyle{(1)}} \psi_n^{\scriptscriptstyle{(2)}})
        + \beta_n\left[\phi_\epsilon^{\scriptscriptstyle{(3)}}
        + \frac{n}{h}\phi_n^{\scriptscriptstyle{(3)}}\right].
\end{gathered}
\label{dispersion_relation_quantities}
\end{equation}

In order to obtain the transport coefficients that are preserved through the linearization
process up to the third order in the gradient expansion, one may compare the aforementioned
dispersion relations with their corresponding equations derived from an underlying microscopic theory, 
such as the strongly coupled Conformal Field Theory (CFT), which emerges as the dual theory in the
AdS/CFT correspondence. For the frame-invariant quantities $t_1^{\scriptscriptstyle{(1)}}$ and
$t_1^{\scriptscriptstyle{(3)}}$ that appear in the shear mode, the connection
is direct: the coefficient of $k^2$ gives $t_1^{\scriptscriptstyle{(1)}}$, and that of $k^4$ provides
$t_2^{\scriptscriptstyle{(3)}}$. However, the analysis becomes more intricate for the longitudinal modes
associated with charge diffusion and sound waves. In principle, one could derive microscopic dispersion
relations of the form \cite{Heller:2020uuy}
\begin{equation}
\begin{aligned}
\omega_{\mbox{\scriptsize{diffu}}}(k) & = i c_2^{\scriptscriptstyle{(0)}}k^2 + i c_4^{\scriptscriptstyle{(0)}}k^4,\\
\omega_{\mbox{\scriptsize{sound}}}^{(\pm)}(k) & = c_1^{\scriptscriptstyle{(\pm)}} k
         + ic_2^{\scriptscriptstyle{(\pm)}}k^2 + c_3^{\scriptscriptstyle{(\pm)}} k^3
         + ic_4^{\scriptscriptstyle{(\pm)}}k^4,
\end{aligned}
\label{microscopic_dispersion_relations}
\end{equation}
where the coefficients in the above series have been conveniently expressed such that all $c$'s are real
numbers. Comparing \eqref{microscopic_dispersion_relations} with the dispersion relations
\eqref{diffusion_dispersion_relation} and \eqref{soundwave_dispersion_relation}, we observe that
coefficients $c_1^{\scriptscriptstyle{(\pm)}}$ are related to the zeroth order data,
as they are determined by the speed of sound, which is obtained from derivatives of the fluid equation of state.
For first-order hydrodynamics, the relation $c_2^{\scriptscriptstyle{(0)}} = - \mathscr{D}$
leads to the conductivity $\sigma$, while the quantity $c_2^{\scriptscriptstyle{(\pm)}}$
allows us to find $\gamma_s$, which is a function of the bulk and shear viscosities, $\zeta$ and $\eta$.
In the second-order hydrodynamics, we additionally have the following equation:
\begin{equation}
c_3^{\scriptscriptstyle{(\pm)}} = \mp \frac{1}{8v_s}\left(\Gamma^2 - 4\mathscr{X}\right)
\;\;\Longrightarrow\;\; \mathscr{X} = \frac{1}{4}\left(\Gamma^2 \pm 8v_s c_3^{\scriptscriptstyle{(\pm)}}\right),
\end{equation}
where the value of $\mathscr{X}$ can be employed to obtain the combination
$\psi_\epsilon^{\scriptscriptstyle{(2)}} + (n/h)\psi_n^{\scriptscriptstyle{(2)}}
+ \beta_n\varsigma_s^{\scriptscriptstyle{(2)}}$, which depends on the second-order
frame-invariant quantities $\{f_l^{\scriptscriptstyle{(2)}}, \ell_l^{\scriptscriptstyle{(2)}},
t_l^{\scriptscriptstyle{(2)}}\}$, for $l=1,2$. Finally, the coefficients of $k^4$ in
Eqs.~\eqref{diffusion_dispersion_relation}, \eqref{soundwave_dispersion_relation}
and \eqref{microscopic_dispersion_relations} lead to
\begin{equation}
c_4^{\scriptscriptstyle{(0)}} = \frac{1}{v_s^2}\left(\mathscr{Y}
    - v_{s}^{2}\phi_n^{\scriptscriptstyle{(3)}}\right), \quad 
c_4^{\scriptscriptstyle{(\pm)}} = - \frac{1}{2v_{s}^{2}}\left(\mathscr{Y}
    + v_{s}^{2}\chi_s^{\scriptscriptstyle{(3)}}\right).
\end{equation}
These constitute two equations for $\phi_\epsilon^{\scriptscriptstyle{(3)}}$, $\phi_n^{\scriptscriptstyle{(3)}}$,
and $\chi_s^{\scriptscriptstyle{(3)}}$, which in turn depend on the third-order frame-invariant quantities
$\{f_1^{\scriptscriptstyle{(3)}}, \ell_1^{\scriptscriptstyle{(3)}}, \ell_2^{\scriptscriptstyle{(3)}}, t_1^{\scriptscriptstyle{(2)}},t_2^{\scriptscriptstyle{(3)}}\}$, of which only
$t_2^{\scriptscriptstyle{(3)}}$ can be independently determined from the shear mode.

It is pertinent to highlight here the use of dispersion relations for the determination of third-order
coefficients, as undertaken in Refs.~\cite{Grozdanov:2015kqa,Diles:2019uft} for a conformal-invariant uncharged
fluid, has been criticized by some authors \cite{Ghosh:2020lel,He:2021jna} in recent years.
According to these critics, the dispersion relations up to $k^4$ ``are related to hydrostatic
data at the quartic order and undetermined by cubic order transport data alone''.
Although we posit that an argument based on power counting is sufficient to exclude the presence
of fourth-order coefficients in dispersion relations up to $k^4$, we have conducted
our calculations considering fourth-order gradients in Eqs.~\eqref{expansion_Pi}
and \eqref{expansion_Upsilon}. In doing so, we have explicitly confirmed that fourth-order
transport coefficients are not present in linearized dispersion relations up to the quartic power of $k$.

\section{Final remarks and future perspectives}
\label{conclusions}
 
We extend the methodology of gradient expansion in relativistic hydrodynamics by employing the
Irreducible-Structure (IS) algorithm, which is applied to obtain the constitutive relations for
a fluid with one conserved charge up to third order in gradients in a general hydrodynamic frame.
The IS algorithm facilitates the formulation of the gradient expansion in terms of tensors with
well-defined symmetries under index permutation, namely, tensors $\{\Theta, \sigma, \Omega\}$,
each possessing a clear physical interpretation. The consistency of the IS algorithm follows from
theorems proven in relativistic hydrodynamics \cite{Grozdanov:2015kqa, Diles:2020cjy} and has been
explicitly verified by comparing the results with those obtained using the Grozdanov-Kaplis (GK) algorithm. 

Considering a non-conformal fluid with a single conserved charge in a general hydrodynamic frame, we find 8
transport coefficients at first order, 59 at second order, and 264 at third order. Upon selecting a frame, these
numbers are reduced to 4 transport coefficients at first order (or 3, considering the 2\textsuperscript{nd} law of thermodynamics), 30 at second order, and 147 at third order. It should be noted that choosing a specific frame is
essential for specifying the fluid mechanics of the system, such that the relevant physical information is encapsulated
by the transport coefficients in that fixed hydrodynamic frame. An interesting extension of this framework involves
considering the presence of an external gauge field in the gradient expansion and investigating the conductive
properties of a fluid within the scope of gradient expansion.

Throughout our analysis of the constitutive relations, we explored the frame dependence of the coefficients
$\{\mathcal{E},\mathcal{P}, \mathcal{N}, \mathcal{Q}^\alpha, \mathcal{J}^\alpha, \tau^{\alpha\beta}\}$ that
appear in the decomposition of the energy-momentum tensor $T^{\alpha\beta}$ and of the current $J^\alpha$. We
explored also the consequent frame dependence of the transport coefficients of the non-conformal fluid. Within
the scope of first-order hydrodynamics, the frame dependence of the fundamental hydrodynamic variables
$\{T,\mu,u^\alpha\}$ and the transport coefficients is limited to linear combinations of gradients. In this
specific context, the energy-momentum tensor and the current can be straightforwardly expressed in terms of a
set of frame-invariant transport coefficients. However, this simplification is not applicable to second- and
third-order hydrodynamics, where a plethora of nonlinear terms appear, making the task of formulating complete
expressions for the conserved charges, using solely frame-invariant quantities, notably challenging.

Fortuitously, at the linear level, we successfully identify the frame dependence of the transport coefficients,
and thereby determine the frame-invariant quantities relevant in this regime. The advantage of finding these
coefficients at the linear level is that the dispersion relations derived from the linear equations of motion
in the momentum space become independent of the choice of the frame, allowing us to select a hydrodynamic frame
at our discretion. On the basis of this approach, we obtained the linearized dispersion relations for a charged
fluid in terms of the frame-invariant coefficients. Thus, at least at the linear level, our program for studying
the gradient expansion of a charged fluid in a general frame is comprehensive. It allows us to express the constitutive relations in a general frame, analyze their frame dependence, identify observables (namely, the transport coefficients)
that are frame-invariant, and present the dynamics (in the form of dispersion relations) in a frame-invariant manner.

The challenge of expressing the full nonlinear, higher-order constitutive relations solely in terms of
frame-invariant transport coefficients is as formidable as it is desirable. Resolution to this problem
would signify a substantial enhancement to the gradient expansion approach discussed in this work. In its
current formulation, the notion of nonlinear second- and third-order transport coefficients is contingent
on the choice of the hydrodynamic frame. This issue is certainly one that we intend to address in future research.

The conformal case presents distinct challenges. Employing the mathematical structure of a Weyl manifold to
establish the constitutive relations for a conformal-invariant fluid is a natural strategy within the gradient
expansion formulation, which necessitates the use of covariant fields and operators compatible with the symmetries of
the system. We observe that, in hydrodynamics, there exists an intricate interplay between geometry and fluid flow, as established by \eqref{WeylConnection}. This implies that the conformal Riemann tensor and all its associated terms receive contributions from velocity gradients. Consequently, in the set of third-order conformal tensors, three distinct structures, represented by $\mathfrak{T}^{\scriptscriptstyle{(3)}}_{15}$, $\mathfrak{T}^{\scriptscriptstyle{(3)}}_{16}$ and $\mathfrak{T}^{\scriptscriptstyle{(3)}}_{17}$, survive the linearization process and interact with the expected linear gradients, namely $\mathfrak{T}^{\scriptscriptstyle{(3)}}_{1}$ and $\mathfrak{T}^{\scriptscriptstyle{(3)}}_{2}$.

This intricate mixing is linked to the nuanced issue that in a conformal-invariant fluid, velocity fluctuations are related with metric fluctuations. However, these metric fluctuations are constrained to vanish when considering wave phenomena in the fluid. In previous studies, such complexities were somewhat simplistically bypassed by merely ``turning off'' the curvature tensors during discussions about wave propagation in conformal fluids. Unfortunately, this strategy obscures contributions arising from the transport coefficients associated with curvature, thus masking the very information we seek to understand. These intricacies, specifically related to the Weyl structure within the realm of relativistic charged hydrodynamics, undoubtedly necessitate further exploration, which will be undertaken in subsequent research.

\begin{acknowledgments}
V.T.Z. thanks Coordenação de Aperfeiçoamento de Pessoal de Nível Superior (CAPES), Brazil, Grant No. 88887.310351/2018-00,
and Conselho Nacional de Desenvolvimento Científico e Tecnológico (CNPq), Brazil, Grant No. 311726/2022-4.
A.M.E. thanks CAPES for a master scholarship, whereas S.D. and A.S.M thank CNPq for Grant No. 406875/2023-5.
\end{acknowledgments}

\paragraph*{Note added:} The calculations presented in this work were performed using open source software.
The tensorial structures in the gradient expansion were generated through the SymPy library in Python 3, whereas
supplementary computational tasks were accomplished using the codes of the SageManifolds project in SageMath 9.3.
Regarding textual content, it was refined with the assistance of ChatGPT 4, which conducted a review of style
coherence, and cohesion. The model's recommendations were implemented to improve the quality of the text.

\appendix

\section{Time derivatives of thermodynamic variables}
\label{sec:D_thermo_vairables}

In addition to discussing the choice of hydrodynamic frame to accurately define thermodynamic variables, the description
of fluid motion necessitates the selection of a specific set of thermodynamic variables to be incorporated into the equations of motion. For a fluid with a single type of charge, there are four main thermodynamic quantities: energy density $\epsilon$, number density $n$, pressure $p$, and entropy density $s$. The zeroth order equations of motion provide relations for the gradients of energy density and number density in terms of fluid scalar expansion $\Theta$. It is desirable to employ the equation of state for pressure, denoted $p = p(T,\mu)$, to express these equations in terms of temperature and chemical potential gradients. Given that $s = (\partial p/\partial T)_\mu$, $n = (\partial p/\partial \mu)_{\scriptscriptstyle{T}}$,
and $\epsilon = -p + Ts + \mu n$, we can express $s$, $n$, and $\epsilon$ as functions of $T$ and $\mu$. These can be
combined with Eq.~\eqref{derivada_temporal_n} and the first equation in the set of Eqs.~\eqref{eq_adicionais_ordem_zero}
to yield the following relations:
\begin{align}
\left(\frac{\partial \epsilon}{\partial T}\right)_\mu DT
+ \left(\frac{\partial \epsilon}{\partial \mu}\right)_{\scriptscriptstyle{T}} D\mu
& = - h \Theta,\\
\left(\frac{\partial n}{\partial T}\right)_\mu DT +
\left(\frac{\partial n}{\partial \mu}\right)_{\scriptscriptstyle{T}} D\mu
& = -n \Theta.
\end{align}
The linear structure of this set of equations indicates that a particular criterion must be met for the equations
to be used to eliminate the time derivatives of $\mu$ and $T$ in terms of $\Theta = \nabla_{\perp \alpha} u^\alpha$.
To clarify, we express the foregoing equations in matrix form:
\begin{equation}\label{22matrix}
\begin{pmatrix}
\epsilon_{,\scriptscriptstyle{T}} & \epsilon_{,\mu} \\
n_{,\scriptscriptstyle{T}} & n_{,\mu} 
\end{pmatrix}
D
\begin{pmatrix}
T  \\
\mu
\end{pmatrix}
= -
\begin{pmatrix}
h\\
n
\end{pmatrix}
\Theta,
\end{equation}
where the operator $D$ act on each element of the matrix to the right and we have used a simplified notation for
partial derivatives, $\epsilon_{,\scriptscriptstyle{T}} = \left(\partial \epsilon/\partial T\right)_\mu$, and so on.
This system of equations is solvable if and only if the $2 \times 2$ matrix on the left-hand side is invertible.
Equivalently, this condition implies that the determinant of the matrix must be nonzero:
\begin{equation}
\epsilon_{,\scriptscriptstyle{T}}n_{,\mu} - \epsilon_{,\mu}n_{,\scriptscriptstyle{T}}\neq 0.
\end{equation}
Once this condition is satisfied, one can use the equations of motion of an ideal fluid to
eliminate the time derivatives of the Lorentz scalars $T,\mu$.

In the case of a multiphase fluid with $N$ particle species, we encounter $N+1$ variables $\epsilon,n_1,n_2,...,n_{\scriptscriptstyle{N}}$ that are functions of the $N+1$ scalars $T,\mu_1,\mu_2,...,\mu_{\scriptscriptstyle{N}}$. For each species, there exists a corresponding
continuity equation, resulting in $N+1$ equations that can be represented in matrix form,
generalizing Eq.~\eqref{22matrix}:
\begin{equation}\label{NNmatrix}
    \begin{pmatrix}
\epsilon_{,\scriptscriptstyle{T}} & \epsilon_{,\mu_1} & \cdots &  \epsilon_{,\mu_{\scriptscriptstyle{N}}}\\
n_{1,\scriptscriptstyle{T}} & n_{1,\mu_1} & \cdots & n_{1,\mu_{\scriptscriptstyle{N}}}\\
\vdots & \vdots & \ddots & \vdots \\
n_{\scriptscriptstyle{N},\scriptscriptstyle{T}} & n_{{\scriptscriptstyle{N}},\mu_1} &
\cdots & n_{{\scriptscriptstyle{N}},\mu_{\scriptscriptstyle{N}}}
\end{pmatrix}
D
\begin{pmatrix}
T\\
\mu_1  \\
\vdots \\
\mu_{\scriptscriptstyle{N}}
\end{pmatrix}
= -
\begin{pmatrix}
h\\
n_1\\
\vdots \\
n_{\scriptscriptstyle{N}}
\end{pmatrix}
\Theta. 
\end{equation}
In this scenario, the necessary and sufficient condition for expressing time derivatives in terms of the scalar expansion
is that the matrix on the left-hand side of the above equation be invertible, or equivalently, its determinant must be nonzero.

\section{Equivalences between symmetric gradients}
\label{equiv_symmetry}

It is well established that a rank-two tensor can be Weyl-decomposed into three components, each defined by its symmetry properties under the rotation group: the antisymmetric part, the symmetric and traceless part, and the scalar (or trace)
part. In the realm of hydrodynamics, it transpires that a fundamental gradient is represented as a rank-two tensor,
which we then subject to Weyl decomposition:
\begin{equation}
    \nabla^\alpha u^\beta =  \sigma^{\alpha \beta}+\Omega^{\alpha \beta} + \frac{1}{d-1}\Theta \Delta^{\alpha\beta}.
\end{equation}
Each term in this decomposition has a distinct physical meaning. The symmetric and traceless tensor (TST) part, denoted by
$\sigma$, encodes shear effects; the trace part, $\Theta$, accounts for expansion and compression; and the antisymmetric
part, $\Omega$, represents vorticity. The physical content of these terms provides a strong motivation to use
$\sigma$, $\Theta$, and $\Omega$ as fundamental structures in the building of the gradient expansion. However, a
nontrivial issue arises. It has been demonstrated in \cite{Diles:2020cjy} that the ordering of transverse derivatives
is irrelevant in the construction of the tensorial structures of the hydrodynamic expansion. If one opts not to decompose
the velocity gradient, this equivalence is easily implemented by establishing
$\nabla^\alpha \nabla^\beta u^\gamma \simeq  \nabla^\beta \nabla^\alpha u^\gamma$. On the contrary,
translating this rule of equivalence to the gradients of $\sigma$, $\Theta$, and $\Omega$ presents a challenge.

It should be noted that the second derivative of the velocity field can be expressed as a linear combination of the
gradients of $\sigma$, $\Theta$, and $\Omega$. Together with the irrelevance of derivative ordering mentioned above,
this property suggests an underlying relationship among the gradients of these tensor components. In fact,
using the notation of \cite{Diles:2020cjy}, one can formally establish that
\begin{equation}
    \nabla_{\perp}^\alpha \Omega^{\beta \gamma} \simeq \nabla_{\perp}^\beta \left(\sigma^{\alpha\gamma}
            + \frac{\Delta^{\alpha \gamma}}{d-1}\Theta\right)
            - \nabla_{\perp}^\gamma \left(\sigma^{\alpha\beta}
            + \frac{\Delta^{\alpha \beta}}{d-1}\Theta\right).
\label{omegaequiv}
\end{equation}
In the above equation, the symbol $\simeq$ indicates that the tensors on both sides of the equation are equivalent,
meaning their difference constitutes a linear combination of other gradients already accounted for in the gradient expansion. For example, the divergence of vorticity satisfies $\nabla_{\perp \alpha} \Omega^{\alpha\gamma} \simeq \nabla_{\perp \alpha} \sigma^{\alpha\gamma} + 2\nabla_{\perp}^\gamma\Theta/(d-1)$. Practically speaking, the implication of Eq.~\eqref{omegaequiv}
is that one may elect to omit vorticity gradients from the terms of the hydrodynamic expansion, substituting them by the shear
and expansion gradients. Such an omission of vorticity derivatives is also coherent in the context of conformal fluids, where
the expansion is identically null.

\section{General-frame derivation of dispersion relations}
\label{general-frame-derivation}

In this appendix, we provide an alternative derivation of the dispersion relations
expressed in \eqref{shear_dispersion_relation}, \eqref{diffusion_dispersion_relation} and
\eqref{soundwave_dispersion_relation}. This derivation relies on constitutive
relations written in a general hydrodynamic frame. More specifically, we consider the
decomposition of the dissipative terms $\Pi^{\mu\nu}$ and $\Upsilon^{\mu}$ as follows:
\begin{align}
\Pi^{\alpha\beta} & = \mathcal{E}u^\alpha u^\beta + \mathcal{P}\Delta^{\alpha\beta} +
2\mathcal{Q}^{(\alpha} u^{\beta)} + \tau^{\alpha\beta},
\label{Pi_function}\\
\Upsilon^{\mu} & = \mathcal{N}u^\mu + \mathcal{J}^\mu.
\label{Upsilon_function}
\end{align}
Each term in these equations is expanded up to the third order in the gradient series,
\begin{equation}
\begin{aligned}
\mathcal{E} & = \mathcal{E}_{\scriptscriptstyle{(1)}} + \mathcal{E}_{\scriptscriptstyle{(2)}} +
\mathcal{E}_{\scriptscriptstyle{(3)}},\quad
& \mathcal{Q}^{\mu} & = \mathcal{Q}^{\mu}_{\scriptscriptstyle{(1)}} + \mathcal{Q}^{\mu}_{\scriptscriptstyle{(2)}}
+ \mathcal{Q}^{\mu}_{\scriptscriptstyle{(3)}},\\
\mathcal{P} & = \mathcal{P}_{\scriptscriptstyle{(1)}} + \mathcal{P}_{\scriptscriptstyle{(2)}} +
\mathcal{P}_{\scriptscriptstyle{(3)}},\quad
& \mathcal{J}^{\mu} & = \mathcal{J}^{\mu}_{\scriptscriptstyle{(1)}} + \mathcal{J}^{\mu}_{\scriptscriptstyle{(2)}}
+ \mathcal{J}^{\mu}_{\scriptscriptstyle{(3)}},\\
\mathcal{N} & = \mathcal{N}_{\scriptscriptstyle{(1)}} + \mathcal{N}_{\scriptscriptstyle{(2)}} +
\mathcal{N}_{\scriptscriptstyle{(3)}},\quad
&\tau^{\mu\nu} & = \tau^{\mu\nu}_{\scriptscriptstyle{(1)}} + \tau^{\mu\nu}_{\scriptscriptstyle{(2)}} +
\tau^{\mu\nu}_{\scriptscriptstyle{(3)}}.
\end{aligned}
\end{equation}
Owing to the computational complexity and the extensive nature of the calculations in a general frame,
this appendix focuses on expanding the aforementioned coefficients in terms of gradients of the densities
$\epsilon$ and $n$, as opposed to temperature and chemical potential. Only after deriving the dispersion
relations do we substitute the relationships between the transport coefficients associated with the pair
$(\epsilon, n)$ and those corresponding to $(T, \mu)$.

The equations of motion comprise one vector equation and two scalar equations. By taking the
vanishing divergence of $T^{\alpha\beta}$ and projecting it in directions parallel and
perpendicular to $u^\alpha$, we derive the scalar equation \eqref{paralelo1} and the
vector equation \eqref{perpendicular1}. The remaining scalar equation, Eq.~\eqref{conservacao_corrente1},
emerges from the vanishing divergence of $J^\alpha$. Substituting \eqref{Pi_function} and
\eqref{Upsilon_function} into Eqs.~\eqref{paralelo1}, \eqref{perpendicular1}, and \eqref{conservacao_corrente1},
we arrive at the following equations:
\begin{gather}
\begin{aligned}
D(\epsilon+\mathcal{E}) & + (h + \mathcal{E} + \mathcal{P})\Theta\\
&+ \tau^{\alpha\beta}\sigma_{\alpha\beta}
+\nabla_\alpha\mathcal{Q}^\alpha + \mathcal{Q}_\alpha Du^\alpha= 0,
\label{paralelo_new1}
\end{aligned}\\
\begin{aligned}
(h + \mathcal{E} & + \mathcal{P})  D u^\alpha + \nabla_{\perp}^\alpha(p+\mathcal{P})+
\Delta^{\alpha}_{\;\;\beta}\nabla_\gamma \tau^{\gamma\beta}\\
    & + \mathcal{Q}^\beta \nabla_\beta u^\alpha
    + \Delta^{\alpha}_{\;\;\beta}D \mathcal{Q}^{\beta} + \mathcal{Q}^{\alpha}\Theta= 0,
\label{perpendicular_new1}
\end{aligned}\\
(n + \mathcal{N})\Theta + D(n + \mathcal{N}) + \nabla_\alpha \mathcal{J}^{\alpha} = 0.
\label{conserva_corrente_new1}
\end{gather}

By retaining only the linear terms in the amplitudes, the dissipation structures
in the above equations can be expressed in terms of transport coefficients as follows:
\begin{equation}
\begin{gathered}
\mathcal{E} = \left(\varepsilon_{\scriptscriptstyle{\Theta}}^{\scriptscriptstyle{(1)}}
            + \varepsilon_{\scriptscriptstyle{\Theta}}^{\scriptscriptstyle{(3)}}\nabla_{\perp}^2\right)\Theta
            + \sum_{a} \varepsilon_{a}^{\scriptscriptstyle{(2)}}\nabla_{\perp}^2 a,\\
\mathcal{P} = \left(\pi_{\scriptscriptstyle{\Theta}}^{\scriptscriptstyle{(1)}}
            + \pi_{\scriptscriptstyle{\Theta}}^{\scriptscriptstyle{(3)}}\nabla_{\perp}^2\right)\Theta
            + \sum_{a} \pi_{a}^{\scriptscriptstyle{(2)}}\nabla_{\perp}^2 a,\\           
\mathcal{N} = \left(\nu_{\scriptscriptstyle{\Theta}}^{\scriptscriptstyle{(1)}}
            + \nu_{\scriptscriptstyle{\Theta}}^{\scriptscriptstyle{(3)}}\nabla_{\perp}^2\right)\Theta
            + \sum_{a} \nu_{a}^{\scriptscriptstyle{(2)}}\nabla_{\perp}^2 a,\\
\begin{aligned}
\mathcal{Q}^{\alpha} & = \sum_{a}\left(\theta_{a}^{\scriptscriptstyle{(1)}}
                        + \theta_{a}^{\scriptscriptstyle{(3)}}\nabla_{\perp}^2\right)\nabla_{\perp}^\alpha a\\
                        & + \theta_{\scriptscriptstyle{\Theta}}^{\scriptscriptstyle{(2)}}\nabla_{\perp}^\alpha \Theta
                        + \theta_{\sigma}^{\scriptscriptstyle{(2)}}\Delta^{\alpha}_{\;\;\beta}\nabla_{\perp\gamma} \sigma^{\gamma\beta},
\end{aligned}\\
\begin{aligned}
\mathcal{J}^{\alpha} & = \sum_{a}\left(\kappa_{a}^{\scriptscriptstyle{(1)}}
                        + \kappa_{a}^{\scriptscriptstyle{(3)}}\nabla_{\perp}^2\right)\nabla_{\perp}^\alpha a\\
                        & + \kappa_{\scriptscriptstyle{\Theta}}^{\scriptscriptstyle{(2)}}\nabla_{\perp}^\alpha \Theta
                        + \kappa_{\sigma}^{\scriptscriptstyle{(2)}}\Delta^{\alpha}_{\;\;\beta}\nabla_{\perp\gamma} \sigma^{\gamma\beta},
\end{aligned}\\
\begin{aligned}
\tau^{\alpha\beta} &= \left(\eta_{\sigma}^{\scriptscriptstyle{(1)}}
                    + \eta_{\sigma}^{\scriptscriptstyle{(3)}}\nabla_{\perp}^2\right)\sigma^{\alpha\beta}\\
                    & + \sum_{a} \eta_{a}^{\scriptscriptstyle{(2)}}\nabla_{\perp}^{\langle\alpha}\nabla_{\perp}^{\beta\rangle}a
                    +\eta_{\scriptscriptstyle{\Theta}}^{\scriptscriptstyle{(3)}} \nabla_{\perp}^{\langle\alpha}\nabla_{\perp}^{\beta\rangle}\Theta,
\end{aligned}
\end{gathered}
\label{linear_relations_coef}
\end{equation}
where $a$ takes the value $\epsilon$ and $n$. As a mnemonic strategy, we choose to use the subscripts
$\scriptstyle{\Theta}$, $\sigma$, $\epsilon$, and $n$ throughout this appendix, rather than the numerical indices used
to denote various transport coefficients in Sections \ref{non-conformal} and \ref{conformal-fluid}.

In the case of a $d$-dimensional flat spacetime and considering the linear fluctuations in the hydrodynamic variables
defined in \eqref{linear_perturbations}, Eqs.~\eqref{paralelo_new1}--\eqref{conserva_corrente_new1} reduce to
\begin{gather}
u^\alpha\partial_\alpha(\delta\epsilon+\delta\mathcal{E}) + h\partial_{\perp\alpha}\delta u^\alpha
+ \partial_{\perp\alpha}\delta \mathcal{Q}^\alpha= 0, \label{paralelo_new2}\\
h u^\beta\partial_\beta\delta u^\alpha + u^\beta\partial_\beta\delta \mathcal{Q}^\alpha +
\partial_{\perp}^{\alpha}(\delta p +\delta \mathcal{P}) +
\partial_{\perp\beta}\delta\tau^{\beta\alpha} = 0,\label{perpendicular_new2} \\
u^\alpha\partial_\alpha(\delta n +\delta\mathcal{N}) + n\partial_{\perp\alpha}\delta u^\alpha
+\partial_{\perp\alpha}\delta \mathcal{J}^\alpha = 0. \label{conserva_corrente_new2}
\end{gather}
By Fourier-decomposing the perturbations in Eqs.~\eqref{paralelo_new2}--\eqref{conserva_corrente_new2}
in terms of the plane waves $\exp{(ik_{\alpha}x^{\alpha})}$, we obtain
\begin{gather}
k_{\parallel}(\delta\epsilon +\delta\mathcal{E}) + k_{\perp\alpha}(h\delta u^{\alpha} + \delta \mathcal{Q}^{\alpha}) = 0,
\label{paralelo_new3}\\
k_{\parallel}(h\delta u^{\alpha}+\delta\mathcal{Q}^\alpha) + k_{\perp}^{\alpha}(\delta p + \delta\mathcal{P})
+k_{\perp\beta} \delta\tau^{\beta\alpha} = 0,
\label{perpendicular_new3}\\
k_{\parallel}(\delta n +\delta\mathcal{N}) + k_{\perp\alpha}(n\delta u^{\alpha} + \delta\mathcal{J}^\alpha) = 0,
\label{conserva_corrente_new3}
\end{gather}
where $k_{\parallel}= u^{\alpha}k_{\alpha}$ and $k_{\perp}^{\alpha}=\Delta^{\alpha\beta}k_{\beta}$.
The perturbations in $\mathcal{E}$, $\mathcal{P}$, $\mathcal{N}$, $\mathcal{Q}^{\alpha}$,
$\mathcal{J}^{\alpha}$ and $\tau^{\alpha\beta}$ are given by the Fourier-transformed linearized
versions of equations \eqref{linear_relations_coef}. Furthermore, the perturbation in the
pressure assumes the form:
\begin{equation}
\delta p = \beta_\epsilon \delta\epsilon + \beta_n \delta n.
\end{equation}

Equations \eqref{paralelo_new3}--\eqref{conserva_corrente_new3} constitute a set of two scalar equations
and one vector equation for the variables $\delta\epsilon$, $\delta n$, and $\delta u^{\alpha}$.
Projecting the vector equation \eqref{perpendicular_new3} along the direction transverse to
$k_{\perp}^{\alpha}$ yields
\begin{equation}
\left[k_{\parallel}\left(2h-\theta_{\sigma}^{\scriptscriptstyle{(2)}}k_{\perp}^{2}\right)+
ik_{\perp}^{2}(\eta_{\sigma}^{\scriptscriptstyle{(1)}}-\eta_{\sigma}^{\scriptscriptstyle{(3)}} k_{\perp}^{2})\right]
\delta u_{\scriptscriptstyle{T}}^{\alpha} = 0,
\label{eq_transversal_new1}
\end{equation}
where $\delta u_{\scriptscriptstyle{T}}^{\alpha}$ is defined as
\begin{equation}
\delta u_{\scriptscriptstyle{T}}^{\alpha} =\left(\Delta^{\alpha}_{\;\;\beta}-
\frac{k_{\perp}^{\alpha}k_{\perp\beta}}{k_{\perp}^{2}}\right)\delta u^\beta.
\end{equation}
Employing a similar procedure, but projecting \eqref{perpendicular_new3} along the direction longitudinal
to $k_{\perp}^{\alpha}$, leads to the following equation:
\begin{equation}
\begin{gathered}
\left[\left(\xi_{s}^{\scriptscriptstyle{(3)}}k_{\perp}^{2}
            + \lambda_{s}^{\scriptscriptstyle{(1)}}\right) k_{\perp}^{2} 
            + i\left(\theta_{s}^{\scriptscriptstyle{(2)}} k_{\perp}^{2}
            - 1\right)k_{\parallel}\right] h\delta u_{\scriptscriptstyle{L}}\\
            + \sum_{a} \left[i\left(\gamma_{a}^{\scriptscriptstyle{(2)}} h k_{\perp}^{2}
            + \beta_{a}\right) - \left(\theta_{a}^{\scriptscriptstyle{(3)}} k_{\perp}^{2}
            - \theta_{a}^{\scriptscriptstyle{(1)}}\right)k_{\parallel}\right]k_{\perp}\delta a= 0,
\end{gathered}
\label{eq_longitudinal3}
\end{equation}
where the index $a$ takes the values $\epsilon$ or $n$. The set of combinations of transport coefficients
introduced above are given by
\begin{equation}
\begin{gathered}
\lambda_{s}^{\scriptscriptstyle{(1)}} = - \frac{1}{h}\left(\pi_{\scriptscriptstyle{\Theta}}^{\scriptscriptstyle{(1)}}
        + \frac{d-2}{d-1}\eta_{\sigma}^{\scriptscriptstyle{(1)}}\right),\\
\gamma_{a}^{\scriptscriptstyle{(2)}} = - \frac{1}{h}\left(\pi_{a}^{\scriptscriptstyle{(2)}}
        + \frac{d-2}{d-1}\eta_{a}^{\scriptscriptstyle{(2)}}\right),\\
\theta_{s}^{\scriptscriptstyle{(2)}} = \frac{1}{h}\left(\theta_{\scriptscriptstyle{\Theta}}^{\scriptscriptstyle{(2)}}
        +\frac{d-2}{d-1}\theta_{\sigma}^{\scriptscriptstyle{(2)}}\right),\\
\xi_{s}^{\scriptscriptstyle{(3)}} = \frac{1}{h}\left(\pi_{\scriptscriptstyle{\Theta}}^{\scriptscriptstyle{(3)}}
        + \frac{d-2}{d-1}\left(\eta_{\scriptscriptstyle{\Theta}}^{\scriptscriptstyle{(3)}}
        + \eta_{\sigma}^{\scriptscriptstyle{(3)}}\right)\right).
\end{gathered}
\end{equation}
Additionally, $\delta u_{\scriptscriptstyle{L}}$ denotes the magnitude of the longitudinal
projection vector of the fluid velocity perturbation, defined as
\begin{equation}
\delta u_{\scriptscriptstyle{L}}^{\alpha} =
\frac{\left(k_{\perp\beta}\delta u^{\beta}\right)}{k_{\perp}^{2}}k_{\perp}^{\alpha}.
\end{equation}
Note that the frame-invariant quantities $\gamma_{s}^{\scriptscriptstyle{(1)}}$ and
$\chi_{s}^{\scriptscriptstyle{(3)}}$, defined in \eqref{new_coeficients},
are equal, respectively, to $\lambda_{s}^{\scriptscriptstyle{(1)}}$ and
$\xi_{s}^{\scriptscriptstyle{(3)}}$ in the Landau frame.

In terms of $\delta u_{\scriptscriptstyle{L}}$, $\delta\epsilon$, and $\delta n$, equations
\eqref{paralelo_new3} and \eqref{conserva_corrente_new3} can be reformulated as
\begin{gather}
\begin{gathered}
\left[i h \left({\theta_{s}^{\scriptscriptstyle{(2)}}} k_{\perp}^{2} - 1\right)
        + \left(\varepsilon_{\scriptscriptstyle{\Theta}}^{\scriptscriptstyle{(3)}}k_{\perp}^{2}
        - \varepsilon_{\scriptscriptstyle{\Theta}}^{\scriptscriptstyle{(1)}}\right)
        k_{\parallel} \right] k_{\perp} \delta u_{\scriptscriptstyle{L}}\\
        + i k_{\parallel}\delta\epsilon
        - \sum_a \left(\theta_{a}^{\scriptscriptstyle{(3)}}k_{\perp}^{2}
        - \theta_{a}^{\scriptscriptstyle{(1)}}
        + i\varepsilon_{a}^{\scriptscriptstyle{(2)}}k_{\parallel}\right) k_{\perp}^{2}\delta a = 0,
\end{gathered}
\label{paralelo_new4}\\
\begin{gathered}
\left[ih\left(\kappa_{s}^{\scriptscriptstyle{(2)}} k_{\perp}^{2} - 1\right)
        + \left(\nu_{\scriptscriptstyle{\Theta}}^{\scriptscriptstyle{(3)}}k_{\perp}^{2}
        - \nu_{\scriptscriptstyle{\Theta}}^{\scriptscriptstyle{(1)}}\right)
        k_{\parallel}\right]k_{\perp}\delta u_{\scriptscriptstyle{L}}\\
        + ik_{\parallel}\delta n
        - \sum_a \left(\kappa_{a}^{\scriptscriptstyle{(3)}} k_{\perp}^{2}
        - \kappa_{a}^{\scriptscriptstyle{(1)}}
        + i\nu_{a}^{\scriptscriptstyle{(2)}}k_{\parallel}\right) k_{\perp}^{2}\delta a = 0,
\end{gathered}
\label{conserva_corrente_new4}
\end{gather}
where $\kappa_{s}^{\scriptscriptstyle{(2)}}$ is defined as
\begin{equation}
\kappa_{s}^{\scriptscriptstyle{(2)}} = \frac{1}{h}\left(\kappa_{\scriptscriptstyle{\Theta}}^{\scriptscriptstyle{(2)}}
                                     + \frac{d-2}{d-1}\kappa_{\sigma}^{\scriptscriptstyle{(2)}}\right).
\end{equation}
Together with \eqref{eq_longitudinal3}, equations \eqref{paralelo_new4} and \eqref{conserva_corrente_new4}
comprise a system of three homogeneous equations for the independent variables $\delta u_{\scriptscriptstyle{L}}$, $\delta\epsilon$, and $\delta n$. From the vanishing of the determinant of the matrix of coefficients of this
system of equations, as well as of the term within square brackets in \eqref{eq_transversal_new1}, we respectively
derive the equations for the longitudinal and transverse hydrodynamic modes.

To find the dispersion relations, we adopt the rest reference frame of the fluid,
\begin{equation}
u^\alpha = (1,\, \vec{0})\;\Rightarrow \; k_{\parallel} = u^{0}k_{0} \equiv -\omega
\;\;\mbox{and}\;\; k_{\perp}^{2}=k^{i}k_{i}\equiv k^2.
\end{equation}
In this manner, the algebraic equations for the hydrodynamic modes transform into
\begin{equation}
\begin{gathered}
\omega\left(2h-\theta_{\sigma}^{\scriptscriptstyle{(2)}}k^{2}\right)
        + ik^{2}(\eta_{\sigma}^{\scriptscriptstyle{(1)}}
        - \eta_{\sigma}^{\scriptscriptstyle{(3)}} k^{2}) = 0,\\
        \det\mathcal{M}(\omega, k) = 0, 
\end{gathered}
\label{eqs_fund_rel_disp}
\end{equation}
where the elements of the matrix $\mathcal{M}$ are defined as
\begin{equation}
\begin{aligned}
\mathcal{M}_{11} & = \left[\left(\xi_{s}^{\scriptscriptstyle{(3)}}k^{2}
                    + \lambda_{s}^{\scriptscriptstyle{(1)}}\right) k^{2}
                    + i \left(\theta_{s}^{\scriptscriptstyle{(2)}} k^{2}
                    - 1\right) \omega\right] h,\\
\mathcal{M}_{12} & = \left[i\left(\gamma_{\epsilon}^{\scriptscriptstyle{(2)}} h k^{2}
                    + \beta_{\epsilon}\right)
                    - \left(\theta_{\epsilon}^{\scriptscriptstyle{(3)}} k^{2}
                    - \theta_{\epsilon}^{\scriptscriptstyle{(1)}}\right)\omega\right]k,\\
\mathcal{M}_{13} & = \left[i\left(\gamma_{n}^{\scriptscriptstyle{(2)}} h k^{2}
                    + \beta_{n}\right)
                    - \left(\theta_{n}^{\scriptscriptstyle{(3)}} k^{2}
                    - \theta_{n}^{\scriptscriptstyle{(1)}}\right)\omega\right]k,\\
\mathcal{M}_{21} & = \left[ih\left(\theta_{s}^{\scriptscriptstyle{(2)}} k^{2} - 1\right)
                    + \left(\varepsilon_{\scriptscriptstyle{\Theta}}^{\scriptscriptstyle{(3)}}k^{2}
                    - \varepsilon_{\scriptscriptstyle{\Theta}}^{\scriptscriptstyle{(1)}}\right)\omega\right]k,\\
\mathcal{M}_{22} & = - \left(\theta_{\epsilon}^{\scriptscriptstyle{(3)}} k^{2}
                    - \theta_{\epsilon}^{\scriptscriptstyle{(1)}}\right) k^{2}
                    - i\left(\varepsilon_{\epsilon}^{\scriptscriptstyle{(2)}}k^2 - 1\right)\omega,\\
\mathcal{M}_{23} & = - \left(\theta_{n}^{\scriptscriptstyle{(3)}}k^{2}
                    - \theta_{n}^{\scriptscriptstyle{(1)}}
                    + i\varepsilon_{n}^{\scriptscriptstyle{(2)}}\omega\right) k^{2},\\
\mathcal{M}_{31} & = \left[i\left(h\kappa_{s}^{\scriptscriptstyle{(2)}} k^{2} - n\right)
                    + \left(\nu_{\scriptscriptstyle{\Theta}}^{\scriptscriptstyle{(3)}}k^{2}
                    - \nu_{\scriptscriptstyle{\Theta}}^{\scriptscriptstyle{(1)}}\right)\omega\right]k,\\
\mathcal{M}_{32} & = - \left(\kappa_{\epsilon}^{\scriptscriptstyle{(3)}}k^{2}
                    - \kappa_{\epsilon}^{\scriptscriptstyle{(1)}}
                    + i \nu_{\epsilon}^{\scriptscriptstyle{(2)}}\omega\right)k^{2},\\
\mathcal{M}_{33} & = - \left(\kappa_{n}^{\scriptscriptstyle{(3)}} k^{2}
                    - \kappa_{n}^{\scriptscriptstyle{(1)}}\right) k^{2}
                    - i\left(\nu_{n}^{\scriptscriptstyle{(2)}} k^2 - 1\right)\omega.\\
\end{aligned}
\end{equation}

The solution to the first of Eqs.~\eqref{eqs_fund_rel_disp} gives rise to the shear mode
associated with the diffusion of transverse momentum in the fluid,
\begin{equation}
 \omega_{\mbox{\scriptsize{shear}}}(k) = i\frac{\eta_{\sigma}^{\scriptscriptstyle{(1)}}}{2h}k^2
        - \frac{i}{2h}\left(\eta_{\sigma}^{\scriptscriptstyle{(3)}}
        - \frac{\eta_{\sigma}^{\scriptscriptstyle{(1)}}\theta_{\sigma}^{\scriptscriptstyle{(2)}}}{2h}\right)k^4,
 \label{rel_shear}
\end{equation}
which reduces to \eqref{shear_dispersion_relation}, given that 
$t_1^{\scriptscriptstyle{(1)}} = \eta_{\sigma}^{\scriptscriptstyle{(1)}}$ and
$t_{2}^{\scriptscriptstyle{(3)}} = \eta_{\sigma}^{\scriptscriptstyle{(3)}}
- \eta_{\sigma}^{\scriptscriptstyle{(1)}}\theta_{\sigma}^{\scriptscriptstyle{(2)}}/2h$,
in accordance with Eqs.~\eqref{shear_viscosity_relation} and \eqref{frame_invariants_3order}.
The solutions to the second of Eqs.~\eqref{eqs_fund_rel_disp} yield the dispersion
relation for the charge diffusion mode, given by
\begin{equation}
\omega_{\mbox{\scriptsize{diffu}}}(k) = -i\mathscr{D}k^2 -
\frac{i}{v_{s}^{2}}\left\{\mathscr{D}\left[v_{s}^{2}(\mathscr{A} - \theta_{s}^{\scriptscriptstyle{(2)}})
+\mathscr{B}\right] + \mathscr{C}\right\}k^4,
\label{rel_diffusion}
\end{equation}
and the dispersion relations for the sound wave modes, which can be expressed as
\begin{equation}
\begin{split}
\omega_{\mbox{\scriptsize{sound}}}^{(\pm)}(k) = & \pm v_{s} k - \frac{i}{2}\Gamma k^2
\mp \frac{1}{8v_s}\left(\Gamma^2 + 4\mathscr{B}\right) k^3\\
& -\frac{i}{2v_{s}^{2}}\left[v_{s}^{2}\left(\Gamma\mathscr{A}+\mathscr{E}\right)-\mathscr{D}\mathscr{B}\right]k^4,
\end{split}
\label{rel_sound}
\end{equation}
where the first-order diffusion and sound-wave damping constants,
\begin{equation}
\begin{gathered}
\mathscr{D}  = \frac{\sigma}{v_{s}^{2}}\left(\alpha_{n}^{\scriptscriptstyle{(1)}}\beta_\epsilon
                 - \alpha_{\epsilon}^{\scriptscriptstyle{(2)}}\beta_n\right)\\
\Gamma = \lambda_{s}^{\scriptscriptstyle{(1)}}
         + \frac{\sigma\beta_n}{v_{s}^{2}}\left(\alpha_{\epsilon}^{\scriptscriptstyle{(1)}}
         + \frac{n}{h}\alpha_{n}^{\scriptscriptstyle{(1)}}\right)
         + \vartheta_{\scriptscriptstyle{\Theta}}^{\scriptscriptstyle{(1)}},
\end{gathered}
\end{equation}
correspond to those presented in Section \ref{dispersion_relations}, but naturally find here in
terms of the original transport coefficients, contrasting with their representation in terms of
frame-invariant quantities in Eqs.~\eqref{new_coeficients} and \eqref{dispersion_relation_quantities}.
The quantities $\mathscr{A}$, $\mathscr{B}$, $\mathscr{C}$, and $\mathscr{E}$ are explicitly defined
in the following equations:
\begin{gather}
\mathscr{A} = \theta_{s}^{\scriptscriptstyle{(2)}} + \varepsilon_{\epsilon}^{\scriptscriptstyle{(2)}}
            + \nu_{n}^{\scriptscriptstyle{(2)}}
            - \frac{1}{h}\left(\theta_{\epsilon}^{\scriptscriptstyle{(1)}}
            \varepsilon_{\scriptscriptstyle{\Theta}}^{\scriptscriptstyle{(1)}}
            + \theta_{n}^{\scriptscriptstyle{(1)}}\nu_{\scriptscriptstyle{\Theta}}^{\scriptscriptstyle{(1)}}\right),\\
\begin{split}
\mathscr{B} & = \mathscr{D}\bigg[\Gamma
            - v_{s}^{2}\bigg(\frac{\varepsilon_{\scriptscriptstyle{\Theta}}^{\scriptscriptstyle{(1)}}}{h}\bigg)\bigg]
            - h \left[\gamma_{\epsilon}^{\scriptscriptstyle{(2)}} + \vartheta_{\epsilon}^{\scriptscriptstyle{(2)}}
            + \frac{n}{h}\left(\gamma_{n}^{\scriptscriptstyle{(2)}} + \vartheta_{n}^{\scriptscriptstyle{(2)}}\right)\right]\\
            & + \Big(\theta_{\epsilon}^{\scriptscriptstyle{(1)}} + \frac{n}{h}\theta_{n}^{\scriptscriptstyle{(1)}}\Big)
            \left(\lambda_{s}^{\scriptscriptstyle{(1)}}
            + \vartheta_{\scriptscriptstyle{\Theta}}^{\scriptscriptstyle{(1)}}\right)
            - \sigma\left(\alpha_{s}^{\scriptscriptstyle{(2)}}\beta_n
            + \alpha_{n}^{\scriptscriptstyle{(1)}}\lambda_{s}^{\scriptscriptstyle{(1)}}\right),
\end{split}\\
\begin{split}
\mathscr{C} & = \sigma\left(\beta_n \alpha_{\epsilon}^{\scriptscriptstyle{(3)}}
            - \beta_\epsilon \alpha_{n}^{\scriptscriptstyle{(3)}}\right)
            + \sigma h\left(\alpha_{n}^{\scriptscriptstyle{(1)}}\gamma_{\epsilon}^{\scriptscriptstyle{(2)}}
            - \alpha_{\epsilon}^{\scriptscriptstyle{(1)}}\gamma_{n}^{\scriptscriptstyle{(2)}}\right)\\
            & + \sigma\alpha_{s}^{\scriptscriptstyle{(2)}}\left(\beta_\epsilon \theta_{n}^{\scriptscriptstyle{(1)}}
            - \beta_n \theta_{\epsilon}^{\scriptscriptstyle{(1)}}\right)
            + \sigma\lambda_{s}^{\scriptscriptstyle{(1)}}
            \left(\alpha_{\epsilon}^{\scriptscriptstyle{(1)}}\theta_{n}^{\scriptscriptstyle{(1)}}
            - \alpha_{n}^{\scriptscriptstyle{(1)}}\theta_{\epsilon}^{\scriptscriptstyle{(1)}}\right),
\end{split}\\
\begin{split}
\mathscr{E} & = -\frac{1}{h}\left(\beta_\epsilon \nu_{n}^{\scriptscriptstyle{(2)}}
                - \beta_n \nu_{\epsilon}^{\scriptscriptstyle{(2)}}\right)
                \varepsilon_{\scriptscriptstyle{\Theta}}^{\scriptscriptstyle{(1)}}
                + \frac{1}{h}\left(\beta_\epsilon \varepsilon_{n}^{\scriptscriptstyle{(2)}}
                - \beta_n \varepsilon_{\epsilon}^{\scriptscriptstyle{(2)}} \right)
                \nu_{\scriptscriptstyle{\Theta}}^{\scriptscriptstyle{(1)}}\\
                & + \gamma_{\epsilon}^{\scriptscriptstyle{(2)}}
                \varepsilon_{\scriptscriptstyle{\Theta}}^{\scriptscriptstyle{(1)}} + \gamma_{n}^{\scriptscriptstyle{(2)}}\nu_{\scriptscriptstyle{\Theta}}^{\scriptscriptstyle{(1)}}
                - \left(\varepsilon_{\epsilon}^{\scriptscriptstyle{(2)}}
                + \nu_{n}^{\scriptscriptstyle{(2)}}\right)\lambda_{s}^{\scriptscriptstyle{(1)}}
                + \xi_{s}^{\scriptscriptstyle{(3)}}
                - \vartheta_{\scriptscriptstyle{\Theta}}^{\scriptscriptstyle{(3)}}\\
                &+ \sigma\left(\alpha_{n}^{\scriptscriptstyle{(1)}}\theta_{\epsilon}^{\scriptscriptstyle{(1)}}
                - \alpha_{\epsilon}^{\scriptscriptstyle{(1)}}\theta_{n}^{\scriptscriptstyle{(1)}}\right)
                \left(\frac{\lambda_{s}^{\scriptscriptstyle{(1)}}}{v_{s}^{2}}
                + \frac{\varepsilon_{\scriptscriptstyle{\Theta}}^{\scriptscriptstyle{(1)}}}{h}\right)
                + \frac{\sigma\beta_n}{v_s^2}\Big[\alpha_{s}^{\scriptscriptstyle{(2)}}
                \theta_{\epsilon}^{\scriptscriptstyle{(1)}}\\
                & - \alpha_{\epsilon}^{\scriptscriptstyle{(1)}}\theta_{s}^{\scriptscriptstyle{(2)}}
                - \alpha_{\epsilon}^{\scriptscriptstyle{(3)}}
                + \frac{n}{h} \left(\alpha_{s}^{\scriptscriptstyle{(2)}}
                \theta_{n}^{\scriptscriptstyle{(1)}}
                - \alpha_{n}^{\scriptscriptstyle{(1)}}\theta_{s}^{\scriptscriptstyle{(2)}}
                - \alpha_{n}^{\scriptscriptstyle{(3)}}\right)\Big]\\
                & -\frac{\sigma h}{v_s^2}\left(\alpha_{n}^{\scriptscriptstyle{(1)}}
                \gamma_{\epsilon}^{\scriptscriptstyle{(2)}}
                - \alpha_{\epsilon}^{\scriptscriptstyle{(1)}}\gamma_{n}^{\scriptscriptstyle{(2)}}\right)
                - \sigma\left(\alpha_{n}^{\scriptscriptstyle{(1)}}\varepsilon_{\epsilon}^{\scriptscriptstyle{(2)}}
                - \alpha_{\epsilon}^{\scriptscriptstyle{(1)}}\varepsilon_{n}^{\scriptscriptstyle{(2)}}\right),
\end{split}
\end{gather}
with the various $\alpha$ and $\vartheta$ defined in terms of the conductivity 
$\sigma$ and other transport coefficients as
\begin{equation}
\sigma\alpha_{\scriptscriptstyle{\#}}^{\scriptscriptstyle{(i)}} =
        - \left(\kappa_{\scriptscriptstyle{\#}}^{\scriptscriptstyle{(i)}}
        - \frac{n}{h}\theta_{\scriptscriptstyle{\#}}^{\scriptscriptstyle{(i)}}\right),\qquad
\vartheta_{\scriptscriptstyle{\#}}^{\scriptscriptstyle{(i)}} =
        \frac{\beta_\epsilon\varepsilon_{\scriptscriptstyle{\#}}^{\scriptscriptstyle{(i)}}
+\beta_n\nu_{\scriptscriptstyle{\#}}^{\scriptscriptstyle{(i)}}}{h},
\end{equation}
where the symbol $\scriptstyle{\#}$ in the subscripts represents $s$, $\epsilon$, $n$
or $\scriptstyle{\Theta}$.

The foregoing dispersion relations involve the transport coefficients associated with the densities $\epsilon$
and $n$, in addition to those corresponding to $\sigma$ and $\Theta$. To express these relations in terms of the
coefficients associated with the thermodynamic variables $T$ and $\mu$, which are frequently utilized in gradient
expansions and feature in the list of structures in Section \ref{non-conformal}, it is essential to consider the
subsequent relationships:
\begin{align}
\delta T = & \left(\frac{\partial T}{\partial \epsilon}\right)_n \delta \epsilon +
\left(\frac{\partial T}{\partial n}\right)_\epsilon \delta n,\label{decompoe_dT2}\\
\delta \mu = & \left(\frac{\partial \mu}{\partial \epsilon}\right)_n \delta \epsilon +
\left(\frac{\partial \mu}{\partial n}\right)_\epsilon \delta n.\label{decompoe_dmu2}
\end{align}
Consequently, transport coefficients such as $\varphi_{\epsilon}^{\scriptscriptstyle{(i)}}$
and $\varphi_{n}^{\scriptscriptstyle{(i)}}$ can be expressed in terms of $\varphi_{\scriptscriptstyle{T}}^{\scriptscriptstyle{(i)}}$ and
$\varphi_{\mu}^{\scriptscriptstyle{(i)}}$ as follows:
\begin{align}
\varphi_{\epsilon}^{\scriptscriptstyle{(i)}} = & \left(\frac{\partial T}{\partial \epsilon}\right)_n
        \varphi_{\scriptscriptstyle{T}}^{\scriptscriptstyle{(i)}}
        + \left(\frac{\partial \mu}{\partial \epsilon}\right)_n \varphi_{\mu}^{\scriptscriptstyle{(i)}},
        \label{transport_fe}\\
\varphi_{n}^{\scriptscriptstyle{(i)}} = & \left(\frac{\partial T}{\partial n}\right)_\epsilon
        \varphi_{\scriptscriptstyle{T}}^{\scriptscriptstyle{(i)}}
        + \left(\frac{\partial \mu}{\partial n}\right)_\epsilon \varphi_{\mu}^{\scriptscriptstyle{(i)}},
        \label{transport_fn}
\end{align}
where $i=1$, 2 or 3, contingent upon the order of the expansion.

After extensive and intricate computations involving numerous combinations and simplifications, and taking into
account the definitions provided in Eqs.~\eqref{def_f_and_ell}, \eqref{shear_viscosity_relation}, \eqref{2order_frame_invariants}, and \eqref{frame_invariants_3order}, together with relations
\eqref{transport_fe} and \eqref{transport_fn}, we arrive at the following results:
\begin{equation}
\begin{gathered}
\mathscr{D}\left[v_{s}^{2}(\mathscr{A} - \theta_{s}^{\scriptscriptstyle{(2)}})
+ \mathscr{B}\right] + \mathscr{C} = - \mathscr{Y} + v_{s}^{2}\phi_n^{\scriptscriptstyle{(3)}},\\
\mathscr{B} = - \mathscr{X},\quad
v_{s}^{2}\left(\Gamma\mathscr{A}+\mathscr{E}\right)-\mathscr{D}\mathscr{B} = \mathscr{Y} + v_{s}^{2}\chi_s^{\scriptscriptstyle{(3)}}.
\end{gathered}
\end{equation}
These equations imply that the dispersion relations \eqref{rel_diffusion} and \eqref{rel_sound}
are equal to those presented in Section \ref{dispersion_relations}, specifically
Eqs.~\eqref{diffusion_dispersion_relation} and \eqref{soundwave_dispersion_relation}.

\bibliographystyle{ieeetr}
\bibliography{library.bib}

\end{document}